\documentclass[10pt]{article}
\usepackage{amsmath,latexsym,amsfonts,amssymb}
\usepackage{wrapfig}
\usepackage{amscd,array,epic,eepic,calc,bbm,float}
\usepackage{ifthen,psfrag,verbatim,epsfig,graphicx,enumerate,subfig}
\usepackage{makeidx}
\usepackage{amsthm}
\usepackage{lineno}
\usepackage{xcolor}
\usepackage{multirow}
\usepackage{natbib}
\usepackage{rotating}
\usepackage[titletoc,title]{appendix}

\def\bql{\begin{equation}\label}
\def\eql{\end{equation}\noindent}

\def\ban{\begin{align}}
\def\ean{\end{align}\noindent}

\def\brl{\begin{eqnarray}\label}
\def\erl{\end{eqnarray}\noindent}

\def\bro{\begin{eqnarray*}}
\def\ero{\end{eqnarray*}\noindent}

\def\brr{\begin{array}}
\def\err{\end{array}\noindent}

\def\bdl{\begin{display}\label}
\def\edl{\end{display}\noindent}

\def\bdo{\begin{display}}
\def\edo{\end{display}\noindent}

\def\bth{\begin{theorem}}
\def\eth{\end{theorem}}

\def\bcr{\begin{corollary}}
\def\ecr{\end{corollary}}

\def\bpr{\begin{proposition}}
\def\epr{\end{proposition}}

\def\bfa{\begin{fact}}
\def\efa{\end{fact}}

\def\blm{\begin{lemma}}
\def\elm{\end{lemma}}

\def\bdf{\begin{definition}}
\def\edf{\end{definition}}

\def\bas{\begin{assumptions}}
\def\eas{\end{assumptions}}

\def\bex{\begin{example}\rm}
\def\eex{\end{example}}

\def\bxx{\begin{exercise}\rm}
\def\exx{\end{exercise}}

\def\brm{\begin{remark}\rm}
\def\erm{\end{remark}}

\def\bma{\begin{pmatrix}}
\def\ema{\end{pmatrix}}

\def\bcs{\begin{cases}}
\def\ecs{\end{cases}}

\def\btb{\begin{center}\begin{tabular}}
\def\etb{\end{tabular}\end{center}}

\def\bit{\begin{itemize}}
\def\eit{\end{itemize}}

\def\bew{\par\noindent{\bf Proof }}
\def\qed{\quad\hfill\mbox{\P}}

\def\a{\alpha}
\def\b{\beta}
\def\c{\chi}

\def\e{\epsilon}

\def\g{\gamma}
\def\h{\eta}

\def\m{\mu}
\def\n{\nu}
\def\p{\pi}

\def\r{\rho}
\def\s{\sigma}
\def\t{\tau}

\def\w{\omega}
\def\x{\xi}

\def\FC{{\cal F}}

\def\GG{\Gamma}

\def\zerob{{\bf0}}

\DeclareMathOperator{\cov}{cov}

\DeclareMathOperator{\var}{var}
\DeclareMathOperator{\VaR}{VaR}
\DeclareMathOperator{\ebbS}{ES}

\def\ebb{{\mathbb E}}
\def\nbb{{\mathbb N}}

\def\rbb{{\mathbb R}}
\def\zbb{{\mathbb Z}}
\def\ind{{\mathbbm{1}}}

\def\inv{^{-1}}

\def\nf{\infty}

\def\ov{\overline}

\def\tod{\buildrel\rm d\over\rightarrow}

\def\nt{\noindent}

\def\theequation{\arabic{section}.\arabic{equation}}

\def\theequation{\arabic{section}.\arabic{equation}}

\newtheorem{theorem}{Theorem}
\newtheorem{lemma}[theorem]{Lemma}
\newtheorem{proposition}[theorem]{Proposition}
\newtheorem{corollary}[theorem]{Corollary}
\newtheorem{fact}[theorem]{Fact}
\theoremstyle{definition}
\newtheorem{definition}{Definition}
\newtheorem{example}{Example}
\newtheorem{exercise}{Exercise}

\theoremstyle{remark}
\newtheorem{remark}{Remark}

\usepackage{framed}
\usepackage{comment}
\definecolor{shadecolor}{gray}{0.9}
\specialcomment{extra}{\begin{shaded}}{\end{shaded}}
\newcommand{\diff}{\,\mathrm{d}}
\newtheorem{exmp}{Example}[section]

\usepackage{authblk}

\makeindex

\begin{document}
\title{Elicitability and backtesting: Perspectives for banking regulation}


\author[1]{Natalia Nolde}
\author[2]{Johanna F. Ziegel}
\affil[1]{Department of Statistics, University of British Columbia, Canada}
\affil[2]{Institute of Mathematical Statistics and Actuarial Science, University of Bern, Switzerland}

\renewcommand\Authands{ and }

\maketitle
\setcounter{page}{1}


\begin{abstract}

Conditional forecasts of risk measures play an important role in internal risk management of financial institutions as well as in regulatory capital calculations. In order to assess forecasting performance of a risk measurement procedure, risk measure forecasts are compared to the realized financial losses over a period of time and a statistical test of correctness of the procedure is conducted. This process is known as backtesting. Such traditional backtests are concerned with assessing some optimality property of a set of risk measure estimates. However, they are not suited to compare different risk estimation procedures. We investigate the proposal of comparative backtests, which are better suited for method comparisons on the basis of forecasting accuracy, but necessitate an elicitable risk measure. We argue that supplementing traditional backtests with comparative backtests will enhance the existing trading book regulatory framework for banks by providing the correct incentive for accuracy of risk measure forecasts. In addition, the comparative backtesting framework could be used by banks internally as well as by  researchers to guide selection of forecasting methods. The discussion focuses on three risk measures, Value-at-Risk, expected shortfall and expectiles, and is supported by a simulation study and data analysis.   

\nt{\bf Keywords}: forecasting, backtesting, elicitability, risk measurement procedure, Value-at-Risk, expected shortfall, expectiles 
\end{abstract}

\section{Introduction}

Financial institutions rely on conditional forecasts of risk measures for the purposes of internal risk management as well as regulatory capital calculations. The two ingredients at the heart of risk measurement are the choice of a suitable risk measure and of a forecasting method, with the forecasting method being typically preceded by the choice of a model and estimation method for the (conditional) loss distribution of the underlying portfolio of risky assets. Traditionally, the choice of a risk measure was based on theoretical considerations linked to practical implications. \citet{Emmer2013} give a recent account of the pros and cons of popular risk measures with an attempt to determine the best risk measure in practice. 
On the other hand, \citet{Cont2010} highlight the need to consider the entire ``risk measurement procedure", which includes not just the choice of a risk measure but also how it is then estimated from the data. In particular, the notion of robustness as sensitivity to outliers is used to compare several risk measurement procedures. In the risk management context, this should also be balanced with robustness to deviations from model assumptions as well as responsiveness or sensitivity to tail events. \citet{Davis2013} introduces a notion of consistency of risk measures and how this is relevant in the context of financial risk management. 

The performance of a (trading book) risk measurement procedure can be monitored over time via a comparison of realized losses with risk measure forecasts, a process known as backtesting; see, e.g., \citet{Christoffersen2003} and \citet{McNeil2005}. Based on results of a backtest, the risk measurement procedure is deemed as adequate or not. Traditional backtests perform a statistical test for the null hypothesis:
\[
H_0: \quad \text{``The risk measurement procedure is correct.''}
\]
If the null hypothesis is not rejected, the risk measurement procedure is considered as adequate. For Value-at-Risk (VaR), the \citet[p.~103--108]{BIS2013} has devised a three-zone approach based on a binomial test for the number of exceedances over the VaR threshold. Traditional backtests are concerned with assessing an optimality property of a set of risk measure estimates; for details see Section \ref{sec:calib}. They are not suited to \emph{compare} different risk estimation procedures, and they may be insensitive with respect to increasing information sets; examples of this fact are provided in \citet{HolzmannEulert2014,Davis2013}. Moreover, traditional backtests may not provide banks with the right incentive of developing procedures which aim for accuracy of risk measure forecasts; for an illustration, see Appendix A. In this simulation-based example, we show how optimization with respect to the test statistic of a traditional backtest may lead to unreasonable ordering of forecasting procedures.  

In view of the anticipated revised standardized approach, which ``should provide a credible fall-back in the event that a bank's internal market risk model is deemed inadequate'' \citep[p.~5--6]{BIS2013}, \citet{FisslerZiegelETAL2015} have recently proposed to replace traditional backtests by comparative backtests based on strictly consistent scoring functions. Comparative backtests also naturally lead to a three-zone approach, which will be described in detail in Section \ref{sec:dom}. Furthermore, they allow for conservative tests and are sensitive with respect to increasing information sets. Roughly, this means that a risk measurement procedure that correctly incorporates more risk factors will always be preferred over a simpler procedure that uses less information. However, comparative backtests necessitate an \emph{elicitable} risk measure. Examples of elicitable risk measures are VaR and expectiles, while expected shortfall (ES) is not elicitable. However, ES turns out to be jointly elicitable with VaR, which allows for comparative backtests also for ES; for details and a literature review on elicitable risk measures, see Section \ref{sec:prelim}. 

The paper raises the point of distinguishing between traditional backtesting (current regulatory practice) and comparative backtesting. We highlight the deficiency of the former in giving financial institutions the right incentive for forecast accuracy, and argue that the existing regulatory framework can be enhanced by inclusion of comparative backtesting. On the methodological side, we show that traditional backtesting can be formalized in the form of conditional calibration tests, which provide a unifying framework for many of the existing backtests of popular risk measures. This contributes to our understanding of those often ad hoc procedures and allows us to view them as part of a bigger picture. The paper then provides a detailed investigation of the proposal of comparative backtests. 

In our discussion of traditional and comparative backtests, we are focussing on the following three risk measures: VaR, a popular risk measure that is elicitable; expectiles, the only coherent and elicitable risk measures; and ES, a coherent and comonotonically additive risk measure, which is jointly elicitable together with VaR, and which is the new standard measure in banking regulation. VaR at level $\alpha \in (0,1)$, denoted $\VaR_\alpha$, of a random variable $X$ is defined as 
\[
\VaR_{\alpha}(X) = \inf\{x \;|\; F_X(x) \ge \alpha\},
\]
where $F_X$ is the cumulative distribution function of $X$. From the statistical perspective, $\VaR_\alpha$ is simply the $\alpha$-quantile of  the underlying distribution. Positive values of $X$ are interpreted as losses in this manuscript, hence we are interested in $\VaR_{\alpha}$ for values of $\alpha$ close to one. The \citet[p.103--108]{BIS2013} specifically requests $\VaR_\alpha$ values for $\alpha=0.99$, which we refer to as the standard Basel VaR level. 
ES of an integrable random variable $X$ at level $\nu \in (0,1)$ is given by
\[
\ebbS_{\nu}(X) = \frac{1}{1-\nu}\int_{\nu}^1 \VaR_\a(X) d\a.
\]
The \citet{BIS2014} proposes $\nu=0.975$ as the standard Basel ES level, as $\ebbS_{0.975}$ should yield a similar magnitude of  risk as $\VaR_{0.99}$ under the standard normal distribution. 
As introduced by \citet{NeweyPowell1987}, the $\tau$-expectile $e_{\tau}(X)$ of $X$ with finite mean is the unique solution $x = e_{\tau}(X)$ to the equation
\begin{equation}\label{eq:expectile}
\tau \int_x^{\infty} (y-x)\diff F_X(y) = (1-\tau) \int_{-\infty}^x (x-y) \diff F_X(y).
\end{equation}
As shown in \citet{BelliniKlarETAL2013,Ziegel2014}, $\tau$-expectiles are elicitable coherent risk measures for $\tau \in [1/2,1)$. Expectiles generalize the expectation just as quantiles generalize the median. Considering the level $\tau=0.99855$ leads to a comparable magnitude of risk as $\VaR_{0.99}$ and $\ebbS_{0.975}$ under the standard normal distribution; see \citet{BelliniBernardin2014}.


The paper is organized as follows. Section~\ref{sec:backtesting} contains the theoretical discussion of backtesting risk measures. In Section~\ref{sec:prelim} we define the notion of elicitability, introduce identifiability and review characterizations of consistent scoring functions for VaR, expectiles and (VaR, ES). In Section \ref{sec:calib} we define what we mean by a calibrated risk measurement procedure and describe how this concept is related to the notion of calibration of \citet{Davis2013} and to traditional backtests in general. We move on to comparative backtests in Section \ref{sec:dom}, where we also explain the comparative three-zone approach. Section~\ref{sec:choice} discusses the choice of the scoring function. Section~\ref{sec:numill} contains numerical illustrations of the proposed backtesting methodologies. We first review some of the existing approaches to forecasting risk measures in Section~\ref{sm}. A simulation study is described in Section~\ref{ssim}, while an application to the returns on the NASDAQ Composite index is presented in Section~\ref{sdat}. Section~\ref{scon} concludes the paper with a summary and a discussion of the findings, in particular, in relation to banking regulation. Appendix B contains the necessary background material for computing and estimation of expectiles, and gives a derivation of an extreme value-based estimator; some of the results here are of interest in their own right. Technical results on characterization of consistent scoring functions with positive-homogeneous score differences are delegated to Appendix C. Finally, Appendix D reports results of a simulation study, which investigates the performance of backtesting procedures in the setting where the out-of-sample size is small.  

\setcounter{equation}{0}
\section{Backtesting of risk measures}\label{sec:backtesting}

\subsection{Preliminaries}\label{sec:prelim}

A risk measure $\rho$ is usually defined on some space of random variables. If $\rho$ is law-invariant, it can alternatively be viewed as a map from some collection of probability distributions $\mathcal{P}$ to the real line $\mathbb{R}$. Law-invariance means that for two random variables $X$ and $Y$ that have the same distribution, we have $\rho(X) = \rho (Y)$. All risk measures considered in this manuscript are law-invariant. Therefore, we sometimes abuse notation and write $\rho(F)$ instead of $\rho(X)$, where $F$ is the distribution of $X$. Let $\Theta(X)=(\rho_1(X),\dots,\rho_k(X))$ be a vector of $k \ge 1$ risk measures.

\begin{definition}
A scoring function $S:\mathbb{R}^k\times \mathbb{R} \to \mathbb{R}$ is called \emph{strictly consistent} for $\Theta$ with respect to $\mathcal{P}$ if
\begin{equation}\label{eq:argmin}
\mathbb{E}(S(\Theta(X),X)) < \mathbb{E}(S(r,X))
\end{equation}
for all $r=(r_1,\dots,r_k)\not=\Theta(X)=(\rho_1(X),\dots,\rho_k(X))$ and all $X$ with distribution in $\mathcal{P}$. The scoring function $S$ is \emph{consistent} if equality is allowed in \eqref{eq:argmin}. The vector of risk measures $\Theta$ is called \emph{elicitable} with respect to $\mathcal{P}$ if there exists a strictly consistent scoring function for it. 
\end{definition}

Elicitability is useful for model selection, estimation, generalized regression, forecast ranking, and, as we will detail in this paper, allows for comparative backtesting. 
Elicitable functionals have already been studied in the thesis of \citet{Osband1985}, although the terminology has been coined by \citet{LambertPennockETAL2008}. A comprehensive literature review on elicitability can be found in \citet{Gneiting2011}, where particular emphasis is on the case $k=1$. Recent advances on the case $k \ge 2$ can be found in \citet{FrongilloKash2014,FisslerZiegel2015}.  

The question of elicitability of risk measures has recently received considerable attention. All available results in the case $k=1$ are based on the simple but powerful observation that a necessary requirement of elicitability are convex level sets in a distributional sense \citep{Osband1985}; see also \citet[Theorem 6]{Gneiting2011}. \citet{Weber2006} was the first to study risk measures with convex level sets. \citet{BelliniBignozzi2013} used his results to study elicitability for the broad class of monetary risk measures. Under weak regularity assumptions, they show that elicitable monetary risk measures are so-called shortfall risk measures \citep{FollmerSchied2002}. For more specific classes of risk measures, such as coherent, convex or distortion risk measures, the same result can be shown without any additional regularity assumptions \citep{Ziegel2014,DelbaenBelliniETAL2014,KouPeng2014,WangZiegel2014}. While expected shortfall is itself not elicitable, \citet{FisslerZiegel2015} have shown that the pair $\Theta=(\VaR_{\alpha},\ebbS_{\alpha})$ is elicitable; see also \citet{AcerbiSzekely2014}. 

The classes of (strictly) consistent scoring functions for $\VaR_{\alpha}$, $\tau$-expectiles and $(\VaR_\nu,\ebbS_\nu)$ have been characterized.  The following three propositions state sufficient conditions for (strict) consistency. Under mild regularity assumptions given in the cited literature and up to equivalence, these conditions are also necessary. Here, two scoring functions are called \emph{equivalent} if their difference is a function of the realization $x\in \mathbb{R}$ only. Let $\mathcal{P}_0$ denote the class of all Borel-probability distributions on $\rbb$, and let $\mathcal{P}_1\subseteq \mathcal{P}_0$ denote the class of all distributions with finite mean. 

\begin{proposition}[\citet{Thomson1979,Saerens2000}]
All scoring functions of the form 
\begin{equation}\label{qSVaR}
S(r,x) = (1 - \alpha - \ind\{x > r\})G(r) + \ind\{x > r\}G(x),
\end{equation}
where $G$ is an increasing function on $\rbb$, are consistent for $\VaR_{\alpha}$, $\alpha \in (0,1)$, with respect to $\mathcal{P}_0$.  The scoring functions of the above form are stricly consistent for $\VaR_{\alpha}$ with respect to $\mathcal{P}'\subseteq \mathcal{P}_0$ if $G$ is stricly increasing, $G(X)$ is integrable for all $X$ with distribution in $\mathcal{P}'$, and all distributions in $\mathcal{P}'$ have a unique $\alpha$-quantile.
\end{proposition}

\begin{proposition}[\citet{Gneiting2011}]
All scoring functions of the form 
\begin{equation}\label{qSexp}
S(r,x) = \ind\{x > r\}(1-2\tau)(\phi(r) -\phi(x) - \phi'(r)(r-x)) - (1-\tau)(\phi(r) - \phi'(r)(r-x)),
\end{equation}
where $\phi$ is a convex function with subgradient $\phi'$, are consistent for the $\tau$-expectile,  $\tau \in (0,1)$, with respect to $\mathcal{P}_1$. If $\phi$ is strictly convex, then the scoring functions of the above form are strictly consistent for the $\tau$-expectile relative to the class $\mathcal{P}' \subseteq \mathcal{P}_1$ such that $\phi(X)$ is integrable for all $X$ with distribution in $\mathcal{P}'$.
\end{proposition}

\begin{proposition}[\citet{FisslerZiegel2015}]
All scoring functions of the form 
\begin{equation}\label{qSVaRES}
S(r_1,r_2,x) = \ind\{x > r_1\}\big(-G_1(r_1) + G_1(x) - G_2(r_2)(r_1 - x)\big) + (1-\nu)\big(G_1(r_1) - G_2(r_2)(r_2 - r_1) + \mathcal{G}_2(r_2) \big),
\end{equation}
where $G_1$ is an increasing function, $\mathcal G_2' = G_2$ and $\mathcal G_2$ is increasing and concave, are consistent for $(\VaR_\nu,\ebbS_\nu)$, $\nu \in (0,1)$, with respect to $\mathcal{P}_1$. If $\mathcal G_2$ is strictly increasing and strictly concave, then the above scoring functions are strictly consistent with respect to the class $\mathcal{P}'\subseteq \mathcal{P}_1$ of distributions  which have unique $\nu$-quantiles and $G_1(X)$ is integrable for all $X$ with distribution in $\mathcal{P}'$. 
\end{proposition}

In risk management applications, it may be useful to allow only for strictly positive risk measure predictions. As shown in Section \ref{sec:choice}, this opens up the possibility for attractive choices of homogeneous scoring functions in the above propositions. If $r \in (0,\infty)$ is assumed in  \eqref{qSVaR} or \eqref{qSexp}, then, for strict consistency, we only need that $G$ or $\phi$ are defined on $(0,\infty)$, and that they are strictly increasing or strictly convex on this domain, respectively. In the case of \eqref{qSVaR} this can  be checked by a fairly straightforward computation. For the claim concerning \eqref{qSexp}, it is useful to use the decomposition of the score difference derived in the proof of \citet[Theorem 10]{Gneiting2011}. Furthermore, it is sufficient to require intergrability of $G(X)\ind\{X > 0\}$ or $\phi(X)\ind\{X > 0\}$ for all $X$ with distribution in $\mathcal{P}'$. If we restrict to predictions with $(r_1,r_2) \in \mathbb{R}\times (0,\infty)$ in \eqref{qSVaRES}, $\mathcal{G}_2$ only has to be defined on $(0,\infty)$ and has to be strictly increasing and strictly concave on this domain.

Closely connected to elicitability is the concept of identifiability. In fact, for $k=1$, identifiability implies elicitability under some additional assumptions; see \citet{SteinwartPasinETAL2014}. For $k \ge 2$, it is currently unclear whether such a general result holds; see \citet{FisslerZiegel2015}.
\begin{definition}
The vector of risk measures $\Theta$ is called \emph{identifiable} with respect to $\mathcal{P}$, if there is a function $V:\mathbb{R}^k \times \mathbb{R} \to \mathbb{R}^k$ such that
\[
\mathbb{E}(V(r,X)) = 0 \quad \Leftrightarrow \quad r = \Theta(X),
\]
for all $X$ with distribution in $\mathcal{P}$. 
\end{definition}

Identification functions are not uniquely defined. In fact, one can multiply any identification function for a functional by a function depending only on the prediction $r$ and taking values in the space of invertible $k\times k$-matrices to obtain another identification function for the same functional.

$\VaR_{\alpha}$ for $\alpha \in (0,1)$ is identifiable with respect to the class $\mathcal{P}_V \subset \mathcal{P}_0$ of distributions with unique quantiles with identification function
\begin{equation}\label{eq:VVaR}
V(r,x) = 1 - \alpha - \ind\{x > r\} ,
\end{equation}
the $\tau$-expectile for $\tau \in (0,1)$ is identifiable with respect to $\mathcal{P}_1$ using the identification function
\begin{equation}\label{eq:Vexp}
V(r,x) = |1- \tau - \ind\{x > r\} |(r-x),
\end{equation}
and $(\VaR_\nu,\ebbS_\nu)$ for the level $\nu \in (0,1)$ has identification function
\begin{equation}\label{eq:VVaRES}
V(r_1,r_2,x) = \begin{pmatrix} 1 - \nu - \ind\{x > r_1\} \\ r_1 - r_2 - \frac{1}{1-\nu}\ind\{x > r_1\}(r_1 - x)\end{pmatrix}
\end{equation}
with respect to $\mathcal{P}_1 \cap \mathcal{P}_V$. 


\subsection{Calibration and traditional backtests}\label{sec:calib}
 We fix the following notation. Suppose that $\Theta= (\rho_1,\dots,\rho_k)$ is an identifiable functional with identification function $V$ with respect to $\mathcal{P}$. Let $\{X_t\}_{t \in \nbb}$ be a series of negated log-returns adapted to the filtration $\mathcal{F} = \{\mathcal{F}_t\}_{t \in \nbb}$ and $\{R_t\}_{t \in \nbb}$ a sequence of predictions of $\Theta$, which are $\mathcal{F}_{t-1}$-measurable. Hence, the predictions are based on the information about $\{X_t\}_{t \in \nbb}$ available at time $t-1$ represented by the sigma-algebra $\mathcal{F}_{t-1}$. Let $\mathcal{L}(X_t|\mathcal{F}_{t-1})$ denote the conditional law of $X_t$ given the information $\mathcal{F}_{t-1}$. We assume that all conditional distributions $\mathcal{L}(X_t|\mathcal{F}_{t-1})$ and all unconditional distributions $\mathcal{L}(X_t)$ belong to $\mathcal{P}$ almost surely.

Inspired by the insightful paper of \citet{Davis2013}, we give the following definition. 
\begin{definition}
The sequence of predictions $\{R_t\}_{t \in \nbb}$ is \emph{calibrated for $\Theta$ on average}  if
\[
\mathbb{E}(V(R_t,X_t)) = 0 \quad \text{for all $t \in \nbb$;} 
\]
it is \emph{super-calibrated for $\Theta$ on average} if $\mathbb{E}(V(R_t,X_t)) \ge 0$ component-wise, for all $t \in \nbb$.
The sequence of predictions $\{R_t\}_{t \in \nbb}$ is \emph{conditionally calibrated for $\Theta$} if
\[
\mathbb{E}(V(R_t,X_t)|\mathcal{F}_{t-1}) = 0, \quad \text{almost surely, for all $t \in \nbb$;} 
\]
it is \emph{conditionally super-calibrated for $\Theta$} if $\mathbb{E}(V(R_t,X_t)|\mathcal{F}_{t-1}) \ge 0$ component-wise, almost surely, for all $t \in \nbb$. \emph{Sub-calibration} is defined analogously. 
\end{definition}

If one knows the conditional distributions $\mathcal{L}(X_t|\mathcal{F}_{t-1})$ and strives for the best possible prediction of $\Theta$ based on the information in $\mathcal{F}_{t-1}$, it is natural to use
\begin{equation}\label{eq:optcond}
\Theta(\mathcal{L}(X_t|\mathcal{F}_{t-1}))
\end{equation}
as a predictor, which we term the \emph{optimal $\mathcal{F}$-conditional forecast} for $\Theta$. For the same reason, we call $\Theta(X_t)=\Theta(\mathcal{L}(X_t))$ the \emph{optimal unconditional forecast}. Recall that we freely abuse notation in using $\Theta$ either as a functional defined on a space of random variables or on a space of probability distributions. 

Calibration characterizes optimal forecasts in the following sense. The optimal unconditional forecast is the only deterministic forecast that is calibrated for $\Theta$ on average. However, there may be other forecasts that are calibrated for $\Theta$ on average which are not deterministic and thus different from the optimal unconditional forecast. Likewise, the optimal conditional forecast is the only $\mathcal{F}$-predictable conditionally calibrated forecast for $\Theta$ up to almost sure equivalence. It is clear that conditional calibration implies calibration on average by the tower property of conditional expectations but the converse is generally false. The notions of calibration introduced here are analogous to the notions of cross-calibration for probabilistic forecasts introduced in \citet{StrahlZiegel2015}. 

We have introduced the notions of super- and sub-calibration as they can often be related to over- or under-estimation of the risk measure at hand. However, this depends on the specific identification function, so some care must be taken. We give details for a correct interpretation for VaR, expectiles and (VaR,ES) in Section~\ref{sec:ex}.

For simplicity, we focus on one-step ahead predictions in this paper. Clearly, multi-step ahead predictions are equally important. In some instances the same theory and concepts can be transferred from the former case to the latter. 

Following \citet{FisslerZiegelETAL2015}, we call any backtest that considers a null hypothesis of the type ``The risk measurement procedure is correct'' a \emph{traditional backtest}. Traditional backtests are similar to goodness-of-fit tests, that is, they allow to demonstrate that the risk measurement procedure under consideration is making incorrect predictions, if the respective null hypothesis can be rejected. Despite the somewhat misleading terminology that a traditional backtest is \emph{passed} if the null hypothesis is not rejected, this does \emph{not} mean that in this case, one can be sure that the null hypothesis is correct (with a pre-specified small probability of error) as this would necessitate that we control the power of the test explicitly. This can virtually never be done as the alternative is too broad; see also \citet[p.~103--105]{BIS2013}. As argued by \citet{FisslerZiegelETAL2015}, these issues may put the use of traditional backtest in regulatory frameworks in question. However, they may be useful for model verification just as goodness-of-fit tests have their established role in statistics.


Testing the null hypothesis
\begin{equation}\label{eq:tradH0A}
H_0: \quad \text{The sequence of predictions $\{R_t\}_{t \in \nbb}$ is calibrated for $\Theta$ on average.}
\end{equation}
amounts to performing a traditional backtest. We describe here how tests for average calibration can be constructed but we do not implement them because the stronger notion of conditional calibration appears more adequate in a dynamic risk management context. In our data example in Section \ref{sdat}, for the more flexible models, the null hypothesis of conditional calibration cannot be rejected which indicates that testing for average calibration is superfluous. However, there may be situations where achieving average calibration is already difficult and then the following tests may be useful. 

Given a series of observations $\{X_t\}_{t=1,\dots,n}$ and forecasts $\{R_t\}_{t=1,\dots,n}$, we define $\ov V_n :=  (1/n)\sum_{t=1}^n V(R_t,X_t)$. Let $\hat{\Sigma}_n$ be a heteroscedasticity and autocorrelation consistent (HAC) estimator of the asymptotic covariance matrix $\Sigma_n= \cov(\sqrt{n}\ \ov V_n)$ (see, e.g., \cite{Andrews1991}). Then, one can hope that $\sqrt{n}\ \hat{\Sigma}_n^{-1/2}\ \ov V_n$ is asymptotically standard normal under suitable assumptions on the identification function and the data generating process. For $k=1$, sufficient mixing assumptions are detailed in \citet[Theorem 4]{GiacominiWhite2006} but a multivariate generalization of this result remains to be worked out. \citet[Theorem 4]{GiacominiWhite2006} show that, for $k=1$, the test is consistent against the alternative $|\ebb\big(\ov V_n\big)| \ge \delta > 0$ for all $n$ sufficiently large, for any $\delta > 0$.

Conditional calibration is a stronger notion than average calibration, and it appears more natural in a dynamic risk management context. A traditional backtest for conditional calibration considers the null hypothesis
\begin{equation}\label{eq:tradH0C}
H_0: \quad \text{The sequence of predictions $\{R_t\}_{t \in \nbb}$ is conditionally calibrated for $\Theta$.}
\end{equation}
The requirement $\mathbb{E}(V(R_t,X_t)|\mathcal{F}_{t-1}) = 0$, almost surely, is equivalent to stating that $\mathbb{E}(h_t'V(R_t,X_t)) = 0$ for all $\mathcal{F}_{t-1}$-measurable $\mathbb{R}^k$-valued functions $h_t$. Following \citet{GiacominiWhite2006}, we consider an $\mathcal{F}$-predictable sequence $\{\mathbf{h}_t\}_{t\in\nbb}$ of $q\times k$-matrices $\mathbf{h}_t$ called \emph{test functions} to construct a Wald-type test statistic:
\begin{align}\label{qTh}
T_1=n \Big( \frac{1}{n} \sum_{t=1}^{n} \mathbf{h}_t V(R_t,X_t) \Big)'\ \widehat\Omega_n\inv \ \Big( \frac{1}{n} \sum_{t=1}^{n} \mathbf{h}_t V(R_t,X_t) \Big),
\end{align}
where 
\[
\widehat\Omega_n= \frac{1}{n}\sum_{t=1}^{n} (\mathbf{h}_t V(R_t,X_t)) (\mathbf{h}_t V(R_t,X_t))'
\]
is a consistent estimator of the variance of the $q$-vector $\mathbf{h}_t V(R_t,X_t)$. Ideally, the parameter $q$ should be chosen such that the rows of $\mathbf{h}_t$ generate $\mathcal{F}_{t-1}$. In applications, the choice of the test functions is motivated by the principle that they should represent the most important information available at time point $t-1$. In our simulation study, we obtained good results with $q=1$ or $q=2$; for further details see Section \ref{ssim2}. We call this type of traditional backtests \emph{conditional calibration tests}. In cases where $\mathbf{h}_t=1$, we refer to these tests as \emph{simple} conditional calibration tests. Theorem 1 in~\citet{GiacominiWhite2006} says that, under the null hypothesis~\eqref{eq:tradH0C}, $T_1\tod \c^2_q$ as $n\to\nf$, subject to certain assumptions on the data generating process $\{X_t\}_{t\in\nbb}$ and test function sequence $\{\mathbf{h}_t\}_{t\in\nbb}$. This asymptotic result justifies a level $\eta$ test which rejects $H_0$ when $T_1 > \c^2_{q,1-\eta}$, where $\c^2_{q,1-\eta}$ denotes the $1-\eta$ quantile of the $\c^2_q$ distribution. \citet[Theorem 3]{GiacominiWhite2006} provide conditions such that $T_1\tod \c^2_q$ as $n\to\nf$ for multi-step ahead predictions, while Theorem 2 of \citet{GiacominiWhite2006} considers consistency of the test against global alternatives. The theorems of \citet{GiacominiWhite2006} are formulated in terms of score differences and not identification functions but their proofs solely rely on the martingale difference property of $\mathbf{h}_t V(R_t,X_t)$ and can thus be applied in our context.

Commonly used backtests for $\VaR_\alpha$ and $\ebbS_\nu$ are closely related to conditional calibration tests for specific choices of the test functions $\mathbf{h}_t$. In fact, choosing $\mathbf{h}_t = 1$ in the case of $\VaR_\alpha$, the conditional calibration test for $\VaR_{\alpha}$ is closely related to the standard backtest for $\VaR_\alpha$ based on the number of VaR exceedances \citep[p.103--108]{BIS2013}. In the case of $\ebbS_\nu$, the conditional calibration test for $(\VaR_\nu,\ebbS_\nu)$ is related to the backtest for $\ebbS_\nu$ of \citet{McNeilFrey2000} based on exceedance residuals. We give further details in Examples \ref{ex:tradVaR}, \ref{ex:tradES}, and \ref{ex:tradexp} below. 

The notion of a calibrated risk measure (or statistic) of \citet{Davis2013} is closely related to our notion of a calibrated sequence of predictions. \citet{Davis2013} considers which risk measures are calibrated for which classes of models. That is, he attempts to characterize the largest class of data generating processes such that $\ov V_n$ goes to zero a.s. as $n \to \infty$ if $\{R_t\}_{ t\in\mathbb{N}}$ is a sequence of optimal conditional forecasts for the risk measure. It turns out that for quantiles only minimal assumptions are necessary, whereas assumptions need to be stronger to work with the mean, for example. The focus of our work is more statistical. Choosing $\mathcal{F}$-predictable test functions $\mathbf{h}_t$ encoding the available information at time point $t-1$, we investigate whether and how it is possible to test in finite samples that the sequence $\{R_t\}_{t \in \nbb}$ is conditionally calibrated.

\subsubsection{One-sided calibration tests}\label{sost}

In certain situations, it may be meaningful to assess super- or sub-calibration. For example, the standard backtest for $\VaR_{\alpha}$ described in \citep[p.103--108]{BIS2013} is a test for conditional super-calibration. This is due to the fact that over-estimation of $\VaR_\alpha$ is not a problem as far as the regulator is concerned. Holding more capital then minimally required should always be allowed. 

Suppose we wish to test the hypothesis of conditional super-calibration that $\ebb[V(R_t,X_t)|\FC_{t-1}]\ge \zerob$ component-wise, for all $t$. That is, in the case of a $k$-variate risk measure, we are interested in $H_0 = \bigcap_{i=1}^k H_{0,i}$, where $${\rm H}_{0,i}:\ \ebb[V_i(R_t,X_t)|\FC_{t-1}]\ge 0\quad {\rm for\ all\ } t,\qquad i=1,\ldots,k. $$
For each component $i$ of the risk measure, let  $\mathbf{h}_{i,t}=(h_{i,t,1},\dots,h_{i,t,q_i})$ be an $\FC_{t-1}$-measurable $(q_i\times 1)$-vector of non-negative test functions.  If $h_{i,t,1},\dots,h_{i,t,q_i}$ generate $\FC_{t-1}$ then $H_{0,i} = \bigcap_{\ell=1}^{q_i} H_{0,i,\ell}$, where $${\rm H}_{0,i,\ell}:\ \ebb[V_i(R_t,X_t) h_{i,t,\ell}]\ge 0\quad {\rm for\ all\ } t,\quad i=1,\ldots,k,\quad \ell=1,\ldots,q_i.  $$
We combine all of the test functions into a $(q\times k)$ matrix ${\bf h}_t$ with $q=\sum_{i=1}^k q_i$, which has the following structure:
$${\bf h}_t = \bma \mathbf{h}_{1,t} & 0 & \cdots & 0 \\ 0 & \mathbf{h}_{2,t} & \cdots & 0\\ \vdots & \vdots & \ddots & \vdots \\ 0 & 0 & \cdots & \mathbf{h}_{k,t}\ema.$$
Setting $Z_t = {\bf h}_t V(R_t,X_t)$, the above hypothesis of conditional super-calibration can alternatively be expressed as $H_0 = \bigcap_{m=1}^q H_{0,m}$ with $H_{0,m}:\ \ebb(Z_{t,m})\ge0$ for all $t$. $m=1,\ldots,q$.

From the proof of \citet[Theorems 1 and 3]{GiacominiWhite2006} it follows that under $H_0$ given at \eqref{eq:tradH0C}, 
\begin{equation}\label{eq:tau4}
T_2=(T_{2,1},\dots,T_{2,q})'=\sqrt{n}^{-1}\widehat{\Omega}_n^{-1/2} \sum_{t=1}^{n}Z_t \tod \mathcal{N}(\zerob,I_q),\qquad n\to\nf,
\end{equation} where $I_q$ denotes the $(q\times q)$ identity matrix. Hence, we can obtain an asymptotic test for $H_{0,m}$ with the p-value given by $\p_m = \Phi\Big(\sqrt{n}^{-1} (\widehat{\Omega}_n)_{mm}^{-1/2} \sum_{t=1}^{n}Z_{t,m}\Big)$, $m=1,\ldots,q$. That is, $\pi_m$ is the (asymptotic) probability of obtaining a more extreme outcome than the one observed, assuming the null hypothesis $H_{0,m}$ is true. Let $\p_{(1)},\ldots,\p_{(q)}$ be the ordered p-values. The classical Bonferroni multiple test procedure rejects the global null hypothesis $H_0$ if the smallest of the p-values $\pi_{(1)}<\eta/q$, where $\eta$ is the desired level of the (global) test. As an alternative, following \citet{Hommel1983}, we obtain a level $\eta$ test by rejecting the global hypothesis $H_0$ if for at least one $m$ we have 
\begin{equation}\label{qHommel}
\p_{(m)} \le \dfrac{m\ \eta}{q\ C_q},\qquad C_q = \sum_{r=1}^q 1/r,\qquad m=1,\ldots,q.
\end{equation}
Hommel's rejection rule has the advantage of allowing to detect situations with both small effects in many components and with large effects in few components. Other testing procedures in this context could also be used.



\subsubsection{Examples}\label{sec:ex}

\begin{example}\label{ex:tradVaR}

\citet{Christoffersen1998} calls a sequence of $\VaR_{\alpha}$ forecasts efficient with respect to $\mathcal{F}$ if 
\[
\ebb[\ind\{X_t > R_t\}|\mathcal{F}_{t-1}] = 1-\alpha, \quad \text{almost surely, $t=1,2,\dots$}.
\]
This requirement is the same as the one of conditional calibration of $\{R_t\}_{t \in \nbb}$ by \eqref{eq:VVaR}. In fact, the dynamic quantile test of \citet{Kuester2006} (see also \citet{Christoffersen1998,EngleManganell2004}) has similarities to a conditional calibration test. In analogy to their test, it is natural to  consider test functions
\[
\mathbf{h}_t = (1,V(r_{t-1},x_{t-1}),\cdots,V(r_{t-p},x_{t-p}),r_t)'
\]
for $p \ge 1$. This is also in line with the suggestion in \citet{GiacominiWhite2006}, who use $\mathbf{h}_t = (1,V(r_{t-1},x_{t-1}))'$.

The standard backtest for $\VaR_\alpha$ specified in the Basel documents \citep[p.103--108]{BIS2013} uses the test statistic
\[
\beta = \sum_{t=1}^{n} \ind\{X_t > R_t\},
\]
which is the number of exceedances over the estimated $\VaR_\alpha$, denoted $R_t$, for time point $t$. Under the null hypothesis \eqref{eq:tradH0C} of conditionally calibrated $\VaR_\alpha$-forecasts, for one-step ahead forecasts, $\beta$ is a binomial random variable with parameters $n$ and $1-\alpha$; see \citet{Rosenblat1952,DieboldGuntherETAL1998,Davis2013}. It is remarkable that this result holds under essentially no assumptions on $\{X_t\}_{t \in \nbb}$ or $\{R_t\}_{t \in \nbb}$. However, when moving away from one-step ahead forecasts to multi-step ahead forecasts, things become more intricate and one has to resort to general limit theorems such as presented above for testing if $\beta$ has mean $n(1-\alpha)$. This test is a test for conditional super-calibration with $\mathbf{h}_t = 1$, because for $\VaR_\alpha$, we obtain using \eqref{eq:VVaR}
\[
T_3:=\sum_{t=1}^n \mathbf{h}_t V(R_t,X_t) = \sum_{t=1}^{n}(\ind\{X_t \le R_t\} - \alpha) = \sum_{t=1}^n (\ind\{X_t > R_t\}-  (1- \alpha))=-(\beta - n(1-\alpha)),
\]
thus, testing the null hypothesis that $\beta$ has mean less or equal to $n(1-\alpha)$ is equivalent to testing that $T_3$ has mean greater or equal to zero. This null hypothesis says that the conditional VaR predictions are at least as large as the true conditional VaR. Assuming that it is an incentive of a bank to state VaR estimates that tend to be lower than the true ones, a more prudent null hypothesis from the viewpoint of a regulator would be the opposite one-sided hypothesis that the conditional VaR predictions are at most as large as the true conditional VaR, that is, a test for conditional sub-calibration.

For one-step ahead predictions, alternatively to theory presented in this section, one can exploit the fact that the exceedance indicators $\ind\{X_t > R_t\}$, $t=1,\dots,n$ at the boundary of the null hypothesis, are independent Bernoulli random variables with success probability $1-\alpha$, which allows for an exact test rather than an asymptotic one.
\end{example}

\begin{example}\label{ex:tradES} We consider the vector of risk measures $\Theta(X) = (\rho_1(X),\rho_2(X)) = (\VaR_{\nu}(X),\ebbS_{\nu}(X))$ for some $\nu \in (0,1)$. Let $r_{1,t}$ and $r_{2,t}$ denote forecasts of $\VaR_\nu(X_t)$ and $\ebbS_{\nu}(X_t)$, respectively. 
Assuming $X_t = \m_t +\s_t Z_t$, where $\m_t$ and $\s_t$ are $\FC_{t-1}$-measurable and the $Z_t$'s form an independent and identically distributed (i.i.d.)~sequence of random variables with zero mean and variance one, for backtesting ES, \citet{McNeilFrey2000} introduced the following test statistic based on exceedance residuals:
\begin{equation}\label{eq:tau1}
T_4=\frac{1}{\#\{t: X_t > r_{1,t}\}}\sum_{t=1}^n \frac{X_t - r_{2,t}}{\sigma_t} \ind\{X_t > r_{1,t}\}.
\end{equation}
It turns out that the ES backtest of \citet{McNeilFrey2000} is closely related to a conditional calibration test as follows.  For $n$ reasonably large, we have that $\#\{t: x_t > r_{1,t}\}/n \approx 1-\nu$. Therefore, for the test statistic $T_4$ in~\eqref{eq:tau1}, we obtain
\[
T_4 \approx \frac{1}{n}\sum_{t=1}^n \frac{1}{1-\nu}\frac{x_t - r_{2,t}}{\sigma_t} \ind\{x_t > r_{1,t}\} = \frac{1}{n}\sum_{t=1}^n \mathbf{h}_t V(r_{1,t},r_{2,t},x_t)\]
with 
$\mathbf{h}_t = \sigma_t^{-1}((r_{2,t} - r_{1,t})/(1-\nu),1)$. Replacing $\sigma_t$ by an estimate $\hat{\sigma}_t$ is natural when considering the test of \citet{McNeilFrey2000} as a conditional calibration test. The estimated volatility $\hat{\sigma}_t$ is then simply a part of the $\mathcal{F}_{t-1}$-measurable test function sequence $\{\mathbf{h}_t\}_{t \in \mathbb{N}}$ that supposedly encodes the relevant information of $\mathcal{F}_{t-1}$. Of course, this test is only reasonable if $\sigma_t$ is estimated as part of the forecasting model with the information at time point $t-1$. The recently proposed backtests for ES of \citet{AcerbiSzekely2014} are in the same spirit as the test of \citet{McNeilFrey2000}.

The backtest for ES suggested by \citet{CostanzinCurran2015} tests if the whole tail of the distribution beyond the $\VaR_\nu$-level has been estimated correctly. Strictly speaking, the test is therefore not a test for the accuracy of a sequence of point forecasts for $(\VaR_\nu,\ebbS_\nu)$ but rather a test for the accuracy of a sequence of probabilistic forecasts for tomorrow's loss distribution with emphasis on the left tail. Other tests in this spirit but of comparative type can be found in \citet{GneitingRanjan2011a}.

As ES is only identifiable jointly with VaR, one has to be careful when formulating a one-sided test for ES. Let $(r_1^*,r_2^*) = \Theta(X)$. Then it holds for all $(r_1,r_2)$ that
\[
r_2^* - r_2 \le \mathbb{E}V_2(r_1,r_2,X) \le r_2^* - r_2 + \frac{\nu - F(r_1)}{1-\nu}(r_1^* - r_1).
\]
This shows that, similarly to the VaR case, testing the null hypothesis of sub-calibration for the ES component $\ebb V_2(r_1,r_2,X)\le 0$ is equivalent to testing that $r_2^*\le r_2$. 
Hence, the test of conditional sub-calibration of (VaR, ES) is a test that the conditional VaR and ES predictions are at least as large as their optimal conditional predictions. The Hommel's procedure described in Section~\ref{sost}
can then be applied with p-value $\p_m = 1-\Phi(T_{2,m})$, where the $T_{2,m}$'s are defined in \eqref{eq:tau4}.
\end{example}

\begin{example}\label{ex:tradexp}
One could conceive a backtesting framework for expectiles as well, in a similar spirit to the ES backtesting procedure proposed by \citet{McNeilFrey2000}. Assuming, as in the example above, that $X_t = \m_t +\s_t Z_t$, where $\m_t$ and $\s_t$ are $\FC_{t-1}$-measurable and the $Z_t$'s are i.i.d.~with zero mean and variance one, the conditional $\t$-expectile satisfies 
\[
e_\t(X_t\mid \FC_{t-1}) = \m_t + \s_t e_\t(Z_t)
\]
and we see that the residuals 
\[
\dfrac{X_t - e_\t(X_t\mid \FC_{t-1})}{\s_t} = Z_t - e_\t(Z_t)
\]
form an i.i.d.~sequence of random variables with zero $\t$-expectile. This implies that $V(e_\tau(Z_t),Z_t)$ with $V$ given at \eqref{eq:Vexp} is an i.i.d.~sequence of random variables with mean zero, which can be tested using a bootstrap (as in \citet{EfronTibshirany1993}, Section~16.4). Here it is necessary to replace the true volatility $\sigma_t$ by an estimate. This is analogous to the suggestion of \citet{McNeilFrey2000} for ES. Noticing that the identification function for expectiles at \eqref{eq:Vexp} is positively 1-homogeneous, we obtain that
\[
\mathbb{E}V(e_\tau(Z_t),Z_t) = \mathbb{E}V(e_\tau(X_t),X_t) \sigma_t^{-1} = 0.
\]
This equality suggests that it is natural to perform a conditional calibration test for expectiles with test function $\mathbf{h}_t = \hat{\sigma}_t^{-1}$ and test statistic $T_1$ given at \eqref{qTh}. This yields a valid asymptotic test under the assumptions in \citet[Theorem 1]{GiacominiWhite2006}. These assumptions are weaker than the model assumption $X_t = \m_t +\s_t Z_t$.

In the case of expectiles, as in the case of VaR, a test for conditional super-calibration assesses the null-hypothesis that all conditional expectile estimates are at least as large as the true conditional expectile. 
\end{example}

\subsection{Elicitability, forecast dominance and comparative backtests}\label{sec:dom}

Suppose now that the functional $\Theta= (\rho_1,\dots,\rho_k)$ is elicitable with respect to $\mathcal{P}$. Let $\{X_t\}_{t \in \nbb}$ be a series of negated log-returns adapted to the filtration $\mathcal{F} = \{\mathcal{F}_t\}_{t \in \nbb}$ as well as to the filtration $\mathcal{F}^* = \{\mathcal{F}^*_t\}_{t \in \nbb}$. Let $\{R_t\}_{t \in \nbb}$ and $\{R_t^*\}_{t \in \nbb}$ be two sequences of predictions of $\Theta $, which are $\mathcal{F}$ and $\mathcal{F}^*$-predictable, respectively. We assume that all conditional distributions $\mathcal{L}(X_t|\mathcal{F}_{t-1})$, $\mathcal{L}(X_t|\mathcal{F}_{t-1}^*)$ and all unconditional distributions $\mathcal{L}(X_t)$ belong to $\mathcal{P}$ almost surely. We refer to the predictions $\{R_t^*\}_{t \in \nbb}$ as the standard procedure, while $\{R_t\}_{t \in \nbb}$ is the internal model. The two filtrations $\mathcal{F}$ and $\mathcal{F}^*$ acknowledge the fact that the internal model and the standard model may be based on different information sets. For example, one model may include more risk factors than the other, or, certain expert opinion may be used to adjust one model but not the other. 

\begin{definition}\label{def:dom}
Let $S$ be a consistent scoring function for $\Theta$  with respect to $\mathcal{P}$. Then, $\{R_t\}_{t \in \nbb}$ \emph{$S$-dominates} $\{R_t^*\}_{t \in \nbb}$ \emph{(on average)} if
\[
\mathbb{E}(S(R_t,X_t)- S(R_t^*,X_t)) \le 0, \quad \text{for all $t \in \nbb$.} 
\]
Furthermore, $\{R_t\}_{t \in \nbb}$ \emph{conditionally $S$-dominates} $\{R_t^*\}_{t \in \nbb}$ if
\begin{equation}\label{eq:conddom}
\mathbb{E}(S(R_t,X_t)- S(R_t^*,X_t)|\mathcal{F}^*_{t-1}) \le 0, \quad \text{almost surely, for all $t \in \nbb$.} 
\end{equation}
\end{definition}

The definition of conditional dominance is asymmetric in terms of the role of the standard procedure and the internal procedure. The standard procedure and the information $\mathcal{F}^*$ it is based on are considered as a benchmark of predictive ability, which is why we condition on $\mathcal{F}_{t-1}^*$ and not on $\mathcal{F}_{t-1}$. Any method that dominates the benchmark has superior predictive ability relative to this benchmark.

Clearly, conditional $S$-dominance implies $S$-dominance on average. \citet[Definition 2]{EhmGneitingETAL2015} introduced the notion of dominance of one sequence of predictions over the other if one $S$-dominates the other on average for all consistent scoring functions $S$ for $\Theta$. The notion of dominance is a strong one. That is, in the data examples of \citet{EhmGneitingETAL2015} it was almost never observed that one forecast dominates the other. This makes the concept difficult to employ in an applied decision making context. Furthermore, currently, a clear theoretical understanding of the notion of dominance remains elusive.

There are several reasons why the predictions $\{R_t\}_{t \in \nbb}$ should be preferred over $\{R_t^*\}_{t \in \nbb}$ if the former dominates the latter. Firstly, comparison of forecasts with respect to the described dominance relations is consistent with respect to increasing information sets. That is, if $\mathcal{F}^*_t \subseteq \mathcal{F}_t$ for all $t$ and $\{R_t\}_{t \in \nbb}$, $\{R_t^*\}_{t \in \nbb}$ are the optimal conditional forecasts with repect to their filtrations as defined at \eqref{eq:optcond}, then the internal procedure dominates the standard procedure, both, conditionally and on average \citep[Theorem 1]{HolzmannEulert2014}. The same is true if $\{R_t\}_{t \in \nbb}$ is $\mathcal{F}^*$-conditionally optimal and $\{R_t^*\}_{t \in \nbb}$ is just $\mathcal{F}^*$-predictable \citep[Corollary 2]{HolzmannEulert2014}; see also \citet{Tsyplakov2014}.

Secondly, in the case $k=1$, for most important functionals, including VaR and expectiles, strictly consistent scoring functions are \emph{order sensitive} or \emph{accuracy rewarding} in the following sense. Essentially, if $\Theta(X) < r < r^*$ or $r^* < r < \Theta(X)$ for some random variable $X$, then
\begin{equation}\label{eq:ordersens}
\mathbb{E} (S(\Theta(X),X)) < \mathbb{E} (S(r, X)) < \mathbb{E}(S(r^*,X));
\end{equation}
see \citet{Nau1985,Lambert2012} for details. Therefore, if the risk measure forecasts $\{R_t\}_{t \in \nbb}$ are always closer than $\{R_t^*\}_{t \in \nbb}$ to the optimal $\mathcal{F}^*$-conditional forecast, that is, $\Theta(\mathcal{L}(X_t|\mathcal{F}_t^*)) < R_t < R_t^*$ or $\Theta(\mathcal{L}(X_t|\mathcal{F}_t^*)) > R_t > R_t^*$ for all $t \in \nbb$ almost surely, then $\{R_t\}_{t \in \nbb}$ conditionally dominates $\{R_t^*\}_{t \in \nbb}$. There are different proposals for notions of order sensitivity in the case $k \ge 2$; see, for example, \citet{LambertPennockETAL2008}, but the situation is less clear in this case. 

The condition for conditional $S$-dominance in \eqref{eq:conddom} can be formulated equivalently as
\[
\mathbb{E}((S(R_t,X_t)- S(R_t^*,X_t))h_t) \le 0, \quad \text{for all $h_t \ge 0$, $\mathcal{F}^*_{t-1}$-measureable,}
\]
for all $t \in \nbb$. It is tempting to work with a vector $\mathbf{h}_t$ of $\mathcal{F}^*$-predictable test functions in order to test for conditional $S$-dominance as suggested in the conditional calibration tests. However, we are interested in comparing the standard procedure to the internal procedure and reach a definite answer as to which one is to be preferred. If $\mathbb{E}((S(R_t,X_t)- S(R_t^*,X_t))\mathbf{h}_{t,i}) >  0$ but $\mathbb{E}((S(R_t,X_t)- S(R_t^*,X_t))\mathbf{h}_{t,j}) < 0$ for different components $\mathbf{h}_{t,i}, \mathbf{h}_{t,j}$ of the vector $\mathbf{h}_t$,  no clear preference for either method can be given. Therefore, we do not pursue this approach further.

In comparative backtesting we are interested in the null hypotheses
\begin{align*}
H_0^-:& \text{ The internal model predicts at least as well as the standard model.}\\
H_0^+:& \text{ The internal model predicts at most as well as the standard model.}
\end{align*}
The null hypothesis $H_0^-$ is analogous to the null hypothesis of a correct model and estimation procedure but now adapted to a comparative setting. As mentioned in the introduction, considering a backtest as passed if the null hypothesis cannot be rejected is anti-conservative or aggressive in nature and may therefore be problematic in regulatory practice. On the other hand, the null hypothesis $H_0^+$ is such that the comparative backtest is passed if we can reject $H_0^+$. This means that we can explicitly control the type {\rm I} error of allowing an inferior internal model over an established standard model. 

We assume in the remainder of the paper that the  limit
\begin{equation}\label{eq:asy}
\lambda:= \lim_{n \to \infty} \frac{1}{n}\sum_{t=1}^n \mathbb{E}(S(R_t,X_t) - S(R_t^*,X_t)) \in [-\infty,+\infty]
\end{equation}
exists (while we allow it to take the values $\pm \infty$). It is clear that $S$-dominance on average implies $\lambda\le 0$. If the sequence of score differences $\{S(R_t,X_t) - S(R_t^*,X_t)\}_{t \in \nbb}$ is first-order stationary, then $\lambda \le 0$ implies $S$-dominance on average. Under Assumption \eqref{eq:asy}, we can compare any two sequences of risk measure estimates with respect to their predictive performance. It is a weak assumption as it requires slightly more than that the average expected score differences are eventually of the same sign. It may be weakened at the cost of further technicalities which we have chosen to avoid. If the limit $\lambda$ in~\eqref{eq:asy} is non-positive, then the internal procedure is \emph{at least as good} as the standard procedure, whereas the internal procedure \emph{predicts at most as well} as the standard procedure if $\lambda \ge 0$. Ordering risk measurement procedures is a compromise in the quest for conditional dominance. On the one hand, it is clearly a weaker notion than conditional dominance, but on the other hand it introduces a meaningful total order on all risk measurement procedures given a sensible choice of the scoring function~$S$; see Section \ref{sec:choice}.  

Therefore, we reformulate our comparative backtesting hypotheses as
\begin{align*}
H_0^-:&\; \lambda \le 0\\
H_0^+:&\; \lambda \ge 0.
\end{align*}
The test statistic
\[
\Delta_n\ov S := \frac{1}{n}\sum_{t=1}^n (S(R_t,X_t) - S(R_t^*,X_t)), 
\]
for $n$ large enough, has expected value less or equal to zero under $H_0^-$, whereas under $H_0^+$ its expectation is non-negative. Tests of  $H_0^+$ or $H_0^-$ based on a suitably rescaled version of $\Delta_n\ov S$ are so-called \emph{Diebold-Mariano tests}; see \citet{DieboldMariano1995}. Under certain mixing assumptions detailed in \citet[Theorem 4]{GiacominiWhite2006},
\[\frac{\Delta_n\ov S - \mathbb{E}(\Delta_n\ov S)}{\hat{\sigma}_n/\sqrt{n}}
\]
is asymptotically standard normal with $\hat{\sigma}_n^2$ an HAC estimator of the asymptotic variance, $\sigma_n^2 = \var(\sqrt{n}\Delta_n \ov S)$.  Therefore, using the test statistic
\begin{equation}\label{eq:DMstat}
T_4 = \frac{\Delta_n \ov S}{\hat{\sigma}_n/\sqrt{n}},
\end{equation}
we obtain an asymptotic level $\eta$-test of $H_0^+$ if we reject the null hypothesis when $\Phi(T_4)\le \eta$, and of $H_0^-$ if we reject the null hypothesis when $1-\Phi(T_4)\le \eta$.

Based on the outcome of the tests of $H_0^+$ and $H_0^-$, \citet{FisslerZiegelETAL2015} suggest the following three-zone approach. We fix a significance level $\eta \in (0,1)$, for example, $\eta=0.05$. If $H_0^-$ is rejected at level $\eta$, then   $H_0^+$ will not be rejected at level $\eta$. Similarly, if $H_0^+$ is rejected at level $\eta$, then $H_0^-$ will not be rejected at level $\eta$. Therefore, we say that the internal procedure is in the red region, that is, it fails the comparative backtest if $H_0^-$ is rejected. The internal procedure is in the green region, that is, it passes the backtest, if $H_0^+$ is rejected. The internal procedure needs further investigation, that is, it falls in the yellow region, if neither $H_0^+$, nor $H_0^-$ can be rejected. For an illustration of these decisions, see \citet[Figure 1]{FisslerZiegelETAL2015}. 

There is one important difference between the three-zone approach described in \citet[p.103--108]{BIS2013} for traditional VaR backtests and the three-zone approach of \citet{FisslerZiegelETAL2015} described here. In the former approach, the zones arise from varying the confidence level of the hypothesis test, whereas in the latter approach the confidence level is fixed a priori, and the zones arise to separate cases where there is enough evidence to clearly decide for superiority of one procedure over the other in contrast to cases where there is no clear evidence.

\subsubsection{Choice of the scoring function}\label{sec:choice}

Based on \eqref{qSVaR}, \eqref{qSexp} and \eqref{qSVaRES}, one has a large number of choices for strictly consistent scoring functions for VaR, expectiles and (VaR, ES). In the case of $\VaR_\alpha$, the standard choice is to take $G(r) = r$ in \eqref{qSVaR} leading to the classical asymmetric piecewise linear loss, see \eqref{eq:SVaR1} below, also known as linlin, hinge, tick or pinball loss; see \citet{Koenker2005} for its relevance in quantile regression. In the case of expectiles, one could argue that a natural choice is taking $\phi(r)=r^2$ in~\eqref{qSexp}, which simplifies to the squared error function for the mean (up to equivalence). This is also the scoring function suggested by \citet{NeweyPowell1987} for expectile regression. Consistent scoring functions for (VaR, ES) have only recently been discovered; see \citet{AcerbiSzekely2014,FisslerZiegel2015}. Therefore, there is no natural classical choice for the functions $G_1$, $G_2$ in \eqref{qSVaRES}.

A scoring function $S$ is called \emph{positive homogeneous} of degree $b$ (or \emph{$b$-homogeneous}) if for all $r = (r_1,\dots,r_k)$ and all $x$
\[
S(c r, cx) = c^b S(r,x), \quad \text{for all $c > 0$.} 
\]
\citet{Efron1991} argues that it is a crucial property of a scoring function to be positive homogeneous in estimation problems such as regression. \citet{Patton2011} underlines the importance of positive homogeneity of the scoring function for forecast ranking. Positive homogeneous scoring functions are also favorable because they are so-called ``unit consistent''; see, for example, \citet{AcerbiSzekely2014}.  That is, if $r$ and $x$ are given in say U.S.~dollars with $r = \$10$ and $s=\$5$, then, for a positive homogeneous scoring function $S$, the score $S(r,x)=S(\$10,\$5) = (\$)^b S(10,5)$ will have unit $(\text{U.S. dollars})^b$. In particular, changing the units, from, say, U.S.~dollars to million U.S.~dollars, will not change the ordering of forecasts assessed by this scoring function, and will thus also leave the results of comparative backtests unchanged. Concerning the choice of the degree $b$ of homogeneity, \citet{Patton2006} shows that in the case of volatility forecasts, $b=0$ requires weaker moment conditions than a larger choice of $b$ for the validity of Diebold-Mariano tests which are used in comparative backtesting. Concerning the power of Diebold-Mariano tests, \citet{PattonSheppard2009} find the best overall power for volatility forecasts for the choice $b=0$.  

Appendix C presents results, which characterize positive homogeneous scoring functions for the risk measures that are of interest in this paper. Note that we only allow for predictions $r > 0$ or $r = (r_1,r_2)$ with $r_2 > 0$. As we are interested in risk measures for losses, this is not a real restriction; see also Section \ref{ssim}.

For some orders of homogeneity $b$, there is no strictly consistent scoring function for the risk measures of interest in this paper. In particular, the attractive choice $b=0$ can often not be realized. However, for comparative backtesting we are not interested in absolute values of expected scores but only in \emph{differences} of expected scores. Therefore, it is sufficient to have a scoring function such that the resulting score differences are homogeneous. Such homogeneous score differences of order $b=0$ exist for VaR, expectiles and (VaR,ES) as shown by the results in Appendix C. Examples below list scoring functions, which will be used subsequently in the simulation study and real data analysis.

\begin{example}\label{ex:compVaR}
For the comparative backtests for VaR that we investigate in Section \ref{ssim}, we consider the classical 1-homogeneous choice obtained by choosing $G(r) = r$ in~\eqref{qSVaR} leading to the scoring function
\begin{equation}\label{eq:SVaR1}
S(r,x) = (1-\alpha - \mathbbm{1}\{x > r\})r + \mathbbm{1}\{x > r\}x.
\end{equation}
Guided by the arguments given above, we alternatively consider the 0-homogeneous score differences by choosing $G(r) = \log r$, $r > 0$ which leads to the score
\begin{equation}\label{eq:SVaR2}
S(r,x) = (1-\alpha - \mathbbm{1}\{x > r\})\log r + \mathbbm{1}\{x > r\}\log x,\qquad r>0.
\end{equation}
\end{example}

\begin{example}\label{ex:compexp}
The choice $\phi(r) = r^2$ in~\eqref{qSexp} leads to the strictly consistent scoring function
\begin{equation}\label{eq:Sexp1}
S(r,x) = -\mathbbm{1}\{x > r\}(1-2\tau)(x-r)^2 + (1-\tau)r(r - 2x)
\end{equation}
for the $\tau$-expectile $e_\tau$. Besides this 2-homogeneous choice, in Section \ref{ssim}, we also investigate the 0-homogeneous alternative that arises by choosing $\phi(r) = -\log(r)$, $r > 0$, hence we obtain the scoring function
\begin{equation}\label{eq:Sexp2}
S(r,x) = \mathbbm{1}\{x > r\}(1-2\tau)\Big(\log\frac{x}{r} + 1 - \frac{x}{r}\Big) + (1-\tau)\Big(\log r - 1 + \frac{x}{r}\Big).
\end{equation}
\end{example}

\begin{example}\label{ex:compES}
For $(\VaR_\nu,\ebbS_\nu)$, we consider the $(1/2)$-homogeneous scoring function given by choosing $G_1(x) = 0$, $\mathcal{G}_2(x) = x^{1/2}$, $x > 0$ in \eqref{qSVaRES} for comparative backtesting in Section \ref{ssim}. It is given by
\begin{equation}\label{eq:SVaRES1}
S(r_1,r_2,x) = \mathbbm{1}\{x > r_1\}\frac{x-r_1}{2\sqrt{r_2}} + (1-\nu)\frac{r_1 + r_2}{2\sqrt{r_2}}.
\end{equation}
As for the other risk measures, we also consider the 0-homogeneous alternative by choosing $G_1(x) = 0$, $\mathcal{G}_2(x) = \log x$, $x > 0$ which yields the scoring function
\begin{equation}\label{eq:SVaRES2}
S(r_1,r_2,x) = \mathbbm{1}\{x > r_1\}\frac{x - r_1}{r_2} + (1-\nu)\Big(\frac{r_1}{r_2} - 1 + \log(r_2)\Big).
\end{equation}
\end{example}

\citet{AcerbiSzekely2014} proposed a class of 2-homogeneous scoring functions for $(\VaR_\nu,\ebbS_\nu)$ depending on a parameter $W > 0$. It is strictly consistent when the class $\mathcal{P}$ of distributions is restricted to contain only distributions $F$ with
\[
\ebbS_\nu(F) < W \VaR_\nu(F).
\] 
In practice, it is generally not possible to say what magnitude of $W$ is realistic to cover all possible applications. Therefore, we prefer to work with the homogeneous choices of strictly consistent scoring functions above and, more generally, of the form in Theorem~C.3.

\setcounter{equation}{0}
\section{Numerical illustrations}\label{sec:numill}

\subsection{Forecasting of risk measures}\label{sm}

In this section we discuss a number of estimation procedures for producing conditional forecasts of the three risk measures discussed in this paper, namely the VaR, expectile and ES. Owing to the widespread use of VaR in the banking sector, a great number of methods exist to produce its point forecasts; see, e.g., \citet{Kuester2006} for an extensive review. In contrast, estimation and forecasting of expectiles in the risk measurement context is a relatively recent topic; see, e.g., \citet{KuanYehETAL2009}. However, in many cases, similar methods as those used for VaR forecasting can be adopted for expectiles. 

For illustrative purposes, we consider the following framework for forecasting of the risk measures. Suppose the series of negated log-returns $\{X_t\}_{t\in\nbb}$ can be modeled as \bql{qfm}X_t=\m_t +\s_t Z_t, \eql
where $\{Z_t\}_{t\in\nbb}$ is a sequence of i.i.d.~random variables with zero mean and unit variance, and $\m_t$ and $\s_t$ are measurable with respect to the sigma algebra $\FC_{t-1}$, representing the information about the process $\{X_t\}_{t\in\nbb}$ available up to time $t-1$. In order to capture typical time dynamics of financial time series, one possibility is to assume that the conditional mean $\m_t$ follows an ARMA process, while the condition variance $\s_t^2$ evolves according to a GARCH model specification.

Let $\r$ denote any of the three risk measures we consider. In the above setting, conditionally on the information up to time $t-1$, the one-step ahead forecast of $\r$ is 
\bql{qfR}\r(X_t\mid\FC_{t-1})=\m_t + \s_t \r(Z),\eql
where $Z$ is used to denote a generic random variable with the same distribution as the $Z_t$'s. Following \citet{McNeilFrey2000} and \citet{Diebold2000}, one can adopt a two-stage estimation procedure for the forecast $\r(X_t\mid\FC_{t-1})$. First $\m_t$ and $\s_t$ are estimated via the maximum likelihood procedure under a specific assumption\footnote{An alternative is to use the quasi-maximum likelihood estimation (MLE) procedure in which innovations $Z_t$ are assumed to be standard normal. This is justified by the result in \citet{Bollerslev1992} saying that $\m_t$ and $\s_t$ would be consistently estimated even if the distribution of innovations is not normal, provided that the models for $\m_t$ and $\s_t$ are correctly specified. As pointed out in \citet{Kuester2006}, the correct specification of dynamics for $\m_t$ and $\s_t$ may be difficult to fulfil in practice.} on the distribution of the innovations $Z_t$ in \eqref{qfm}. The second stage involves estimation of $\r(Z)$, the risk measure for i.i.d.~sequence $\{Z_t\}_{t\in\nbb}$, based on the sample of standardized residuals 
\begin{equation}\label{qsr}
\{\hat z_t = (x_t - \hat \m_t)/\hat\s_t\}.  
\end{equation}
We consider the following three approaches to handle the second stage in the forecasting procedure: fully parametric (FP), filtered historical simulation (FHS), and a semi-parametric estimation based on extreme value theory (EVT).

\subsubsection{Fully parametric estimation}\label{sm1}

Under the fully parametric approach, a specific (parametric) model is assumed for the sequence of innovations $\{Z_t\}_{t\in\nbb}$. Examples of typically used probability distributions include the normal, Student's t and a skewed t distribution (see, e.g.,~\citet{Fernandez1998}).
Parameters of the assumed distribution for $Z_t$'s, denoted $F_Z$, can be estimated based on the standardized residuals $\{\hat z_t\}$ in~\eqref{qsr} using, for example, the maximum likelihood method. If the model for $Z_t$'s coincides with the one used to estimate the filter in the first stage, then no additional estimation is required at the second stage with all model parameters coming directly from the first stage estimation. The fitted distribution is used to compute the estimate of a given risk measure. In the case of $\VaR_\a(Z)$, this is given by the $\a$-quantile, $\hat F_Z\inv (\a)$, whereas a $\t$-expectile $e_\t(Z)$ can be computed as discussed in Appendix B.1, where we give analytic expressions for expectiles of several commonly used distributions. Since we consider only continuous distributions $F_Z$, the ES can be computed as $$\ebbS_\nu(Z) = \ebb(Z|Z\ge \VaR_\nu(Z)),$$ where we use numerical integration to evaluate the conditional expectation.

\subsubsection{Filtered historical simulation}\label{sm2}

The method employs a non-parametric estimation of the risk measures based on the standardized residuals $\{\hat z_t\}$ in~\eqref{qsr}, which can be seen as representing a filtered time series; see, e.g., \citet[Chapter~5.6]{Christoffersen2003}. In particular, we draw a sample $\{\hat z_i^\ast; 1\le i \le N\}$ of a large size $N$ (e.g., $N=10,000$) from $\{\hat z_t; 1\le t \le n\}$ and then take the empirical estimate of a given risk functional as the estimate for $\r (Z)$.  The empirical $\a$-quantile gives the VaR estimate $\widehat{\VaR}_\a^{\text{{\tiny FHS}}}(Z)$. The empirical $\t$-expectile $\hat e_\t^{\text{{\tiny FHS}}}(Z)$ is obtained using the least asymmetric weighted squares via iterative minimization of 
$$\sum_{i=1}^N \w_i(\t) (\hat z_i^\ast - e_\t)^2,\qquad  \w_i(\t) = \t\ind\{\hat z_i^\ast > e_\t\} + (1-\t) \ind\{\hat z_i^\ast < e_\t\} \qquad {\rm with\ respect\ to\ } e_\t.$$
The ES is estimated by the empirical version of the conditional expectation given that the residual exceeds the corresponding VaR estimate:
$$\widehat{\ebbS}_\n^{\text{{\tiny FHS}}}(Z) = \dfrac1{\# \{i: i=1,\ldots,N, \hat z^*_i> \widehat{\VaR}_\a^{\text{{\tiny FHS}}}(Z)\}} \sum_{i=1}^N \hat z^*_i \ind\{\hat z^*_i> \widehat{\VaR}_\a^{\text{{\tiny FHS}}}(Z)\}.$$

\subsubsection{EVT-based semi-parametric estimation}\label{sm3}

Risk is naturally associated with extremal events, and hence risk measure estimates rely on accurate estimation of a tail of the underlying distribution. However, inference about the distributional tails is notoriously difficult as there are frequently not enough data points in the tail regions neither to give a proper justification for a parametric model nor to obtain reliable empirical estimates. Hence, unless a sufficiently long time series is available relative to the desired risk level for risk measure estimation, the two methods outlined in Sections~\ref{sm1} and~\ref{sm2} are unlikely to produce accurate forecasts. An alternative is to base estimation on asymptotic results of extreme value theory (EVT). For a detailed account, refer to, e.g., \citet{EKM1997}. 


The main premise is that, for a sufficiently high threshold $u$, conditional excesses of random variable $Z$ satisfy: \begin{equation}\label{qgpd} Z-u\mid Z>u\sim GP(\b_u,\x),\end{equation}
where $GP(\b,\x)$ denotes the generalized Pareto distribution with scale $\b>0$ and shape parameter $\x\in\rbb$. It is common in applications to set the threshold at an upper order statistic; i.e., $u=z_{(k+1)}$ for some $k<n$, where $z_{(1)}>z_{(2)}>\cdots>z_{(n)}$ are the decreasing order statistics of the sample $\{z_1,\ldots, z_n\}$ from $F_Z$. This leads to the following EVT-based estimates of $\VaR_\a(Z)$ and $\ebbS_\nu(Z)$ (see \citet{McNeilFrey2000}):
\begin{equation}\label{qVaRevt}
\widehat{ \VaR}_\a^{\text{{\tiny EVT}}} (Z) = u +\dfrac{\hat\b_u}{\hat\x}\Big(\Big(\dfrac{k}{\a\ n}\Big)^{\hat\x}  -1 \Big),\qquad \hat\x\neq 0,
\end{equation}
and
\begin{equation}\label{qESevt}
\widehat{ \ebbS}_\n^{\text{{\tiny EVT}}} (Z) = \widehat{ \VaR}_\n^{\text{{\tiny EVT}}} (Z) \Bigg( \dfrac1{1-\hat\xi} + \dfrac{\hat\beta - \hat\xi\ u}{(1-\hat\xi) \widehat{ \VaR}_\n^{\text{{\tiny EVT}}} (Z)} \Bigg),
\end{equation}
with $(\hat \b_u,\hat \x)$ being parameter estimates of the GP distribution fitted to excesses over~$u$. In the spirit of the above EVT-based estimators for VaR and ES, we derive an estimator for the $\t$-expectile. The details are provided in Appendix B.2.

In the discussion above we assume that threshold $u$ or equivalently $k$, the number of upper order statistics, is given so as to ensure adequacy of the approximation in~\eqref{qgpd}. However, in practice, an accurate choice has to be made to balance the bias-variance trade-off as a too large value of $u$ increases variability of the parameter estimates of $\b_u$ and $\x$, while insufficiently large $u$ introduces the bias due to invalidity of~\eqref{qgpd}. Various techniques have been proposed to assist with the choice of threshold such as graphical tools based on linearity of the mean excess function. As such methods require judgement at every time step at which conditional forecasts of risk measures are to be made, they are prohibitive for our purposes. Hence, we adopt a pragmatic approach as in \citet{McNeilFrey2000}, and take $k=60$ in samples of size $n=500$.  


\subsection{Simulation study}\label{ssim}

In practice, traditional backtesting is perhaps the most commonly used way to evaluate and subsequently choose among a number of competing forecasting procedures. While traditional backtesting is certainly suitable to capture some aspects of forecasting procedures, it does not provide information on the relative performance of different procedures with respect to the accuracy of forecasts, a seemingly natural criterion for a forecasting method. The aim of the present simulation study is to illustrate the use of the methodologies for traditional and comparative backtests discussed in the paper as well as to highlight the different messages delivered by the two types of backtests.

\subsubsection{Set-up and forecasting methods}\label{ssim1}
The data $\{X_t\}_{t\in\zbb}$ used for the analysis are generated from an AR(1)-GARCH(1,1) process:
\begin{align}\label{qsimDGP}
X_t = \m_t + \e_t,\qquad \m_t = -0.05 + 0.3 X_{t-1},\qquad \e_t = \s_t Z_t,\qquad \s_t^2 = 0.01 + 0.1\e_{t-1}^2 + 0.85 \s_{t-1}^2,
\end{align}
where innovations $\{Z_t\}_{t\in\zbb}$ form a sequence of independent random variables with a common skewed~t distribution  (see Example~B.6) with shape parameter $\n=5$ and skewness parameter $\g=1.5$. 

Quality of a forecasting procedure is determined by various factors. In a parametric or semi-parametric set-up, potential model misspecification as well as estimation uncertainty in small samples can be detrimental for prediction. Non-parametric methods, while requiring no assumptions on the underlying model, are also subject to sampling variability and have strong limitations when dealing with extreme or tail events. The forecasting procedures we consider in the simulation study aim to cover a spectrum of models and estimation methods. 
We assume that the underlying process follows an AR(1)-GARCH(1,1) dynamics with innovations $\{Z_t\}_{t\in\zbb}$ coming from one of the following three distributions: the normal, the Student's t and the skewed t distribution as in Example~B.6. We then consider the following estimation procedures:
\bit
\item fully parametric estimation (Section~\ref{sm1}) with the methods abbreviated as "n-FP", "t-FP" and "st-FP"  under the assumption of normal, t and skewed t distributed innovations, respectively; 
\item filtered historical simulation (Section~\ref{sm2}) with the methods abbreviated as "n-FHS", "t-FHS" and "st-FHS";
\item EVT-based estimation (Section~\ref{sm3}) with the methods abbreviated as "n-EVT", "t-EVT" and "st-EVT". 
\eit
In addition to the above-mentioned methods, we supplement results with the optimal forecasts (abbreviated as "opt"), which uses the knowledge of the data generating process. Estimation is conducted using the moving window of size 500, and forecasts are evaluated based on the out-of-sample size of 5000 verifying observations. 


\subsubsection{Backtesting of risk measure forecasts}\label{ssim2}

Table~\ref{tabsimsum} contains an overview of the one-step ahead forecasts obtained under the procedures described in the previous section. In particular, we report the average forecasts based on the series of moving estimation windows for each of the three considered risk measures, denoted $\ov{\VaR}_\a$, $\ov{e}_\t$ and $\ov{\ebbS}_\n$. The $\a$ levels for VaR are chosen in accordance with typical values used for internal risk management (such as $\a=0.90$ and $\a=0.95$) as well as the standard Basel VaR level $\a=0.99$. For expectiles and ES, the levels are selected in such a way that the risk measure forecasts agree under the standard normal model. 


In order to link to the previously used approaches to assess the quality of VaR forecasts (and to make comparisons between the methods), we computed the percentage of times the observations exceeded the $\VaR_\a$ forecasts, commonly referred to as the percentage of violations. Based on the values reported under the column "\% Viol." in Table~\ref{tabsimsum} , we observe that some of the misspecified models were actually able to hit nearly exactly the expected proportion of violations by matching the risk measure level $(1-\a)$. This is the case, for instance, for "n-EVT" and "t-EVT" methods at $\a=0.99$. Although large deviations from the risk measure confidence level do suggest substantial method deficiencies (as in the case of "n-FP" and "t-FP" methods), these values also highlight that the deviations from the $(1-\a)$ level alone are unlikely to provide a good basis for differentiating the methods' performance in terms of prediction.

Table~\ref{tabTB} illustrates the traditional backtesting methodology presented in Section~\ref{sec:calib}. Test statistics $T_1$ in~\eqref{qTh} and $T_2$ in~\eqref{eq:tau4} are used, respectively, for two-sided and one-sided conditional calibration tests. The one-sided tests for $\VaR_\a$ and $\t$-expectile are tests for super-calibration with p-values given by $\Phi(T_2)$. In the case of $(\VaR_\n,\ebbS_\n)$, we make use of the Hommel's procedure with the adjusted p-values computed as $\tilde \pi = q\ C_q\min\{\pi_{(m)}/m; m=1,2\}$ and capped at one, where $\pi_m = 1-\Phi(T_{2,m})$ for the one-sided tests of sub-calibration; see~\eqref{qHommel}. (The classical Bonferroni multiple test procedure resulted in qualitatively similar conclusions.) For the simple conditional calibration tests, we set $\boldsymbol{h}_t = 1$. The test functions that were found to work well in this simulation study for general conditional calibration tests are 
\begin{equation}\label{qtf2}
{\bf h}_t = \bcs (1,r_t)' & {\rm for\ }\VaR_\a,\\
\hat\s_t^{-1} & {\rm for\ expectile\ } e_\t ,\\
\hat\s_t^{-1}((r_{2,t}-r_{1,t})/(1-\nu),1) & {\rm for\ } (\VaR_\n,\ebbS_\n)
\ecs
\end{equation}
in the case of two-sided tests, and  
\begin{equation}\label{qtf1}
{\bf h}_t = \bcs (1,|r_t|)' & {\rm for\ }\VaR_\a,\\
\hat\s_t^{-1} & {\rm for\ expectile\ } e_\t ,\\
\bma 1 & |r_{1,t}| & 0 & 0\\ 0 & 0 & 1 & \hat\s_t^{-1}\ema' & {\rm for\ } (\VaR_\n,\ebbS_\n)
\ecs
\end{equation}
in the case of one-sided tests. The choice of test functions is important as it affects the properties of the test. For example, we found that inclusion of the lagged values of the identification function as in Example~\ref{ex:tradVaR} resulted in tests which rejected all of the models including the optimal forecaster for $\VaR_{0.99}$ in the two-sided conditional calibration tests. A possible explanation for this phenomenon is that for a chosen test function the distribution of the test statistic becomes heavily skewed, making convergence to the asymptotic distribution slow.  Another contributing factor, suggested by a referee, could be the instability of the $\hat\Omega^{-1}$ estimate in~\eqref{qTh} due to high correlation of lagged values of the identification function. As discussed in \citet{GiacominiWhite2006}, the choice of the test function with too few or too many components will also have direct implications on the power of the tests. 

As expected, the numerical results in Table~\ref{tabTB} show that the backtesting decisions based on the general conditional calibration tests are more conservative in comparison to the corresponding simple conditional calibration tests, subject to a sensible choice of the test function. This is particularly visible for one-dimensional risk measures (VaR and expectiles) when performing the two-sided tests. The two-sided conditional calibration tests for these two risk measures suggest the importance of the correct specification of the likelihood used in fitting the AR(1)-GARCH(1,1) filter. The entirely parametric methods with misspecified models (here "n-FP" and "t-FP") fail traditional backtests even when testing for simple conditional calibration (with the exception of $\VaR_{0.90}$). The general conditional tests are able to pick-up the misspecified likelihoods at least in some instances; for example, when forecasting $\VaR_{0.90}$ and using the (symmetric) t distribution instead of the true asymmetric underlying model, and similarly for $\tau$-expectiles with $\tau=0.96561$ and $\tau=0.98761$. The general conditional two-sided calibration tests also detect the differences in the second stage of risk measure forecasting when different methods are applied to filtered series of innovations. For instance, at the highest risk measure levels, the EVT-based methods tend to pass the conditional backtests in contrast to their empirical and in some cases even parametric (correctly specified) counterparts; see panels for $\VaR_{0.99}$ and $0.99855$-expectile. This is true even under a misspecified likelihood model in the AR(1)-GARCH(1,1) filter. 

We also note that the tests for one-dimensional risk measures appear to have better power properties than the tests for the two-dimensional risk measure, $(\VaR_\n,\ebbS_\n)$, although a more thorough investigation into finite sample properties of these tests would be necessary to draw more definitive conclusions. It can also be observed that the one-sided tests are less conclusive than their two-sided analogues. This is perhaps not a surprise as it may well happen that a method is not good at predicting the risk measure but gives a correct bound and thus should not be rejected by a one-sided calibration test. 

\begin{sidewaystable}
\scriptsize
\caption{Risk measure forecasts and method comparisons based on the sample average of consistent scoring functions in the simulation study; see Section~\ref{ssim} for details. The average scores are scaled by one minus the risk measure level to avoid very small values for presentation purposes. "\% Viol." column shows the percentage of times observations exceeded the corresponding forecasts of $\VaR_\a$. The values in brackets indicate method ranks based on their average scores. } \label{tabsimsum}
\begin{tabular*}{1\textwidth}{@{\extracolsep{\fill}} | l | c  c   c c |  c c c | c c c | } \hline
Method & $\ov{\VaR}_\a$ & \% Viol. & $\ov{S}$ [eq.~\eqref{eq:SVaR1}] & $\ov{S}$ [eq.~\eqref{eq:SVaR2}] & $\ov{e}_\t$ &  $\ov{S}$ [eq.~\eqref{eq:Sexp1}] & $\ov{S}$ [eq.~\eqref{eq:Sexp2}] & $\ov{\ebbS}_\n$ & $\ov{S}$ [eq.~\eqref{eq:SVaRES1}] & $\ov{S}$ [eq.~\eqref{eq:SVaRES2}] \\\hline

 & \multicolumn{4}{c|}{$\a=0.90$} & \multicolumn{3}{c|}{$\t=0.96561$} & \multicolumn{3}{c|}{$\n=0.754$}\\\hline
 
n-FP    & 0.440 & 9.4 & 0.7496 ( ~9 ) & -0.4325 ( 7 ) & 0.440 & 1.0149 ( 9 ) & -1.0526 ( 9 ) & 0.440 & 0.6685 ( 10 ) & -0.8119 ( 9 ) \\ 
n-FHS  & 0.406 & 10.2 & 0.7484 ( ~8 ) & -0.4288 ( 9 ) & 0.542 & 1.0006 ( 7 ) & -1.3076 ( 7 ) & 0.450 & 0.6626 ( 5 ) & -0.8361 ( 4 ) \\ 
n-EVT  & 0.406 & 10.2 & 0.7477 ( ~7 ) & -0.4304 ( 8 ) & 0.553 & 1.0039 ( 8 ) & -1.3188 ( 5 ) & 0.449 & 0.6655 ( 9 ) & -0.8270 ( 8 ) \\ 
t-FP    & 0.348 & 12.2 & 0.7527 ( 10 ) & -0.3944 ( 10 ) & 0.424 & 1.0200 ( 10 ) & -0.904 ( 10 ) & 0.421 & 0.6645 ( 7 ) & -0.8040 ( 10 ) \\ 
t-FHS  & 0.413 & 10.0 & 0.7473 ( ~6 ) & -0.4350 ( 5 ) & 0.550 & 0.9899 ( 5 ) & -1.3055 ( 8 ) & 0.456 & 0.6622 ( 4 ) & -0.8356 ( 5 ) \\ 
t-EVT  & 0.410 & 10.3 & 0.7471 ( ~5 ) & -0.4329 ( 6 ) & 0.562 & 0.9944 ( 6 ) & -1.3137 ( 6 ) & 0.457 & 0.6654 ( 8 ) & -0.8289 ( 7 ) \\ 
st-FP  & 0.417 & 9.9 & 0.7442 ( ~2 ) & -0.4391 ( 2 ) & 0.559 & 0.9865 ( 4 ) & -1.3378 ( 3 ) & 0.461 & 0.6606 ( 2 ) & -0.8460 ( 3 ) \\ 
st-FHS & 0.412 & 10.1 & 0.7451 ( ~4 ) & -0.4387 ( 3 ) & 0.550 & 0.9808 ( 2 ) & -1.3342 ( 4 ) & 0.455 & 0.6606 ( 3 ) & -0.8488 ( 2 ) \\ 
st-EVT & 0.410 & 10.2 & 0.7449 ( ~3 ) & -0.4363 ( 4 ) & 0.561 & 0.9844 ( 3 ) & -1.3409 ( 2 ) & 0.457 & 0.6642 ( 6 ) & -0.8350 ( 6 ) \\ 
opt    & 0.424 & 9.5 & 0.7431 ( ~1 ) & -0.4454 ( 1 ) & 0.565 & 0.9643 ( 1 ) & -1.4257 ( 1 ) & 0.467 & 0.6575 ( 1 ) & -0.8704 ( 1 ) \\ \hline

& \multicolumn{4}{c|}{$\a=0.95$} & \multicolumn{3}{c|}{$\t=0.98761$} & \multicolumn{3}{c|}{$\n=0.875$}\\\hline

n-FP    & 0.586 & 5.9 & 0.9925 ( ~8 ) & -0.1055 ( 9 ) & 0.586 & 1.9845 ( 10 ) & -0.4650 ( 10 ) & 0.587 & 0.8177 ( 10 ) & -0.3975 ( 10 ) \\ 
n-FHS  & 0.632 & 5.0 & 0.9910 ( ~7 ) & -0.1123 ( 7 ) & 0.801 & 1.8718 ( 7 ) & -0.8939 ( 5 ) & 0.667 & 0.8121 ( 8 ) & -0.4261 ( 7 ) \\ 
n-EVT  & 0.628 & 5.1 & 0.9930 ( ~9 ) & -0.1080 ( 8 ) & 0.810 & 1.8756 ( 8 ) & -0.8935 ( 6 ) & 0.670 & 0.8121 ( 7 ) & -0.4259 ( 8 ) \\ 
t-FP    & 0.518 & 7.3 & 1.0106 ( 10 ) & -0.0555 ( 10 ) & 0.631 & 1.9008 ( 9 ) & -0.6419 ( 9 ) & 0.716 & 0.8137 ( 9 ) & -0.4233 ( 9 ) \\ 
t-FHS  & 0.631 & 5.1 & 0.9902 ( ~5 ) & -0.1148 ( 5 ) & 0.822 & 1.8428 ( 5 ) & -0.8929 ( 7 ) & 0.675 & 0.8112 ( 5 ) & -0.4292 ( 5 ) \\ 
t-EVT  & 0.630 & 5.1 & 0.9910 ( ~6 ) & -0.1128 ( 6 ) & 0.826 & 1.8506 ( 6 ) & -0.8885 ( 8 ) & 0.677 & 0.8117 ( 6 ) & -0.4274 ( 6 ) \\ 
st-FP  & 0.639 & 4.9 & 0.9858 ( ~2 ) & -0.1227 ( 2 ) & 0.832 & 1.8313 ( 4 ) & -0.9156 ( 3 ) & 0.688 & 0.8096 ( 3 ) & -0.4356 ( 3 ) \\ 
st-FHS & 0.632 & 5.0 & 0.9887 ( ~3 ) & -0.1161 ( 3 ) & 0.821 & 1.8164 ( 2 ) & -0.9174 ( 2 ) & 0.675 & 0.8096 ( 2 ) & -0.4357 ( 2 ) \\ 
st-EVT & 0.630 & 5.1 & 0.9890 ( ~4 ) & -0.1154 ( 4 ) & 0.825 & 1.8221 ( 3 ) & -0.9153 ( 4 ) & 0.677 & 0.8100 ( 4 ) & -0.4341 ( 4 ) \\ 
opt    & 0.649 & 4.7 & 0.9834 ( ~1 ) & -0.1267 ( 1 ) & 0.837 & 1.7481 ( 1 ) & -1.0189 ( 1 ) & 0.696 & 0.8070 ( 1 ) & -0.4503 ( 1 ) \\ \hline

& \multicolumn{4}{c|}{$\a=0.99$} & \multicolumn{3}{c|}{$\t=0.99855$} & \multicolumn{3}{c|}{$\n=0.975$}\\\hline

n-FP    & 0.859 & 2.5 & 1.8649 ( 10 ) & 0.7041 ( 10 ) & 0.859 & 8.4605 ( 10 ) & 2.1097 ( 10 ) & 0.863 & 1.1638 ( 10 ) & 0.3969 ( 10 ) \\ 
n-FHS  & 1.193 & 1.1 & 1.7398 ( ~8 ) & 0.4992 ( 7 ) & 1.492 & 6.1819 ( 7 ) & 0.0652 ( 6 ) & 1.218 & 1.1268 ( 8 ) & 0.2453 ( 8 ) \\ 
n-EVT  & 1.189 & 1.0 & 1.7115 ( ~5 ) & 0.4801 ( 5 ) & 1.480 & 6.1153 ( 5 ) & 0.0651 ( 5 ) & 1.243 & 1.1240 ( 7 ) & 0.2381 ( 7 ) \\ 
t-FP    & 0.948 & 1.8 & 1.7605 ( ~9 ) & 0.5679 ( 9 ) & 1.186 & 6.0364 ( 3 ) & 0.2244 ( 9 ) & 1.781 & 1.1472 ( 9 ) & 0.2847 ( 9 ) \\ 
t-FHS  & 1.207 & 1.1 & 1.7392 ( ~7 ) & 0.5025 ( 8 ) & 1.629 & 6.7232 ( 9 ) & 0.0771 ( 8 ) & 1.246 & 1.1205 ( 5 ) & 0.2334 ( 6 ) \\ 
t-EVT  & 1.203 & 1.0 & 1.7064 ( ~4 ) & 0.4755 ( 4 ) & 1.546 & 6.1387 ( 6 ) & 0.0658 ( 7 ) & 1.266 & 1.1208 ( 6 ) & 0.2328 ( 5 ) \\ 
st-FP  & 1.214 & 0.9 & 1.6987 ( ~3 ) & 0.4734 ( 3 ) & 1.583 & 5.9688 ( 2 ) & -0.0491 ( 2 ) & 1.287 & 1.1156 ( 2 ) & 0.2195 ( 2 ) \\ 
st-FHS & 1.209 & 1.1 & 1.7339 ( ~6 ) & 0.4991 ( 6 ) & 1.614 & 6.4895 ( 8 ) & 0.0236 ( 3 ) & 1.245 & 1.1161 ( 3 ) & 0.2221 ( 4 ) \\ 
st-EVT & 1.202 & 0.9 & 1.6929 ( ~2 ) & 0.4651 ( 2 ) & 1.543 & 6.0779 ( 4 ) & 0.0306 ( 4 ) & 1.265 & 1.1164 ( 4 ) & 0.2215 ( 3 ) \\ 
opt    & 1.227 & 0.9 & 1.6614 ( ~1 ) & 0.4369 ( 1 ) & 1.574 & 4.9567 ( 1 ) & -0.3749 ( 1 ) & 1.297 & 1.1066 ( 1 ) & 0.1887 ( 1 ) \\ 

\hline
\end{tabular*}
\end{sidewaystable}

\begin{sidewaystable}
\scriptsize
\caption{P-values for traditional backtests in the simulation study; see Section~\ref{ssim} for details. The one-sided tests for $\VaR_\a$ and $\tau$-expectile are tests of super-calibration, and of sub-calibration for $(\VaR_\n,\ebbS_\n)$. The test functions used in general conditional calibration tests are given in~\eqref{qtf2} and~\eqref{qtf1}. Values in boldface are significant at 5\% level. } \label{tabTB}
\begin{tabular*}{1\textwidth}{@{\extracolsep{\fill}} | l | c   c   c  c ||  c c c c || c c c c | } \hline
 &   \multicolumn{4}{c||}{$\VaR_\a$} &  \multicolumn{4}{c||}{$\t$-expectile} &  \multicolumn{4}{c|}{$(\VaR_\n,\ebbS_\n)$}  \\\hline

&  \multicolumn{2}{c|}{two-sided}&  \multicolumn{2}{c||}{one-sided}&  \multicolumn{2}{c|}{two-sided}&  \multicolumn{2}{c||}{one-sided}&  \multicolumn{2}{c|}{two-sided}&  \multicolumn{2}{c|}{one-sided}\\\hline

Method  & simple & general  & simple & general  & simple & general   & simple & general   & simple & general  & simple & general  \\\hline

 & \multicolumn{4}{c||}{$\a=0.90$} & \multicolumn{4}{c||}{$\t=0.96561$} & \multicolumn{4}{c|}{$\n=0.754$}\\\hline
 
n-FP    & 0.146 & {\bf 0.018} & 0.927 & 1.000 & {\bf 0.000} & {\bf 0.000} & {\bf 0.000} & {\bf 0.000} & {\bf 0.000} & {\bf 0.000} & {\bf 0.000} & {\bf 0.000} \\ 
n-FHS  & 0.576 & 0.058 & 0.288 & 0.863 & 0.887 & {\bf 0.048} & 0.443 & 0.193 & 0.881 & 0.184 & 0.712 & 0.744 \\ 
n-EVT  & 0.608 & 0.056 & 0.304 & 0.911 & 0.684 & {\bf 0.042} & 0.658 & 0.364 & 0.754 & 0.672 & 1.000 & 0.629 \\ 
t-FP    & {\bf 0.000} & {\bf 0.000} & {\bf 0.000} & {\bf 0.000} & {\bf 0.000} & {\bf 0.000} & {\bf 0.000} & {\bf 0.000} & 0.086 & {\bf 0.006} & {\bf 0.041} & {\bf 0.011} \\ 
t-FHS  & 0.962 & {\bf 0.006} & 0.481 & 1.000 & 0.728 & {\bf 0.030} & 0.636 & 0.330 & 0.936 & 0.512 & 0.960 & 0.256 \\ 
t-EVT  & 0.514 & {\bf 0.011} & 0.257 & 0.772 & 0.360 & {\bf 0.023} & 0.820 & 0.542 & 0.880 & 0.475 & 0.815 & {\bf 0.008} \\ 
st-FP  & 0.740 & 0.090 & 0.630 & 1.000 & 0.429 & 0.084 & 0.786 & 0.546 & 0.569 & 0.824 & 1.000 & 0.991 \\ 
st-FHS & 0.851 & 0.091 & 0.425 & 1.000 & 0.708 & 0.123 & 0.646 & 0.400 & 0.909 & 0.796 & 0.956 & 0.744 \\ 
st-EVT & 0.674 & 0.066 & 0.337 & 1.000 & 0.377 & 0.098 & 0.812 & 0.596 & 0.935 & 0.706 & 0.851 & {\bf 0.032} \\ 
opt    & 0.228 & 0.294 & 0.886 & 1.000 & 0.234 & 0.458 & 0.883 & 0.850 & 0.401 & 0.337 & 0.732 & 1.000 \\ 

\hline

& \multicolumn{4}{c||}{$\a=0.95$} & \multicolumn{4}{c||}{$\t=0.98761$} & \multicolumn{4}{c|}{$\n=0.875$}\\\hline

n-FP    & {\bf 0.006} & {\bf 0.004} & {\bf 0.003} & {\bf 0.009} & {\bf 0.000} & {\bf 0.000} & {\bf 0.000} & {\bf 0.000} & {\bf 0.000} & {\bf 0.000} & {\bf 0.000} & {\bf 0.000} \\ 
n-FHS  & 0.948 & {\bf 0.042} & 0.526 & 1.000 & 0.702 & 0.067 & 0.351 & 0.158 & 0.912 & 0.349 & 0.997 & 0.609 \\ 
n-EVT  & 0.797 & 0.075 & 0.398 & 1.000 & 0.868 & 0.062 & 0.434 & 0.208 & 0.720 & 0.549 & 1.000 & 0.762 \\ 
t-FP    & {\bf 0.000} & {\bf 0.000} & {\bf 0.000} & {\bf 0.000} & {\bf 0.000} & {\bf 0.000} & {\bf 0.000} & {\bf 0.000} & {\bf 0.000} & {\bf 0.000} & 1.000 & 1.000 \\ 
t-FHS  & 0.700 & 0.053 & 0.350 & 1.000 & 0.793 & {\bf 0.027} & 0.603 & 0.325 & 0.951 & 0.492 & 0.864 & 0.368 \\ 
t-EVT  & 0.654 & 0.106 & 0.327 & 0.981 & 0.713 & {\bf 0.033} & 0.643 & 0.363 & 0.699 & 0.771 & 1.000 & 0.845 \\ 
st-FP  & 0.794 & 0.261 & 0.603 & 1.000 & 0.568 & 0.066 & 0.716 & 0.467 & 0.655 & 0.898 & 0.907 & 0.249 \\ 
st-FHS & 0.897 & 0.111 & 0.449 & 1.000 & 0.729 & 0.073 & 0.635 & 0.393 & 0.908 & 0.690 & 0.904 & 0.875 \\ 
st-EVT & 0.797 & 0.180 & 0.398 & 1.000 & 0.643 & 0.077 & 0.679 & 0.435 & 0.599 & 0.968 & 1.000 & 1.000 \\ 
opt    & 0.284 & 0.552 & 0.858 & 1.000 & 0.315 & 0.523 & 0.843 & 0.798 & 0.311 & 0.624 & 0.263 & 0.194 \\ \hline

& \multicolumn{4}{c||}{$\a=0.99$} & \multicolumn{4}{c||}{$\t=0.99855$} & \multicolumn{4}{c|}{$\n=0.975$}\\\hline

n-FP    & {\bf 0.000} & {\bf 0.000} & {\bf 0.000} & {\bf 0.000} & {\bf 0.000} & {\bf 0.000} & {\bf 0.000} & {\bf 0.000} & {\bf 0.000} & {\bf 0.000} & {\bf 0.000} & {\bf 0.000} \\ 
n-FHS  & 0.420 & {\bf 0.007} & 0.210 & 0.630 & 0.377 & {\bf 0.045} & 0.188 & 0.100 & 0.653 & 0.231 & 0.549 & 0.538 \\ 
n-EVT  & 1.000 & 0.186 & 0.500 & 1.000 & 0.300 & 0.080 & 0.150 & 0.085 & 0.886 & 0.226 & 0.804 & 0.577 \\ 
t-FP    & {\bf 0.000} & {\bf 0.000} & {\bf 0.000} & {\bf 0.000} & {\bf 0.003} & {\bf 0.010} & {\bf 0.002} & {\bf 0.001} & {\bf 0.000} & {\bf 0.000} & 1.000 & 1.000 \\ 
t-FHS  & 0.679 & {\bf 0.029} & 0.339 & 1.000 & 0.783 & {\bf 0.013} & 0.391 & 0.212 & 0.697 & 0.717 & 1.000 & 1.000 \\ 
t-EVT  & 0.888 & 0.140 & 0.444 & 1.000 & 0.509 & 0.067 & 0.254 & 0.145 & 0.995 & 0.498 & 0.807 & 1.000 \\ 
st-FP  & 0.454 & 0.221 & 0.773 & 1.000 & 0.601 & {\bf 0.048} & 0.301 & 0.169 & 0.695 & 0.419 & 0.597 & 0.511 \\ 
st-FHS & 0.584 & {\bf 0.018} & 0.292 & 0.876 & 0.826 & {\bf 0.026} & 0.413 & 0.238 & 0.843 & 0.758 & 1.000 & 1.000 \\ 
st-EVT & 0.554 & 0.270 & 0.723 & 1.000 & 0.552 & 0.087 & 0.276 & 0.162 & 0.962 & 0.564 & 0.868 & 1.000 \\ 
opt    & 0.364 & 0.576 & 0.818 & 1.000 & 0.825 & 0.491 & 0.588 & 0.513 & 0.131 & 0.571 & 0.073 & 0.101 \\

\hline
\end{tabular*}
\end{sidewaystable}


In addition to risk measure average forecasts, Table~\ref{tabsimsum} also reports the average scores along with the corresponding method rankings using two different (consistent) scoring functions for each of the three considered risk measures. As the scoring functions we use require risk measure forecasts to be positive, we set the scores across all methods to zero in those few cases where forecasts are negative. Note that in the case of $(\VaR_\n,\ebbS_\n)$, only the forecasts for $\ebbS_\n$ are restricted to be positive. 

The method rankings based on the average scores appear to be reasonable, and suggest some more general conclusions with respect to method selection on the basis of forecasting accuracy. Similar to the results of traditional backtesting, the numerical values in Table~\ref{tabsimsum} provide further support to the observation that the choice of the likelihood model in fitting the AR(1)-GARCH(1,1) filter has an appreciable influence on the accuracy of forecasts, perhaps more than previously thought in the context of using the quasi-maximum-likelihood methods. Within each likelihood model, at lower levels for risk measure, fully parametric and FHS approaches tend to demonstrate better predictive performance, whereas at higher levels EVT-based methods seem to have an advantage, in particular, in the case of VaR. When the likelihood model is misspecified in fitting the AR(1)-GARCH(1,1) filter, the non-parametric methods such as FHS and semi-parametric methods such as EVT-based estimation allow for greater flexibility to diminish the effects of model misspecification than the fully parametric approaches do. While in many cases, rankings obtained from each pair of consistent scoring functions coincide, there also exist some discrepancies. This is not a surprise in the presence of misspecified models and estimation uncertainty as already pointed out by \citet{Patton2014}. For models for which the mean score is finite, the weak law of large numbers suggests convergence of the sample average (score) to the true mean (score) as the out-of-sample size tends to infinity. However, the convergence can be fairly slow. We found that in our simulation study, the out-of-sample size of at least 1000 data points is necessary to achieve some stability in rankings. Hence, in finite sample situations, one has to be aware of the effects of sampling variability on the final rankings of the forecasting methods. Appendix D discusses results of a study where only 250 verifying observations were considered to perform backtesting. In small samples, results of both traditional and comparative backtesting may be greatly distorted by unrepresentative samples even when the underlying data generating process is stationary.

\begin{figure}
\centering
\includegraphics[width=1\linewidth]{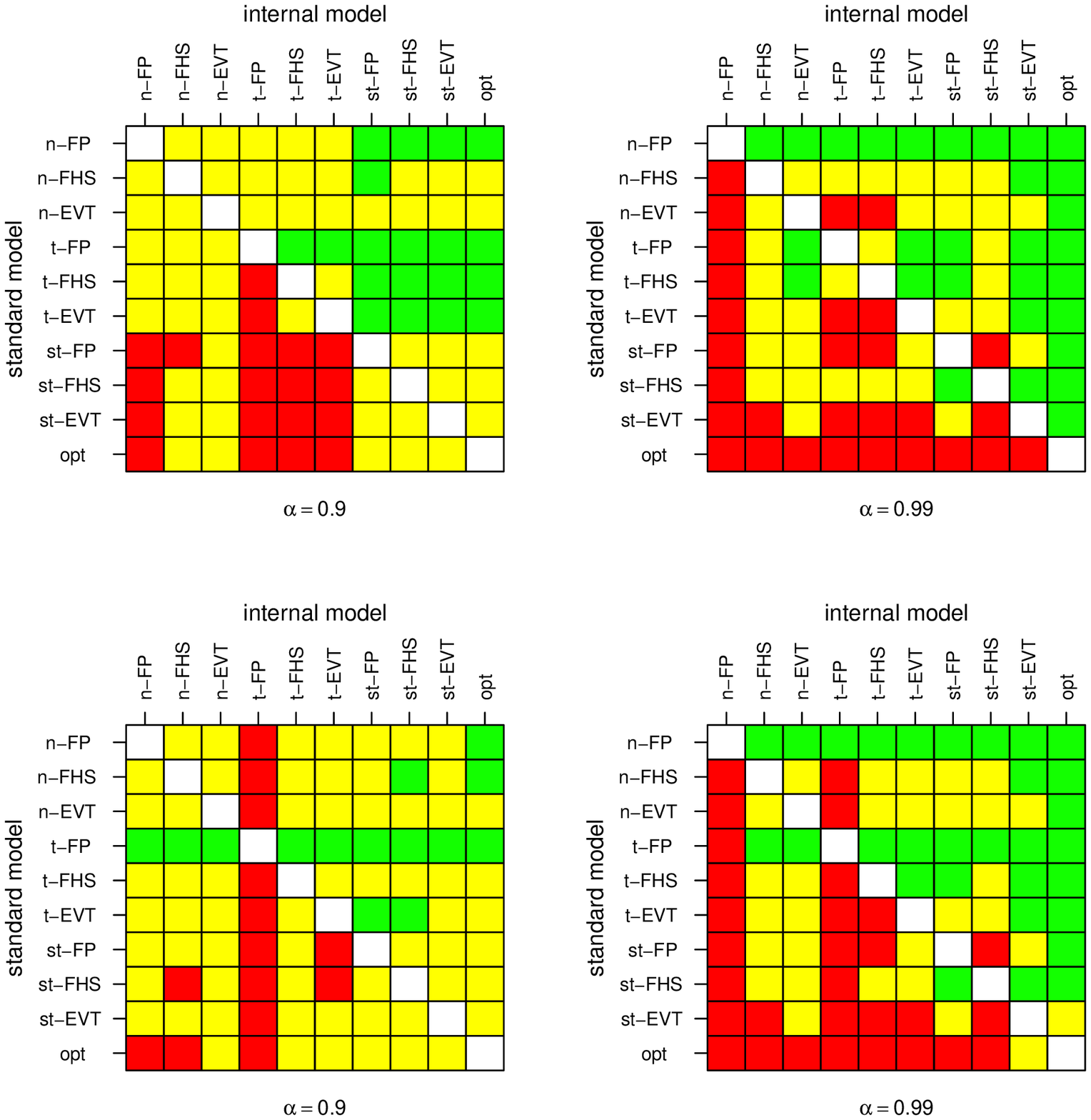}
\caption{Traffic light matrices for $\VaR_\a$ forecasts at the test confidence level $\eta=0.05$. The top and bottom rows are based on the scoring functions in \eqref{eq:SVaR1} and \eqref{eq:SVaR2}, respectively.}
\label{fig:TLMVaR}
\end{figure}
\begin{figure}
\centering
\includegraphics[width=1\linewidth]{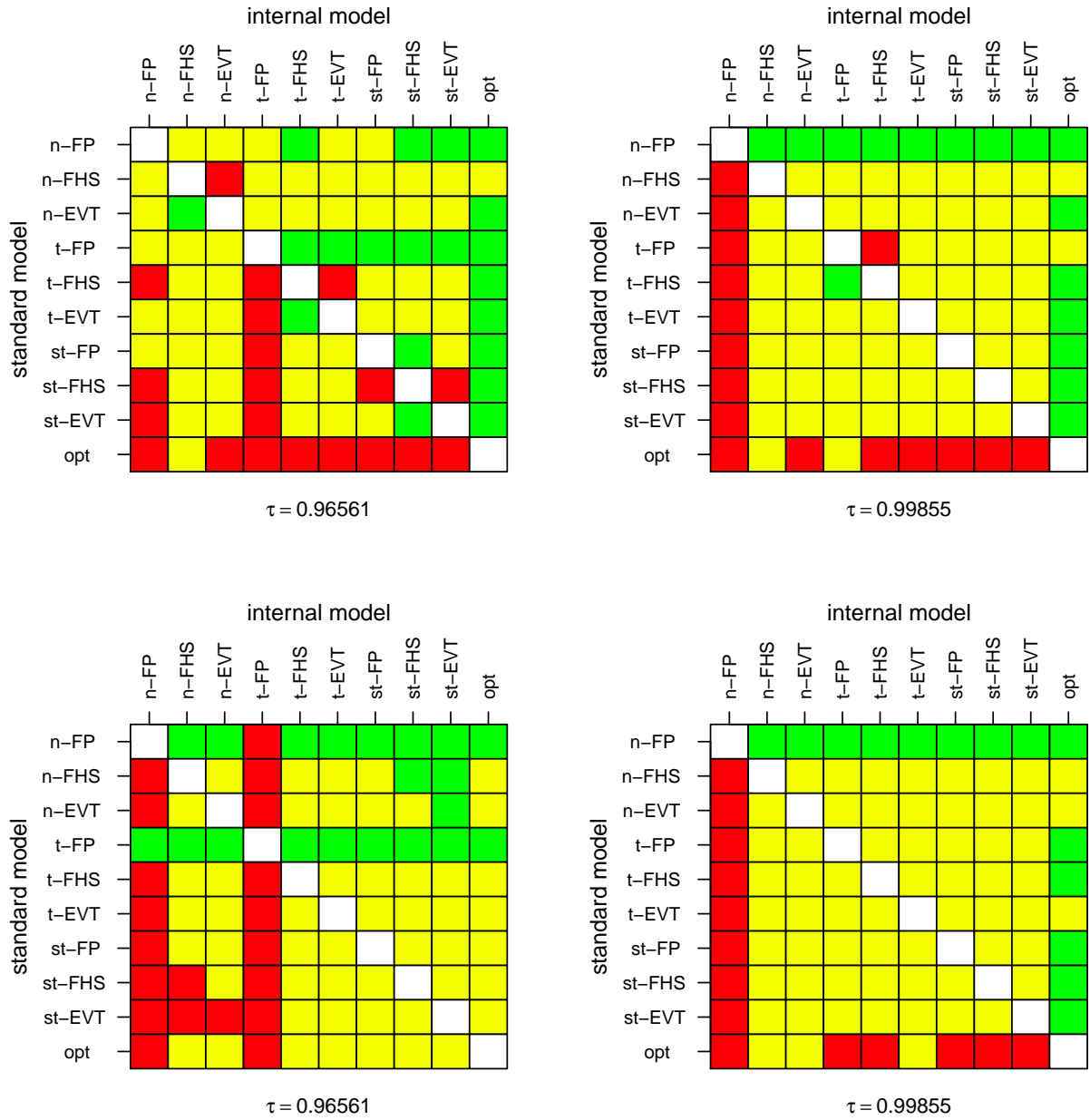}
\caption{Traffic light matrices for $\tau$-expectile forecasts at the test confidence level $\eta=0.05$. The top and bottom rows are based on the scoring functions in \eqref{eq:Sexp1} and \eqref{eq:Sexp2}, respectively.}
\label{fig:TLMexp}
\end{figure}
\begin{figure}
\centering
\includegraphics[width=1\linewidth]{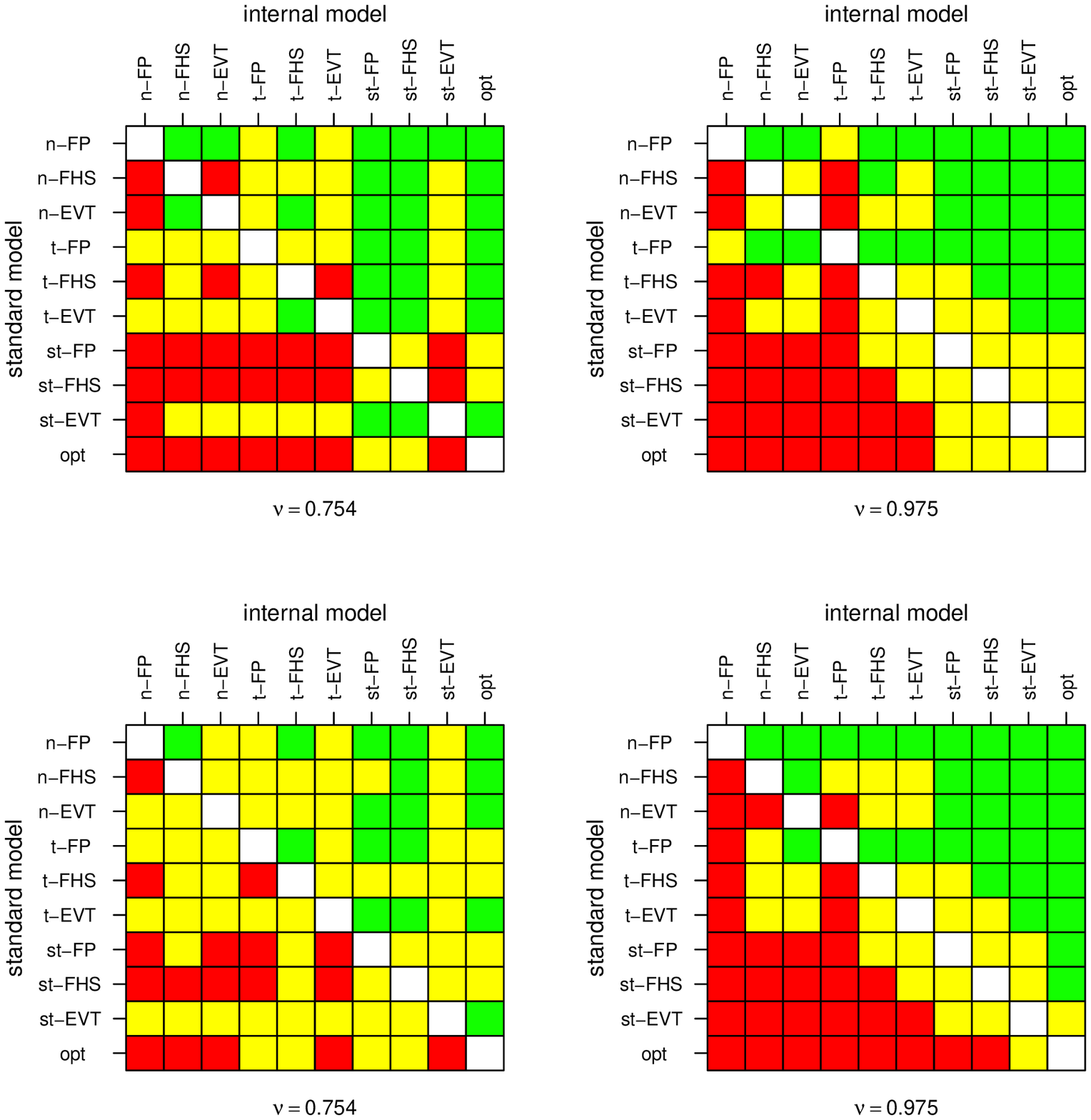}
\caption{Traffic light matrices for $(\VaR_\n,\ebbS_\n)$ forecasts at the test confidence level $\eta=0.05$. The top and bottom rows are based on the scoring functions in \eqref{eq:SVaRES1} and \eqref{eq:SVaRES2}, respectively.}
\label{fig:TLMVaRES}
\end{figure}

Finally, Figures~\ref{fig:TLMVaR} -~ \ref{fig:TLMVaRES} display the traffic light matrices for the three risk measures and two forms of consistent scoring functions for each. These plots complement the method rankings on the basis of just the average scores with the tests of predictive ability at the test level $\h=5\%$. Along the vertical axis we consider hypothetical "standard" models with the investigated "internal" models displayed along the horizontal axis. The red and green cells correspond to situations in which the comparative backtest is failed or passed, while yellow cells indicate cases where no conclusive evidence is available to pass or fail the comparative backtest.  The rows in each figure correspond to different scoring functions used to compare the methods. 

Inconclusive traffic light matrices can result if all methods are performing reasonably well, or, if the chosen scoring function has poor discrimination ability. 
In the case of VaR, as the discrimination ability of both chosen scoring functions seems good at level $\alpha=0.99$, it is likely that at $\alpha=0.90$ several models show a reasonable predictive ability. This is in line with the largely inconclusive traditional backtests at level $\alpha=0.90$. 
At $\a=0.90$, the scoring function in \eqref{eq:SVaR1} is better at identifying models with the correctly specified likelihood than the scoring function in \eqref{eq:SVaR2}, for which with just a few exceptions only the "t-FP" method fails the comparative backtests as an internal method against all the other possible standard methods. 
At $\a=0.99$, the two scoring functions result in a good agreement with "n-FP" being the worst forecaster (i.e., failing the comparative backtests against all the other methods), the optimal method passing comparative backtests against all other methods (the exception is "st-EVT" under the scoring function in \eqref{eq:SVaR2}). 

The situation is less clear for the $\tau$-expectile. At level $\tau=0.96561$, the "n-FP" method fails the comparative backtest against most of the other methods under both scoring functions; the use of the scoring function in \eqref{eq:Sexp2} suggests failing the "t-FP" method as well. The "st-EVT" method would pass the comparative backtest against the models with the normal likelihood and "t-FP". At level $\tau = 0.99855$, both scoring functions do not discriminate the methods much except for flagging the optimal forecaster as better than most other methods and failing the "n-FP" method. Expectiles have been used much less as a risk measure and it may be possible that the present methods are indeed suboptimal for expectile prediction at high levels. Again, this is in line with the results of the traditional backtests, in particular, the conditional two-sided tests. 

For $(\VaR_\n,\ebbS_\n)$, the large number of conclusive comparative backtesting results indicates that we can discriminate well between methods, and, as in the case of VaR it appears less important which method to use at a lower level than at a higher level. In particular, we once again see that the methods with the correctly specified likelihood show superior predictive performance. According to the scoring function in \eqref{eq:SVaRES1}, the "st-EVT" method fails the comparative backtest against its parametric and non-parametric counterparts "st-FP" and "st-FHS" at lower levels of $\n$. No definitive conclusions with respect to these models can be drawn at $\n=0.975$. 

\subsection{Data analysis}\label{sdat}

We have fitted an AR(1)-GARCH(1,1) model to the negated log-returns of the NASDAQ Composite index using a moving estimation window of 500 data points. The time series we consider is from Feb. 8, 1971 until May 18, 2016, which gives us an out-of-sample size $n=$10,920 to perform backtesting. Table~\ref{tabdata} summarizes results of traditional and comparative backtesting for six forecasting methods (refer to Section~\ref{ssim} for details on these methods) and, as before, for the three risk measures (VaR, expectile and the (VaR, ES) pair) at their standard Basel levels.

\begin{figure}
\centering
\includegraphics[width=0.7\linewidth]{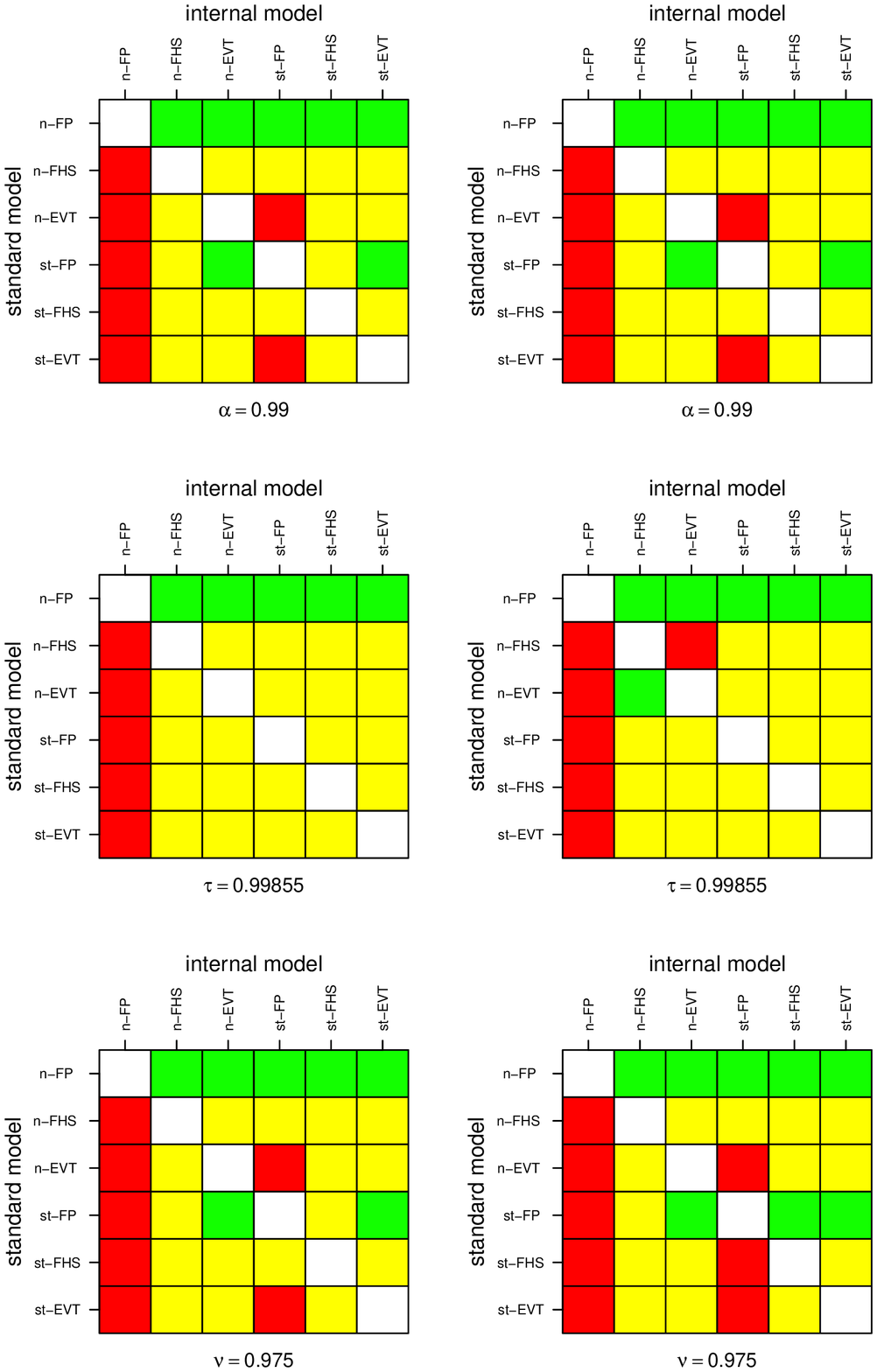}
\caption{Traffic light matrices for $\VaR_\a$ (top row) based on scoring functions in~\eqref{eq:SVaR1} (left) and~\eqref{eq:SVaR2} (right), for $\tau$-expectile (middle row) based on scoring functions in~\eqref{eq:Sexp1} (left) and~\eqref{eq:Sexp2} (right), and for $(\VaR_\n,\ebbS_\n)$ (bottom row) based on scoring functions in~\eqref{eq:SVaRES1} (left) and~\eqref{eq:SVaRES2} (right) at the test confidence level $\eta=0.05$, for the data analysis in Section~\ref{sdat}. }
\label{fig:TLMdata}
\end{figure}

In the case of $\VaR_{0.99}$, the traditional backtests based on the two-sided simple conditional calibration tests are passed only under the n-EVT and st-EVT methods. So, here, the choice of the likelihood function in fitting the AR(1)-GARCH(1,1) filter seems to have a lower impact than the choice of the  method at the second stage of forecasting applied to the fitted residuals. At this relatively high risk measure level, the EVT-based methods outperform their other competitors based on both the traditional backtests and the average scores. It should also be noted that the two scoring functions have lead to the same rankings of the forecasting procedures. The fully parametric methods (n-FP and st-FP) show the worst performance in terms of their predictive ability. n-FP falls into the red region against all other methods, whereas st-FP fails against the EVT methods and cannot win against the FHS methods; see the traffic light matrices in Figure~\ref{fig:TLMdata} (top row).  

On the other hand, for the 0.99855-expectile, the tests of simple conditional calibration are rejected (at 5\% level) for all the methods that use the normal likelihood. Those methods that use the skewed-t likelihood also tend to rank higher; although, in terms of significance, most methods fall into the yellow region (apart from the n-FP method). The ranking of forecasts is different for the two scoring functions used. The 0-homogeneous choice at \eqref{eq:Sexp2} clearly ranks the methods using the normal likelihood lower than those using the skewed-t likelihood in agreement with the results of the simple conditional calibration tests which is an argument in favour of using \eqref{eq:Sexp2} rather than \eqref{eq:Sexp1}.

For both $\VaR_{0.99}$ and  0.99855-expectile, the conditional calibration tests with the test functions as in the simulation study, lead to the failure of the corresponding traditional backtest; see Table~\ref{tabdata} for the expectile. This may seem overly-conservative for practical purposes, and suggests either re-examining suitability of the GARCH-type filter for these data, or the use of a more appropriate test function. For $\VaR_{0.99}$, we performed the conditional calibration tests also with the test function $\mathbf{h}_t = (1,V(r_{t-1},x_{t-1}))'$ (see Example~\ref{ex:tradVaR}) and the resulting p-values are reported in Table~\ref{tabdata}. They lead to conclusions similar to those based on the simple conditional calibration tests. This example underlines the importance of further studies on appropriate choices of test functions.

The results for $(\VaR_\n,\ebbS_\n)$ with $\n=0.975$ suggest better performance when a more flexible model such as the skewed-t is used to fit the AR(1)-GARCH(1,1) filter, although the use of EVT-based methods has a potential to compensate for likelihood mis-specifications. Again, fully parametric methods (n-FP and st-FP) fall into the red region in the comparative backtests against most of the other more flexible alternatives; see bottom panels in Table~\ref{tabdata} and Figure~\ref{fig:TLMdata}. The outcomes show one interesting aspect which is not in contradiction with the theory but may be puzzling and merit further investigation in future studies: The conditional calibration test rejects all methods using a normal likelihood but the scoring functions rank the n-EVT method as the best or second best performing method. It seems that the test function used in the conditional calibration test is sensitive to the likelihood function used in fitting the AR(1)-GARCH(1,1) filter whereas the scoring functions are more sensitive to the method at the second stage giving preference to the EVT methods.

\begin{table}
\caption{Summary of traditional and comparative backtesting based on the negated log-returns on the NASDAQ Composite index with an AR(1)-GARCH(1,1) filter fitted over moving estimation window of 500 observations, and the out-of-sample size of $n=$10,920; refer to Section~\ref{sdat} for details. The second column reports the average risk measure forecasts. ``\%~Viol." gives the percentage of $\VaR_{0.99}$ forecast exceedances. The simple CCT and general CCT columns contain the p-values for two-sided simple and general conditional calibration tests, respectively. The final two columns show the average scores, scaled by one minus the risk measure confidence level for presentation purposes, based on the specified scoring functions along with the corresponding method ranks (in brackets).} \label{tabdata}
\begin{tabular*}{1\textwidth}{@{\extracolsep{\fill}} l  c  c  c  c  c  c  } \hline
Method & $\ov{\VaR}_{0.99}$ & \% Viol. & simple CCT & general CCT & $\ov{S}$ [eq.~\eqref{eq:SVaR1}] & $\ov{S}$ [eq.~\eqref{eq:SVaR2}] \\\hline

n-FP    & 2.363 & 2.3 & {\bf 0.000} & {\bf 0.000} & 3.8497 ( 6 ) & 1.3017 ( 6 ) \\ 
n-FHS  & 2.758 & 1.3 & {\bf 0.017} & {\bf 0.028} & 3.5842 ( 3 ) & 1.1604 ( 3 ) \\ 
n-EVT  & 2.774 & 1.2 & 0.112 & 0.152 & 3.5675 ( 2 ) & 1.1550 ( 2 ) \\ 
st-FP  & 2.739 & 1.3 & {\bf 0.004} & {\bf 0.012} & 3.5976 ( 5 ) & 1.1669 ( 5 ) \\ 
st-FHS & 2.785 & 1.2 & {\bf 0.046} & 0.108 & 3.5904 ( 4 ) & 1.1609 ( 4 ) \\ 
st-EVT & 2.811 & 1.1 & 0.181 & 0.290 & 3.5654 ( 1 ) & 1.1517 ( 1 ) \\  \hline &&&&& \\ \hline 
 & \multicolumn{2}{c}{$\ov{e}_{0.99855}$}  & simple CCT & general CCT & $\ov{S}$ [eq.~\eqref{eq:Sexp1}] & $\ov{S}$ [eq.~\eqref{eq:Sexp2}] \\\hline
n-FP    & \multicolumn{2}{c}{2.363} & {\bf 0.000} & {\bf 0.000} & 25.9030 ( 6 ) & 0.9660 ( 6 ) \\ 
n-FHS  & \multicolumn{2}{c}{2.986} & {\bf 0.049} & {\bf 0.002} & 19.7333 ( 2 ) & 0.2933 ( 4 ) \\ 
n-EVT  & \multicolumn{2}{c}{2.966} & {\bf 0.023} & {\bf 0.001} & 19.8196 ( 5 ) & 0.3084 ( 5 ) \\ 
st-FP  & \multicolumn{2}{c}{3.041} & 0.163 & {\bf 0.011} & 19.8159 ( 4 ) & 0.2509 ( 1 ) \\ 
st-FHS & \multicolumn{2}{c}{3.078} & 0.227 & {\bf 0.011} & 19.7533 ( 3 ) & 0.2589 ( 2 ) \\ 
st-EVT & \multicolumn{2}{c}{3.037} & 0.112 & {\bf 0.006} & 19.6963 ( 1 ) & 0.2687 ( 3 ) \\ \hline &&&&& \\\hline 
 & \multicolumn{2}{c}{$\ov{\ebbS}_{0.975}$}  & simple CCT & general CCT & $\ov{S}$ [eq.~\eqref{eq:SVaRES1}] & $\ov{S}$ [eq.~\eqref{eq:SVaRES2}] \\\hline

n-FP    & \multicolumn{2}{c}{2.375} & {\bf 0.000} & {\bf 0.000} & 1.7020 ( 6 ) & 1.0492 ( 6 ) \\ 
n-FHS  & \multicolumn{2}{c}{2.777} & {\bf 0.022} & {\bf 0.035} & 1.6587 ( 4 ) & 0.9637 ( 4 ) \\ 
n-EVT  & \multicolumn{2}{c}{2.813} & 0.261 & {\bf 0.015} & 1.6560 ( 1 ) & 0.9607 ( 2 ) \\ 
st-FP  & \multicolumn{2}{c}{2.810} & {\bf 0.001} & 0.248 & 1.6622 ( 5 ) & 0.9691 ( 5 ) \\ 
st-FHS & \multicolumn{2}{c}{2.816} & 0.139 & 0.067 & 1.6582 ( 3 ) & 0.9617 ( 3 ) \\ 
st-EVT & \multicolumn{2}{c}{2.857} & 0.327 & 0.117 & 1.6563 ( 2 ) & 0.9597 ( 1 ) \\ \hline

\end{tabular*}
\end{table}

\setcounter{equation}{0}
\section{Discussion}\label{scon}

In the paper, we have discussed two approaches to backtesting risk measure forecasts. We differentiate between traditional backtesting, which gives a ``yes" or ``no" answer to the question of whether a method is acceptable or not, and comparative backtesting, specifically aimed at comparing the predictive performance of different forecasting methods. In general, there appears to be a need for both traditional and comparative backtesting methodologies. The former poses a requirement of identifiability on the risk measure functional, and serves the purpose of categorizing methods based on whether the backtest is passed or not, albeit with a somewhat limited ability to fail misspecified models. However, traditional backtesting does not provide a statistically justifiable basis for method comparisons often sought when assessing the performance of say a newly proposed forecasting procedure against an existing one, or when defending an internal procedure against some standard procedure. Comparative backtesting provides a methodology to serve exactly these purposes. For methods that are deemed acceptable under a traditional backtest, comparative backtesting allows to rank methods according to their predictive performance based on a chosen consistent scoring function, provided that the risk measure under consideration is an elicitable functional. 

Traditional backtesting, which we formalize in the form of conditional calibration tests, provides a unifying framework for currently available backtests of risk measures. To assess performance of different calibration tests in a controlled environment, a simulation study was conducted. It emerged that in fact many methods based on misspecified models may pass traditional backtests. And while the outcome of the backtest is the same in all such cases (a pass), differences in risk measure forecasts under different methods will ultimately lead to different capital requirements. One practical implication of this is that such backtests may create a wrong incentive of minimizing the capital, subject to passing the backtest, rather than aiming for a more accurate forecasting method. From the simulation study, we have also seen that general conditional calibration tests have a slightly better ability at detecting methods with misspecified models  in comparison to the corresponding simple conditional calibration tests, with the latter being able to flag only the most under-performing methods. However, for the real data,  often, simple and general conditional calibration tests produced similar results, suggesting that in practice the use of simple conditional calibration tests may suffice. General conditional calibration tests offer a more refined alternative, but require the choice of a test function. Further research is necessary to gain more insight into the choice of the test function for different risk measures and how this choice affects the outcomes of the tests.


In light of the above mentioned limitations of traditional backtests, regulators may additionally apply a comparative backtest in cases where a traditional backtest is passed. This necessitates a standard model against which the bank's internal model is to be tested. Such a standard model should not be confused with the standardized approaches currently used by regulators for trading book risk management of banks that either are not able to go for the (internal) model-based approach or do not pass the regulatory backtesting. These standardized approaches do not produce risk measure forecasts, and hence could not be incorporated into the comparative backtesting framework. However, comparative backtests will create the correct incentive for the banks to develop risk measure forecasting methods that aim for accuracy of forecasts and hence can adequately quantify the risks. If the Basel committee were to introduce comparative backtesting, a foresting method to serve as the ``standard model" should be chosen among flexible methods that have low model risk and are known to do well under fairly broad range of circumstances. One such possibility could be the filtered historical simulation with a GARCH filter fitted using a flexible likelihood model such as the skew-t in our numerical examples.

In summary, our recommendation to the Basel committee would be to adopt a two-stage backtesting framework. At stage~{\rm I}, a calibration test is applied in line with the current practice. In terms of implementation, the easiest option is to use the two-sided simple conditional calibration test. Conditionally on passing the stage~{\rm I} test, stage~{\rm II} will then assess the bank's ``internal model" against the regulator's ``standard model" via a comparative backtest. From the regulatory point of view, the statistical significance of the comparative backtests can be nicely summarized by means of traffic light matrices highlighting which methods pass or fail against a standard procedure, and when not enough evidence is available to make a conclusive statement. Provided that the regulatory risk measure is elicitable, comparative backtests require a choice of a consistent scoring function for that risk measure. In the case of backtesting ES, the current regulatory risk measure for banks' trading books, the 0-homogeneous scoring function in equation~\eqref{eq:SVaRES2} would be a reasonable choice as it is unit consistent and has milder moment restrictions on the underlying stochastic process than other positive homogeneous alternatives. Additionally, based on the data analysis, it yields results in rankings which are in better agreement with the outcomes of the calibration tests and leads to slightly more conclusive results in terms of the traffic light matrix entries versus the considered 1/2-homogeneous alternative.

It is worth noting that the comparative backtesting methodology can also be used by financial institutions internally to select better performing methods among competing alternatives. The same would apply to academic literature seeking to compare different forecasting methods, with the comparison done on the basis of forecast accuracy, in addition to calibration.


There are still many open problems and follow-up questions that require further investigation to create a fuller understanding of the usability of the presented backtesting methodologies. In the context of traditional backtesting, we found conditional calibration tests to be better at detecting model mis-specifications. However, these conditional tests require the user to choose a set of test functions. An exploration of potential test function choices and their influence on finite sample properties of the tests in a broader context than covered in our simulation study would be beneficial to guide practical applicability of these backtests. A choice problem also arises in the context of comparative backtesting where it is possible to make use of any member of the family of consistent scoring functions for a given risk measure functional. Here, different aspects of the resulting backtests can be assessed. One particular aspect to consider is the existence of the mean score (or difference in scores) for the underlying process. Financial time series tend to have fairly heavy tails and this would place restrictions on the choice of a suitable scoring function. From this perspective, the proposed scoring functions with 0-homogeneous score differences allow to study heavier-tailed processes than the $b$-homogeneous choices (with $b>0$). Finally, we have not explored the potentially promising possibility of using conditional comparative backtests. There are many open questions on how they should be formulated and implemented to be informative in practice.

Some of the risk measures used in practice are in fact non-elicitable. A prominent example here is the ES. In such cases the notion of joint elicitability may open the door to the ability to conduct backtests, in this case for multivariate risk measure functionals. We have explored the joint elicitability of VaR and ES, and, on the basis of our simulation study, the backtesting results show a good ability to identify and differentiate among methods relying on correct and misspecified model formulations. However, further research is needed to provide a clearer interpretation of  both traditional and comparative backtests. For example, in the case of the pair (VaR, ES), the question would be whether it is a poor forecasting of VaR or ES or both that caused a (traditional or comparative) backtest to fail.

\vspace{0.5cm}
{\large {\bf Acknowledgements.}} 
We would also like to thank Prof.~Paul Embrechts for a number of inspiring discussions, as well as RiskLab at ETH Zurich for its hospitality when we began working on this project. Financial support of the Natural Sciences and Engineering Research Council of Canada (N.~Nolde) and Swiss National Foundation via grant 152609 (J.~F.~Ziegel) is gratefully acknowledged. 

\bibliographystyle{plainnat} 
\bibliography{biblio}{}

\appendix
\renewcommand{\theequation}{\Alph{section}.\arabic{equation}}
\section*{Appendix}
\section{Backtesting and forecast comparisons: an example}\label{sec:motiv.ex}

In \citet[Section 1.2]{Gneiting2011}, a simulation study is reported to illustrate how using inconsistent scoring functions for forecast comparison may lead to grossly misguided conclusions. The simulation study of \citet{Gneiting2011} is concerned with forecasting a ``best'' point estimate of a random variable in the sense of a most probable value or a mean, which may not be so relevant to risk management where the interest lies in the extreme events. Therefore, we would like to report a small simulation study that shows that similar problems with forecast comparison may occur in a scenario where the interest is in forecasting $\VaR_{\alpha}$ or $\ebbS_{\nu}$ at high levels $\alpha$ or $\nu$, respectively.

Suppose we observe realizations $x_t$, $t = 1,2,\dots$ of a risky asset $\{X_t\}_{t \in \nbb}$ following a GARCH(1,1)-model given by
\begin{equation}\label{eq:simmodel}
X_t = \sigma_t Z_t,
\end{equation}
where $\sqrt{2}Z_t$ has a Student's $t$ distribution with four degrees of freedom and
\[
\sigma_t^2 = 0.20 X_{t-1}^2 + 0.75 \sigma_{t-1}^2 + 0.05.
\]
We are interested in $\VaR_{\alpha}$ or $\ebbS_{\nu}$ predictions at levels $\alpha = 0.99$ and $\nu=0.975$, respectively. In both cases we consider four forecasters that give predictions for timepoint $t$ based on the information available at timepoint $t-1$. The magician predicts 
\[
R_{V,\alpha,t}^{(1)}=\sigma_t \sqrt{2} t_4^{-1}(\alpha),
\]
for $\VaR_{\alpha}(X_t)$, and
\[
R_{E,t}^{(1)}=\sigma_t\sqrt{2}\frac{g_{4}(t_{4}^{-1}(\nu))}{1-\nu}\Big(\frac{4 + (t_{4}^{-1}(\nu))^2}{3}\Big),
\]
for $\ebbS_{\nu}(X_t)$, where $t_4^{-1}$ is the quantile function of the $t$ distribution with four degrees of freedom and $g_4$ its density function. The predictions of the magician are best possible as they correspond to $\VaR_\alpha$ and $\ebbS_\nu$ of the $X_t$ conditional on the information available at $t-1$. The other three forecasters are called historian-$n$, where $n=250$, $500$ or $1000$. They predict $R_{V,\alpha,t}^{(i)}$, $i=2,3,4$, as the empirical $\alpha$-quantile of the last $n$ observations for $\VaR_{\alpha}(X_t)$ and $R_{E,t}^{(i)}$, $i=2,3,4$, the empirical expected shortfall for $\ebbS_\nu(X_t)$, that is the mean of the observations exceeding the empirical $\nu$-quantile.

In the case of $\VaR_{\alpha}$, there are at least two possibilities for comparing the forecasters' performance. Firstly, we can compare percentages of exceedances over $R_{V,t}^{(i)}$, $i=1,\dots,4$, respectively, and secondly we can evaluate the forecasts using the performance criterion
\begin{equation}\label{eq:perf}
\overline{S}^{(i)} = \frac{1}{n}\sum_{t=1}^n S(R_{V,\alpha,t}^{(i)},X_t)
\end{equation}
for $n$ observed time points, where $S$ is a (strictly) consistent scoring function for $\VaR_{\alpha}$; see~(2.3). Here, the different forecasts $\{R^{(i)}_{V,t}\}_{t \in \nbb}$, $i=1,\dots,4$, are sorted according to their predictive performance, that is, the lower $\overline{S}^{(i)}$, the better the forecasting procedure. The first procedure has been employed for example in \citet[Section 2.3.6]{McNeil2005}. The second one has recently been promoted by \citet{Gneiting2011} and is commonplace in the econometrics literature; see \citet{DieboldMariano1995}.

The results are given in Table \ref{tab:1} for the choice $G(r) = r$ in~(2.3). In summary, for a very long time series of 95,000 observations, both, the percentage of VaR exceedances as well as the mean score $\bar{S}$ identify the magician as the most accurate forecaster. However, for a more realisitic sample size of 5,000 observations using the percentage of VaR exceedances would deem the historian-1000 a better forecaster than the magician. The mean score is not affected by this problem and still clearly identifies the magician as the best.
\begin{table}
\begin{center}
\begin{tabular}{l|cc|cc}
&\multicolumn{2}{c|}{Length: 95,000}&\multicolumn{2}{c}{Length: 5,000}\\\hline
Forecaster & \% of VaR exceedances & Mean score $\bar{S}$ & \% of VaR exceedances & Mean score $\bar{S}$\\\hline
Magician & \textbf{1.04} & \textbf{0.0309} & 1.08 & \textbf{0.0275}\\
Historian-250 & 1.57 & 0.0427 & 1.42 & 0.0303\\
Historian-500 & 1.34 & 0.0428 & 1.20 & 0.0309\\
Historian-1000 & 1.16 & 0.0429 & \textbf{0.96} & 0.0302\\[1ex]
\end{tabular}
\caption{Ordering forecasters by percentage of VaR$_{0.99}$ exceedances or by the mean score of a consistent scoring function for a time series simulated from the model given at \eqref{eq:simmodel}.\label{tab:1}}
\end{center}
\end{table}

For backtesting ES, \citet{McNeilFrey2000} introduced the following test statistic based on exceedance residuals
\begin{equation}
T_5=\frac{1}{\#\{t: X_t > R_{V,\nu,t}^{(i)}\}}\sum_{t=1}^T \frac{X_t - R_{E,t}^{(i)}}{\sigma_t} \ind\{X_t > R_{V,\nu,t}^{(i)}\}.
\end{equation}
If the model \eqref{eq:simmodel} is correct, the non-zero summands of $T_5$ are an i.i.d. sample of a mean zero random variable and this can be used for a hypothesis test. One may be tempted to use this test statistic also for forecast comparison calling a forecaster more accurate the closer the mean exceedance residuals are to zero; see, for example, \citet{ChunShapiroETAL2012}. In practice, also $\sigma_t$ has to be replaced by an estimate. In our simulation study, we used the true~$\sigma_t$. 

A second possibility to compare the performance of the ES predictions is by using one of the consistent scoring functions given in (2.5) for the pair $(\VaR_\nu,\ebbS_\nu)$ in the performance criterion \eqref{eq:perf}. We choose $G_1(r_1)=r_1$ and $G_2(r_2) = \exp(r_2)/(1+\exp(r_2))$ as in \citet{FisslerZiegelETAL2015}; see also Section 2.3.1 for a discussion of the choice of scoring functions. The results are given in Table \ref{tab:2}. For the unrealistically long time series with 95,000 observations, the historian-500 is preferred over the magician in terms of mean exceedance residuals, while for 5,000 observations, the historian-250 is preferred by the exceedance residuals. In both cases, the mean score $\bar{S}$ correctly identifies the magician as the most skillful forecaster. 

\begin{table}
\begin{center}
\begin{tabular}{l|cc|cc}
&\multicolumn{2}{c|}{Length: 95,000}&\multicolumn{2}{c}{Length: 5,000}\\\hline
Forecaster & Mean exceedance residuals & Mean score $\bar{S}$ & Mean exceedance residuals & Mean score $\bar{S}$\\\hline
Magician & -0.0102 & \textbf{-0.0610} & 0.1437 & \textbf{-0.658}\\
Historian-250 & 0.1067 & 0.0253 & \textbf{0.0585} & 0.492\\
Historian-500 & \textbf{0.0084} & 0.0246 & -0.2021 & 0.457\\
Historian-1000 & -0.2227 & 0.0348 & -0.4456 & 0.466\\
\end{tabular}
\caption{Ordering forecasters by mean exceedance residuals and by the mean score of a consistent scoring function for a time series simulated from the model given at \eqref{eq:simmodel} of length 95,000.\label{tab:2}}
\end{center}
\end{table}

This small simulation example illustrates that the test statistics used for traditional backtesting are not suitable to compare different risk measurement procedures but that a performance criterion as defined at \eqref{eq:perf} with a consistent scoring function should be used to obtain a meaningful ordering. We would like to emphasize that this problem is not a defect of the particular test statistic used for the traditional backtest, that is percentages of VaR exceedances or exceedance residuals for ES. Traditional backtests are simply not designed for comparisons between models. This fact may be problematic when traditional backtests are used in regulatory frameworks as this may create an incentive to optimize the test statistic used rather than optimizing predictive performance. Our simulation example shows that both goals may lead to different choices of the risk measurement procedure. This problem may be resolved using comparative backtests instead. In the main paper, we provide more details on the differences between traditional and comparative backtests and how they relate to identifiability and elicitability, respectively. 

\setcounter{equation}{0}

\section{Expectiles}\label{sec:expectile}
\subsection{Model-based computation}

Let $X$ be an random variable with finite mean. If $X$ is not constant, then expectile $e_{\cdot}(X)$ is the (generalized) inverse of the function
\[
G_X\colon \mathbb{R} \to [0,1], \quad z \mapsto \frac{\int_{-\infty}^z|z - y| dF_X(y)}{\int_{-\infty}^{\infty}|z - y|dF_X(y)} = \frac{\ebb(|z - X|\mathbbm{1}\{X \le z\})}{\ebb(|z - X|)};
\]
see \citet{AbdousRemillard1995}. Alternatively, the function $G_X(\cdot)$ can be written as
\bql{qGv2} G_X(z)=\dfrac{zF_X(z)-M_X(z)}{2(zF_X(z)-M_X(z))+\ebb(X)-z},\eql
where $M_X(z)=\int_{-\nf}^z y dF_X(y)$ is the partial moment of $X$. For simplicity of notation, we will omit the subscript indicating the underlying random variable in the notation of function $G_X$, when there is no ambiguity. The function $G$ is right-continuous, increasing, and $G(-\infty) = 0$, $G(\infty) = 1$. Note that there is no need to know the normalizing constant of the density in order to calculate $G$. Quantiles and expectiles both characterize the distribution of $X$. However, they are quite different in nature.

Expectiles are related to the functional $\Omega$ of \citet{KeatingShadwick2002}, defined as
\[
\Omega\colon \mathbb{R} \to [0,\infty), \quad r \mapsto \Omega(r):=\frac{\int_r^{\infty}|y-r|dF(y)}{\int_{-\infty}^r|y - r|dF(y)}.
\]
In particular, we have
\begin{equation}\label{eq:magical}
G(r) = \frac{1}{1 + \Omega(r)}, \quad \Omega(r) = \frac{1}{G(r)} - 1.
\end{equation}


In the following examples we present the function $G_X(\cdot)$ for a number of probability distributions. Examples~\ref{ex:n}, \ref{ex:t} and~\ref{ex:st} are used for computing model-based expectiles in the simulation study.

\begin{exmp}[Normal distribution]\label{ex:n}
If $X \sim \mathcal{N}(\mu,\sigma^2)$, then
\[
G_X(z) = \frac{\sigma\varphi\big(\frac{z-\mu}{\sigma}\big) + (z - \mu)\Phi\big(\frac{z-\mu}{\sigma}\big)}{2\sigma\varphi\big(\frac{z-\mu}{\sigma}\big) + (z - \mu)\big(2\Phi\big(\frac{z-\mu}{\sigma}\big) - 1\big)}
\]
where $\varphi$ and $\Phi$ denote the density and distribution function of a standard normal random variable. The $\tau$-expectile of $X$ is given by $\mu_{\tau}(X) = G_X^{-1}(\tau)$.
\end{exmp}
\begin{exmp}[Exponential distribution]
If $X \sim EXP(\lambda)$ for some $\lambda > 0$, then
\[
G_X(z) = \frac{z - \frac{1}{\lambda} + \frac{1}{\lambda}e^{-\lambda z}}{z - \frac{1}{\lambda} + \frac{2}{\lambda}e^{-\lambda z}}.
\]
\end{exmp}
\begin{exmp}[Student's $t$ distribution]\label{ex:t}
If $X$ has a $t$ distribution with $\nu > 1$ degrees of freedom, then
\[
G_X(z) = \frac{\frac{\nu+z^2}{\nu-1}g_{\nu}(z) + z t_{\nu}(z)}{2\frac{\nu+z^2}{\nu-1}g_{\nu}(z) + z \big(2 t_{\nu}(z)-1\big)},
\]
where $g_\nu$ and $t_\nu$ denote the density and the cumulative distribution function of the $t$ distribution.
\end{exmp}
\begin{exmp}[Pareto distribution]
For a Pareto distribution with density
\[
f(x) = \frac{\alpha}{x^{\alpha+1}}, \quad x \ge 1,
\]
where $\alpha > 1$, we obtain for $z \ge 1$
\[
G_X(z)= \frac{\alpha (1-z)  + z F(z)}{\alpha (1-z) +  z (2F(z) - 1)},
\]
where $F$ is the cumulative distribution function of $f$.
\end{exmp}

\begin{exmp}[Generalized Pareto distribution]\label{ex:gp} The cumulative distribution function of a generalized Pareto distribution with scale $\s>0$ and shape parameter $\x\in\rbb$, denoted $GP(\s,\x)$, is given by
$$H(y) = 1 - \Big(1+\x y/\s \Big)^{-1/\x},\qquad x\ge0\ {\rm and\ } 1+\x x/\s\ge0.$$
Hence, if $X\sim GP(\s,\x)$ then
$$G_X(z)=\dfrac{z-(\s+\x z) H(z)}{\s+z(1+\x)-2(\s+\x z) H(z)},\qquad z\ge0\ {\rm and\ } 1+\x z/\s\ge0,$$
provided $\x<1$.
\end{exmp}

\begin{exmp}[Skewed t distribution] \label{ex:st} Consider a skewed Student's t distribution with density of the form:
\bql{qfst}f(x)=\dfrac{2}{1/\g+\g}\Big(g_\n(\g x)\ind{\{x\le0\}} +g_\n(x/\g)\ind{\{x>0\}}  \Big),\qquad x\in\rbb,\eql
where $\n>0$ and $\g>0$ are the shape and skewness parameters, respectively, and as before $g_\n$ is the density of the Student's t distribution with $\n$ degrees of freedom; see \citet{Hansen1994} and \citet{Fernandez1998}. The inverse expectile function in this case is given by
$$G_X(z)=\dfrac{zF(z) +\dfrac1{\n-1}\Big(\dfrac{\n}{\g^2}+z^2\Big)f(z)}
{2\Big(zF(z) +\dfrac1{\n-1}\Big(\dfrac{\n}{\g^2}+z^2\Big)f(z)\Big) + \ebb(X) -z}=:G_X^-(z) \quad{\rm for\ } z<0,$$
and $$G_X(z)=\dfrac{zF(z) +\dfrac1{\n-1}(\n\g^2+z^2)f(z) - \dfrac{\n}{\n-1}(\g^2-1/\g^2)f(0)}
{2\Big(zF(z)+\dfrac1{\n-1}(\n\g^2+z^2)f(z) - \dfrac{\n}{\n-1}(\g^2-1/\g^2)f(0)\Big) + \ebb(X) -z}=:G_X^+(z) \quad{\rm for\ } z\ge0,$$
where $f$ and $F$ are the density and cumulative distribution function, respectively, of the skewed Student's~t distribution introduced above. Note $G_X(0)=1/(1+\g^4)$, and hence $e_\t(X)=(G_X^-)\inv(\t)$ for $\t<1/(1+\g^4)$ and $e_\t(X)=(G_X^+)\inv(\t)$ for $\t\ge1/(1+\g^4)$.

Using moments expressions of an asymmetric Student's t distribution in \citet{Zhu2010}, it can be shown that if random variable $X$ has a skewed t distribution with density in \eqref{qfst}, then the mean and variance of $X$ are given by
$$\ebb(X)=2K_\n\dfrac{\n}{\n-1}\Big(\g-\dfrac1{\g}\Big) $$
and
$$Var(X)=\Big[\dfrac{\n}{\n-2}\Big(1-3\dfrac{\g^2}{(1+\g^2)^2}\Big)-4K_\n^2\Big(\dfrac{\n}{\n-1}\Big)^2\Big( 1-\dfrac{2}{1+\g^2} \Big)^2\Big]\Big(\g+\dfrac1{\g}\Big)^2,\qquad \n>2, $$
respectively, where $K_\n=\GG((\n+1)/2)/[\sqrt{\p\n}\ \GG(\n/2)]$. These expressions can be used to compute an expectile of a skewed t distribution with mean zero and variance one.
\end{exmp}

\begin{exmp}[Asymmetric Student-t distribution] \citet{Zhu2010} introduced a more general class of asymmetric Student-t (AST) distributions which includes the one mentioned in the previous example. This class of models allows for different shape parameters in the upper and lower tails of the distribution. The density of an AST distribution with skewness parameter $\a\in(0,1)$ and lower and upper tail shape parameters $\n_1>0$ and $\n_2>0$, respectively, is equal to
\bql{qfast} f(x) = \dfrac{\a}{\a^\ast}K(\n_1)\Big[1+\dfrac1{\n_1}\Big(\dfrac{x}{2\a^\ast} \Big)^2 \Big]^{-\frac{\n_1+1}{2}}\ind{\{x\le0\}} + \dfrac{1-\a}{1-\a^\ast}K(\n_2)\Big[1+\dfrac1{\n_2}\Big(\dfrac{x}{2(1-\a^\ast)} \Big)^2 \Big]^{-\frac{\n_2+1}{2}}\ind{\{x>0\}},
\eql
where $K(\n)=\GG((\n+1)/2)/(\GG(\n/2)\sqrt{\n\p})$ and $\a^\ast=\a K(\n_1)/[\a K(\n_1)+(1-\a)K(\n_2)]$. 
The inverse expectile function $G_X(\cdot)$ can be obtained from 
$$G_X(z)=\dfrac{zF(z)-M(z)}{2(zF(z)-M(z))+\ebb(X)-z} $$
with the partial moment function $M(\cdot)$ given by
$$M(z)=\bcs-4(\a^\ast)^2\dfrac{\n_1}{\n_1-1}\Big[1+\dfrac1{\n_1}\Big(\dfrac{z}{2\a^\ast} \Big)^2 \Big]f(z), & z\le0\\
-4(1-\a^\ast)^2\dfrac{\n_2}{\n_2-1}\Big[1+\dfrac1{\n_2}\Big(\dfrac{z}{2(1-\a^\ast)} \Big)^2 \Big]f(z)-4B\Big[(\a^\ast)^2\dfrac{\n_1}{\n_1-1}+(1-\a^\ast)^2\dfrac{\n_2}{\n_2-1}\Big], & z>0
\ecs,$$
where $B=\a K(\n_1)+(1-\a) K(\n_2)$. As before, $f$ and $F$ denote density and cumulative distribution function of random variable $X$ here having the AST distribution. Expected value and variance of $X$ are 
$$\ebb(X)=4B\Big[ -(\a^\ast)^2\dfrac{\n_1}{\n_1-1} + (1-\a^\ast)^2\dfrac{\n_2}{\n_2-1}\Big] $$
and
$$Var(X)=4\Big[\a(\a^\ast)^2\dfrac{\n_1}{\n_1-2}+(1-\a)(1-\a^\ast)^2\dfrac{\n_2}{\n_2-2} \Big] -\ebb(X)^2;$$
see Equations~(14) and~(15) in~\citet{Zhu2010}.
\end{exmp}

\subsection{EVT expectile estimator}\label{sec:evt_expectile}


Given the series of standardized residuals $\{\hat z_t;\ t=1,\ldots,n\}$ in~(3.3), we now discuss a semi-parametric estimation of the expectile of $Z_t$'s based on asymptotic results of EVT.

In order to obtain an estimator of the expectile for the i.i.d. series $\{Z_t\}$, we first derive an estimator of the function $G_Z(z)$  in~\eqref{qGv2}, whose inverse will then give us an estimator of the $\t$-expectile, $e_\t(Z)$, of $Z_t$'s.

Recall the Omega ratio:
\begin{equation}\label{qor}\Omega_Z(z)=\dfrac{\int_z^\nf |z-y|dF_Z(y)}{\int_{-\nf}^z |z-y|dF_Z(y)}. \end{equation}
We first assume that the $\t$-expectile of $Z_t$, given by $e_Z(\t)=G_Z\inv(\t)=\Omega_Z\inv(1/\t-1)$, exceeds the chosen threshold $u$. This would be the case for large values of $\t$, which are of interest from the risk measurement perspective placing emphasis on the far upper tail of the loss distribution. 

The integral in the numerator of~\eqref{qor} can be written as:
\begin{align*}
\int_z^\nf |z-y|dF_Z(y) = \ebb[(Z-z)\ind{\{Z>z\}}]=\ov{F}_Z(z)\ \ebb(Z-z\mid Z>z).
\end{align*}
The threshold stability property of the GP distribution says that if $X-u\mid X>u\sim GP(\b_u,\x)$ then, for any $v\ge u$, $X-v\mid X>v\sim GP(\b_u+\x(v-u),\x)$. Also, if $X\sim GP(\b,\x)$ then $\ebb(X)=\b/(1-\x)$, provided $\x<1$ (see, e.g., \citet{EKM1997}, Theorem~3.4.13(a)). Combining these two facts we find
$$\ebb(Z-z\mid Z>z)=\dfrac{\b_u+\x(z-u)}{1-\x},\qquad \x<1,\ z>u, $$
which can be estimated by replacing parameters $\b_u$ and $\x$ with their estimates.

Turning to the integral in the denominator of~\eqref{qor}, write
\begin{align*}
\int_{-\nf}^z (z-y) dF_Z(y)& = z F_Z(z) - \ebb(Z\ind{\{Z\le z\}})\\
& = z F_Z(z) - \ebb(Z\ind{\{Z\le u\}}) - \ebb(Z\ind{\{u<Z\le z\}}).
\end{align*}
The first expectation above can be estimated empirically: $$\hat\ebb(Z\ind{\{Z\le u\}})=\dfrac1{n}\sum_{t=k+1}^n \hat z_{(t)}=\dfrac1{n}\sum_{t=1}^n \hat z_t\ind{\{\hat z_t\le u\}}=:\ov z_u.$$
For the second expectation, we have
$$\ebb(Z\ind{\{u<Z\le z\}}) = \ov F_Z(u) \ebb(Z\ind{\{Z\le z\}}\mid Z>u) = \ov F_Z(u)\ebb((Z-u)\ind{\{Z\le z\}}\mid Z>u) +u\ov F_Z(u), $$
where, using the peaks-over-threshold tail estimator (see, e.g., \cite{EKM1997}, equation~(6.45)),
\begin{align*}
\ebb((Z-u)\ind{\{Z-u\le z-u\}}\mid Z>u)&=\int_0^{z-u} y\ \dfrac1{\b_u}(1+\x y/\b_u)^{-1/\x -1} dy\\
&=\dfrac{\b_u}{1-\x}\Big\{1-\Big(1+\dfrac{z-u}{\b_u} \Big) \Big(1+\x\dfrac{z-u}{\b_u} \Big)^{-1/\x} \Big\}.
\end{align*}
Combining above derivations gives the following estimator for the Omega ratio:
$$\widehat\Omega_Z(z)= \dfrac{\dfrac{k}{n}\ \dfrac{\hat\b_u}{1-\hat\x}\ \Big(1+\hat\x\ \dfrac{z-u}{\hat\b_u} \Big)^{-1/\hat\x+1}}{z+\dfrac{k}{n}\Big(1+\hat\x\ \dfrac{z-u}{\hat\b_u} \Big)^{-1/\hat\x}\Big(\dfrac{\hat\x}{1-\hat\x}(z-u) + \dfrac{\hat\b_u}{1-\hat\x} -u\Big) - c},\qquad z>u=\hat z_{(k+1)},$$
where $$c=\ov z_u +\dfrac{k}{n}\Big(u+\dfrac{\hat\b_u}{1-\hat\x} \Big). $$
Based on the relationship between the function $G_Z$ and the Omega ratio, we have $\widehat G_Z(z)=1/(1+\widehat\Omega_Z(z))$, and hence the EVT-based estimator for the expectile of $Z_t$'s is given implicitly by the inverse $\hat e_\t^{\text{{\tiny EVT}}}(Z)=\widehat G_Z\inv(\t)$, provided $\widehat G_Z(u)>\t$ and $\hat\x<1$. 

If $\widehat G_Z(u)\le\t$, the empirical estimator of $e_\t(Z)$ can be used.

\setcounter{equation}{0}
\section{Characterization of positive homogeneous scoring functions}

In this section we characterize strictly consistent scoring functions for three risk measures VaR, expectiles and (VaR,ES) so that the resulting score differences are positive homogeneous. For VaR, we consider the class of scoring functions given by 
\begin{equation}\label{eq:qSVaR}
S(r,x) = (1 - \alpha - \ind\{x > r\})G(r) + \ind\{x > r\}G(x),
\end{equation}
where $G$ is a strictly increasing function; compare Proposition 1. 

\begin{theorem}\label{thm:VaRhom} {\rm(Value-at-Risk)}
\begin{enumerate}
\item Let $b > 0$. The only scoring functions $S:\mathbb{R} \times \mathbb{R} \to \mathbb{R}$ of the form \eqref{eq:qSVaR} that are positively homogeneous of degree $b$ are obtained by choosing $G(x) = (c_0\mathbbm{1}\{x \ge 0\} - c_0'\mathbbm{1}\{x < 0\})|x|^b$ with constants $c_0,c_0' > 0$. 
\item Let $b < 0$. The only scoring functions $S:(0,\infty) \times \mathbb{R} \to \mathbb{R}$ of the form \eqref{eq:qSVaR} that are positively homogeneous of degree $b$ are obtained by choosing $G(x) = -c_0 x^b$, $x > 0$ with some constant $c_0 > 0$.\newline They cannot be extended to yield a strictly consistent scoring function on $\mathbb{R} \times \mathbb{R}$. 
\item There is no positively homogeneous scoring function of degree $b=0$ of the form \eqref{eq:qSVaR}. 
\item Choosing $G(x) = c_0 + c_1 \log x$, $x > 0$ with $c_0 \in \mathbb{R}$ and $c_1 > 0$ in \eqref{eq:qSVaR} is the only way for obtaining a scoring function $S$ of the form \eqref{eq:qSVaR} such that the score difference $(0,\infty) \times (0,\infty) \times \mathbb{R}, (r,r',x) \mapsto S(r,x) - S(r',x)$ is positively homogeneous of degree $b=0$.
\end{enumerate}
\end{theorem}

\bew
Let $b \in \mathbb{R}$. If a scoring scoring function for $\VaR_\alpha$ as given in~(2.3) is positively homogeneous of order $b$, then for all $r \in (0,\infty)$, $x \in \mathbb{R}$, $c \in (0,\infty)$, we obtain
\[
S(cr,cx) = (1-\alpha - \mathbbm{1}\{x > r\})G(cr) + \mathbbm{1}\{x > r\}G(cx) = c^b(1-\alpha - \mathbbm{1}\{x > r\})G(r) + c^b\mathbbm{1}\{x > r\}G(x) = c^bS(r,x).
\]
Choosing $x=0$, $r =1$, we obtain 
\[
G(c) = c^b G(1), \quad \text{for all $c \in (0,\infty)$.}
\]
Therefore, the function $G$ is strictly increasing on $(0,\infty)$ if $b \not=0$ and $G(1) > 0$ for $b > 0$ and $G(1) < 0$ for $b < 0$. For $b = 0$, the function $G$ is constant and thus, there is no stricly consistent score of order $b=0$. Considering score differences, we obtain for all $r,r' \in (0,\infty)$, $x \in \mathbb{R}$, $c \in (0,\infty)$ that $S(cr,cx) - S(cr',cx) = S(r,x) - S(r',x)$, that is, for $x = 0$, $r' = 1$ and all $r,c \in (0,\infty)$
\[
G(cr)-G(1) = G(c)-G(1) + G(r) - G(1).
\]
As $G$ is required to be strictly increasing, the only solutions to this functional equation on $(0,\infty)$ are $G(r) = c_0 + c_1\log r$ with constants $c_0 \in\mathbb{R}$ and $c_1 > 0$.  
\qed

For expectiles, we consider the class of scoring functions given by
\begin{equation}\label{eq:qSexp}
S(r,x) = \ind\{x > r\}(1-2\tau)(\phi(r) -\phi(x) - \phi'(r)(r-x)) - (1-\tau)(\phi(r) - \phi'(r)(r-x)),
\end{equation}
where $\phi$ is a strictly convex twice differentiable function; compare Proposition 1.

\begin{theorem}\label{thm:exphom} {\rm(Expectiles)} 
\begin{enumerate}
\item Let $b > 1$. The only scoring functions $S:\mathbb{R} \times \mathbb{R} \to \mathbb{R}$ of the form \eqref{eq:qSexp} that are positively homogeneous of degree $b$ are obtained by choosing $\phi(x) = (c_1\mathbbm{1}\{x \ge 0\} + c_1'\mathbbm{1}\{x < 0\}) |x|^b$, $x \in\rbb$ with constants $c_1,c_1' > 0$.
\item Let $b < 1$, $b \not=0$. The only scoring functions $S:(0,\infty) \times \mathbb{R} \to \mathbb{R}$ of the form \eqref{eq:qSexp} that are positively homogeneous of degree $b$ are obtained by choosing $\phi(x) = c_1x^b/(b(b-1))$, $x > 0$ with some constant $c_1 > 0$.\newline They cannot be extended to yield a strictly consistent scoring function on $\mathbb{R} \times \mathbb{R}$. 
\item There is no positively homogeneous scoring function of degree $b \in \{0,1\}$ of the form \eqref{eq:qSexp}.
\item Choosing $\phi(x) = c_0 - c_1 \log x + c_2 x$, $x > 0$ with $c_0,c_2 \in \mathbb{R}$ and $c_1 > 0$ in \eqref{eq:qSexp} is the only way for obtaining a scoring function $S$ of the form \eqref{eq:qSexp} such that the score difference $(0,\infty) \times (0,\infty) \times \mathbb{R}, (r,r',x) \mapsto S(r,x) - S(r',x)$ is positively homogeneous of degree $b=0$. 
\item Choosing $\phi(x) = c_0 + c_1 x \log x + c_2 x$, $x > 0$ with $c_0,c_2 \in \mathbb{R}$ and $c_1 > 0$ in \eqref{eq:qSexp} is the only way for obtaining a scoring functions $S$ such that the score difference $(0,\infty) \times (0,\infty) \times \mathbb{R}, (r,r',x) \mapsto S(r,x) - S(r',x)$ is positively homogeneous of degree $b=1$.
\end{enumerate}
\end{theorem}

\bew
Let $b \in \mathbb{R}$. The relation $S(cr,cx) = c^b S(r,x)$ has to hold for all $r,c \in (0,\infty)$, $x \in \mathbb{R}$. Using the form of the scoring functions in~(2.4), the relation for $x = 0$ and $r = 1$ implies
\[
\phi(c) - c\phi'(c) = c^b(\phi(1) - \phi'(1)).
\]
We find that $\phi''(c) = -(\phi(1) - \phi'(1))bc^{b-2}$. If $\phi(1) = \phi'(1)$ or $b=0$, we obtain that $\phi$ is linear, hence not strictly convex and therefore, there is no striclty consistent score of order $b=0$. If $b = 1$ and $\phi(1) \not=\phi'(1)$, we obtain that $\phi(x) = c_0 + c_1 x\log x + c_2 x$ for $x \in (0,\infty)$ with $c_0, c_2 \in \mathbb{R}$ and $c_1 > 0$. However, the corresponding scoring function is not homogeneous of order $b=1$, which can be shown by plugging in the explicit expression for $\phi$ in~(2.4). For $b \not\in \{0,1\}$, we obtain that $\phi(x) = c_0 + c_1x^b/(b(b-1)) + c_2 x$ for $x \in (0,\infty)$ with $c_0, c_2 \in \mathbb{R}$ and $c_1 > 0$. By plugging in this form of $\phi$ in~(2.4), we find that we obtain a homogeneous scoring function of order $b \not\in\{0,1\}$ for $c_0 = c_2 = 0$. For $b < 1$, the function $\phi$ cannot be extended to a convex function on all of $\mathbb{R}$. For $b > 1$, we can use the homogeneity relation with $x=-2$ and $r = -1$ to obtain $\phi(-c) - c\phi'(-c) = c^b(\phi(-1) - \phi'(-1))$ for all $c \in (0,\infty)$ which yields the claim with the same arguments as above.

Considering score differences for $b=1$, we obtain for all $r,r' \in (0,\infty)$, $x \in \mathbb{R}$, $c \in (0,\infty)$ that $S(cr,cx) - S(cr',cx) = S(r,x) - S(r',x)$, that is, for $x = 0$, $r' = 1$ and all $r,c \in (0,\infty)$
\[
\phi(cr) - c\phi(r) - cr(\phi'(cr) - \phi'(r)) = \phi(c) - c\phi'(c) - c\phi(1) + c\phi'(1).
\]
Differentiating with respect to $r$ at $r = 1$, we obtain $\phi''(x) = c_1 x^{-1}$ for some $c_1 > 0$, hence $\phi(x) = c_0 + c_1x\log x + c_2 x$ for $x \in (0,\infty)$ with $c_0, c_2 \in \mathbb{R}$ and $c_1 > 0$. One can check that the resulting score difference is in fact homogeneous of order $b=1$. Considering score differences for $b=0$, $x = 0$, $r' = 1$ and all $r,c \in (0,\infty)$, we obtain
\[
\phi(cr) - c\phi(r) - cr(\phi'(cr) - \phi'(r)) = \phi(c) - c\phi'(c) - c\phi(1) + c\phi'(1).
\]
Again, differentiating with respect to $r$ at $r = 1$ yields that $\phi''(x)= c_1 x^{-2}$ for some $c_1 > 0$, hence $\phi(x) = c_0 - c_1\log x + c_2 x$ for $x \in (0,\infty)$ with $c_0, c_2 \in \mathbb{R}$ and $c_1 > 0$ which yields homogeneous score differences of order $b=0$.

\qed

For $(\VaR_\n,\ebbS_\n)$, we consider the class of scoring functions given by 
\begin{equation}\label{eq:qSVaRES}
S(r_1,r_2,x) = \ind\{x > r_1\}\big(-G_1(r_1) + G_1(x) - G_2(r_2)(r_1 - x)\big) + (1-\nu)\big(G_1(r_1) - G_2(r_2)(r_2 - r_1) + \mathcal{G}_2(r_2) \big),
\end{equation}
where $G_1$ is an increasing function, $\mathcal G_2$ is twice differentiable, strictly increasing and strictly concave, and $\mathcal G_2' = G_2$; compare Proposition 3 in the main paper. The next result gives a characterization of positively homogeneous scoring functions for $(\VaR_\n,\ebbS_\n)$. It includes the 0-homogeneous case, which is considered in~\citet{PattonZiegel2016}. 

\begin{theorem}\label{thm:VaREShom} {\rm(Value-at-Risk and expected shortfall)} 
\begin{enumerate}
\item Let $b \in (0,1)$. The only scoring functions $S:\mathbb{R} \times (0,\infty) \times \mathbb{R} \to \mathbb{R}$ of the form \eqref{eq:qSVaRES} that are positively homogeneous of degree $b$ are obtained by choosing $G_1(x) = (d_1\mathbbm{1}\{x \ge 0\} - d_1'\mathbbm{1}\{x < 0\})|x|^b - c_0$ and $\mathcal{G}_2(x)= c_1 x^b + c_0$, $x > 0$ with constants $c_0 \in \mathbb{R}$, $d_1,d_1' \ge 0$, $c_1 > 0$. 
\item Let $b \in (-\infty,0)$. The only scoring functions $S:\mathbb{R} \times (0,\infty) \times \mathbb{R} \to \mathbb{R}$ of the form \eqref{eq:qSVaRES} that are positively homogeneous of degree $b$ are obtained by choosing $G_1(x) = - c_0$ and $\mathcal{G}_2(x)= - c_1 x^b + c_0$, $x > 0$ with constants $c_0 \in \mathbb{R}$, $c_1 > 0$. 
\item There is no positively homogeneous scoring function of degree $b = 0$ or $b \ge 1$ of the form \eqref{eq:qSVaRES}.
\item Choosing $G_1(x) = d_0\mathbbm{1}\{x \ge 0\} + d_0'\mathbbm{1}\{x < 0\}$ and $\mathcal{G}_2(x) = c_1 \log x + c_0$, $x > 0$ with $d_0,d_0',c_0 \in \mathbb{R}$, $d_0' \le d_0$ and $c_1 > 0$ in \eqref{eq:qSVaRES} is the only way for obtaining a strictly consistent scoring function $S$ such that the score difference $\mathbb{R} \times (0,\infty) \times \mathbb{R} \times (0,\infty)\times \mathbb{R}, (r_1,r_2,r_1',r_2',x) \mapsto S(r_1,r_2,x) - S(r_1',r_2',x)$ is positively homogeneous of degree $b=0$. 
\item For $b \ge 1$, there is no scoring function of the form \eqref{eq:qSVaRES} such that the score differences are positively homogeneous of degree $b$. 
\end{enumerate}
\end{theorem}

\bew
For $r_1 = x = 0 \in \mathbb{R}$, $c,r_2 > 0$, we obtain from $S(cr_1,cr_2,cx) = c^b S(r_1,r_2,x)$ that
\[
G_1(0) - cr_2G_2(cr_2) + \mathcal{G}_2(cr_2) = c^b (G_1(0) - r_2 G_2(r_2) + \mathcal{G}_2(r_2)).
\]
Differentiating with respect to $r_2$ at $r_2 = 1$ yields
\[
G_2'(c) = G_2'(1) c^{b-2},
\]
hence, for $b \not= 1$, we obtain $G_2(x) = -c_1 x^{b-1}/(b-1) + c_2$, $x > 0$. As $G_2$ is strictly decreasing, we have to have that $c_1 > 0$. The condition $G_2 > 0$ shows that $c_2=0$ for $b < 1$ and there is no solution for $b \ge 1$. Then for $b \in (-\infty,1)\backslash\{0\}$, $\mathcal{G}_2(x) = -c_1 x^b/((b-1)b) + c_0$, $x > 0$ with $c_0 \in \mathbb{R}$, and for $b = 0$, $\mathcal{G}_2(x) = c_1 \log x + c_0$, $x > 0$ with $c_0 \in \mathbb{R}$.

With $x = 0$, $r_1 = r_2 = 1$, we obtain for all $c > 0$ that
\[
G_1(c) = c^b (G_1(1) + \mathcal{G}_2(1)) - \mathcal{G}_2(c).
\]
For $b \not=0$, this implies that $G_1(x) = d_1 x^b - c_0$, $x > 0$ with $d_1 \ge 0$ as $G_1$ has to be increasing. If $b \in (-\infty,0)$, we cannot extend $G_1$ to an increasing function on $\mathbb{R}$ unless $d_1 = 0$. For $b = 0$, we obtain that $G_1(x) = G_1(x) - c_1\log x$, $x > 0$, which is not increasing, thus there is no strictly consistent scoring function with order of homogeneity $b=0$.
With $x = r_1 = -1$ and $r_2 = 1$, we find for all $c > 0$ that
\[
G_1(-c) = c^b(G_1(-1) - 2G_2(1) + \mathcal{G}_2(1)) + 2cG_2(c) - \mathcal{G}_2(c).
\]
This implies for $b \not=0$ that $G_1(-x) = d_1' x^b - c_0$, $x > 0$ with $d_1' \le 0$ for $b \in (0,1)$ and $d_1' \ge 0$ for $b \in (-\infty,0)$. Again, in the case $b \in (-\infty,0)$, we obtain that only the choice $d_1'=0$ can be extended to an increasing function on all of $\mathbb{R}$. It is easy to check that the stated functions yield a homogeneous scoring function of order $b \in (-\infty,1)\backslash\{0\}$.

Considering score differences with $r_2,c \in (0,\infty)$, $r_2' = 1$, $r_1 = r_1'=x=0$, we obtain the condition
\[
-cr_2 G_2(cr_2) + \mathcal{G}_2(cr_2) = c^b(-r_2G_2(r_2) +\mathcal{G}_2(r_2) + G_2(1) - \mathcal{G}_2(1)) - cG_2(c) + \mathcal{G}_2(c).
\]
Differentiating with respect to $r_2$ at $r_2 = 1$ yields
\[
G_2'(c) = G_2'(1) c^{b-2},
\]
hence there is again no solution for $b \ge 1$, and for $b = 0$, we obtain $\mathcal{G}_2(x) = c_1 \log x + c_0$, $x > 0$ with $c_1 > 0$ and $c_0 \in \mathbb{R}$. For $b = 0$ and $c \in (0,\infty)$, $r_1 = r_2 \in (0,\infty)$, $r_1'=r_2' = 1$, $x = 0$, we obtain the functional equation
\[
G_1(cr_1) + \mathcal{G}_2(cr_1) = G_1(c) + \mathcal{G}_2(c) + G_1(r_1) + \mathcal{G}_2(r_1) - G_1(1) - \mathcal{G}_1(1).
\]
As $G_1 + \mathcal{G}_2$ is strictly increasing, the only solutions to this equation are of the form $G_1(x) + \mathcal{G}_2(x) = d_1 \log x + d_0$, $x  > 0$ with $d_1 > 0$ and $d_0 \in \mathbb{R}$. Plugging in the form of $\mathcal{G}_2$, we find that $G_1(x) = (d_1 - c_1)\log x + d_0 - c_0$, $x > 0$. The function $G_1$ can only be extended to an increasing function on all of $\mathbb{R}$ if $d_1 = c_1$. For $c \in (0,\infty)$, $r_1 = x = -1$, $r_1' = r_2 = r_2'=1$, we obtain the condition
\[
G_1(-c) - 2cG_2(c) - G_1(c) = G_1(-1) - 2G_2(1) - G_1(1),
\]
which shows that $G_1(x) = d_0'$, $x< 0$ for some $d_0' \in \mathbb{R}$ with $d_0'\le d_0 - c_0$.

\qed

\setcounter{equation}{0}
\section{Backtesting with small out-of-sample sizes}\label{sec:sim250}

\begin{figure}
\centering
\includegraphics[width=1\linewidth]{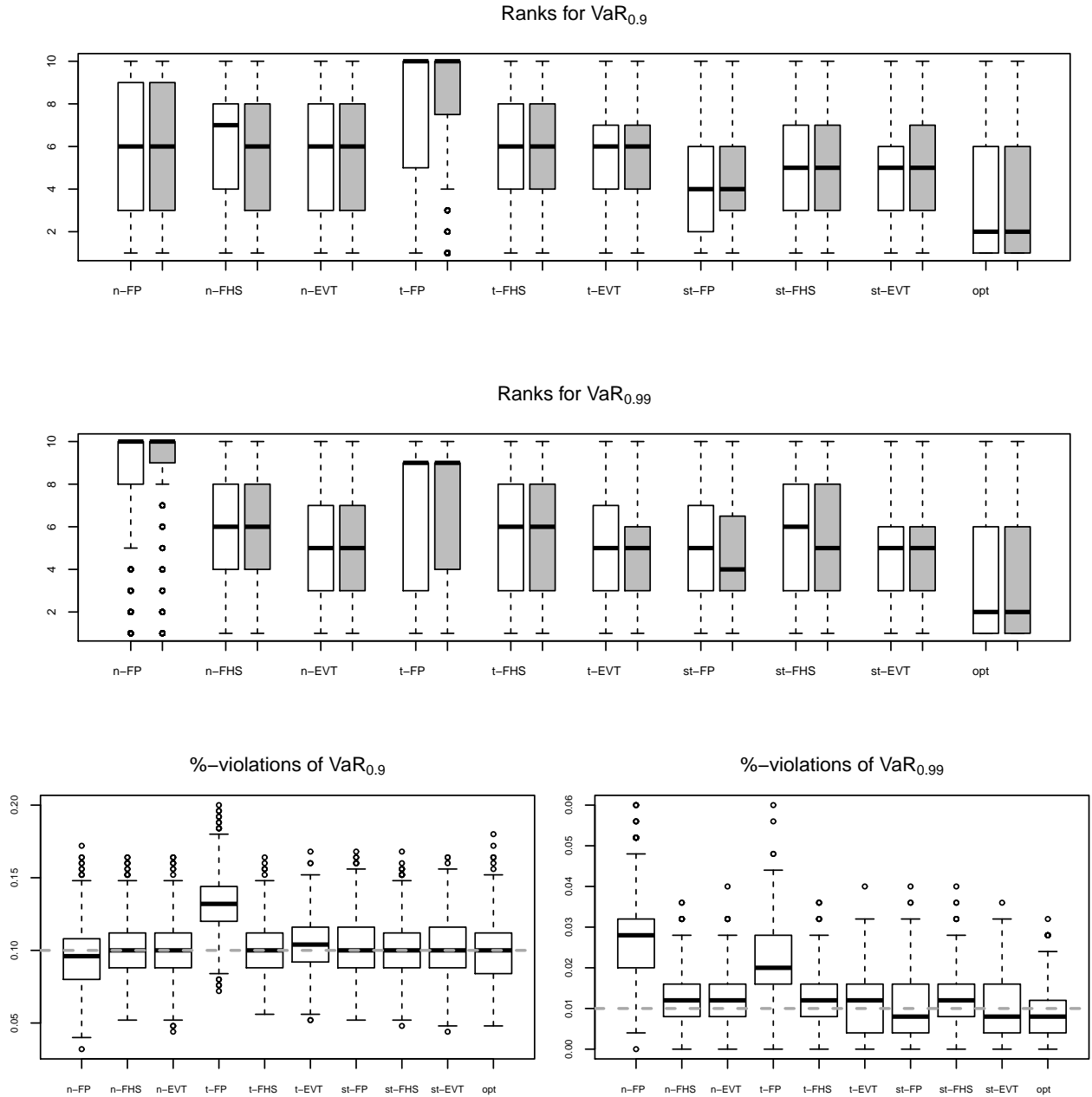}
\caption{The top two panels display boxplots of ranks of various forecasting methods for $\VaR_\alpha$ based on average scores computed using scoring functions in~(2.19) (white background) and in~(2.20) (grey background). The bottom panel contains boxplots of percentage of violations of $\VaR_\alpha$ under different forecasting methods. Dashed grey lines indicate $\alpha$, the risk measure level. The underlying data come from the simulation study with 1000 samples generated to have the out-of-sample size of 250 verifying observations to assess predictions. See Section~\ref{sec:sim250} for details.}
\label{fig:sim250VaR}
\end{figure}

\begin{figure}
\centering
\includegraphics[width=1\linewidth]{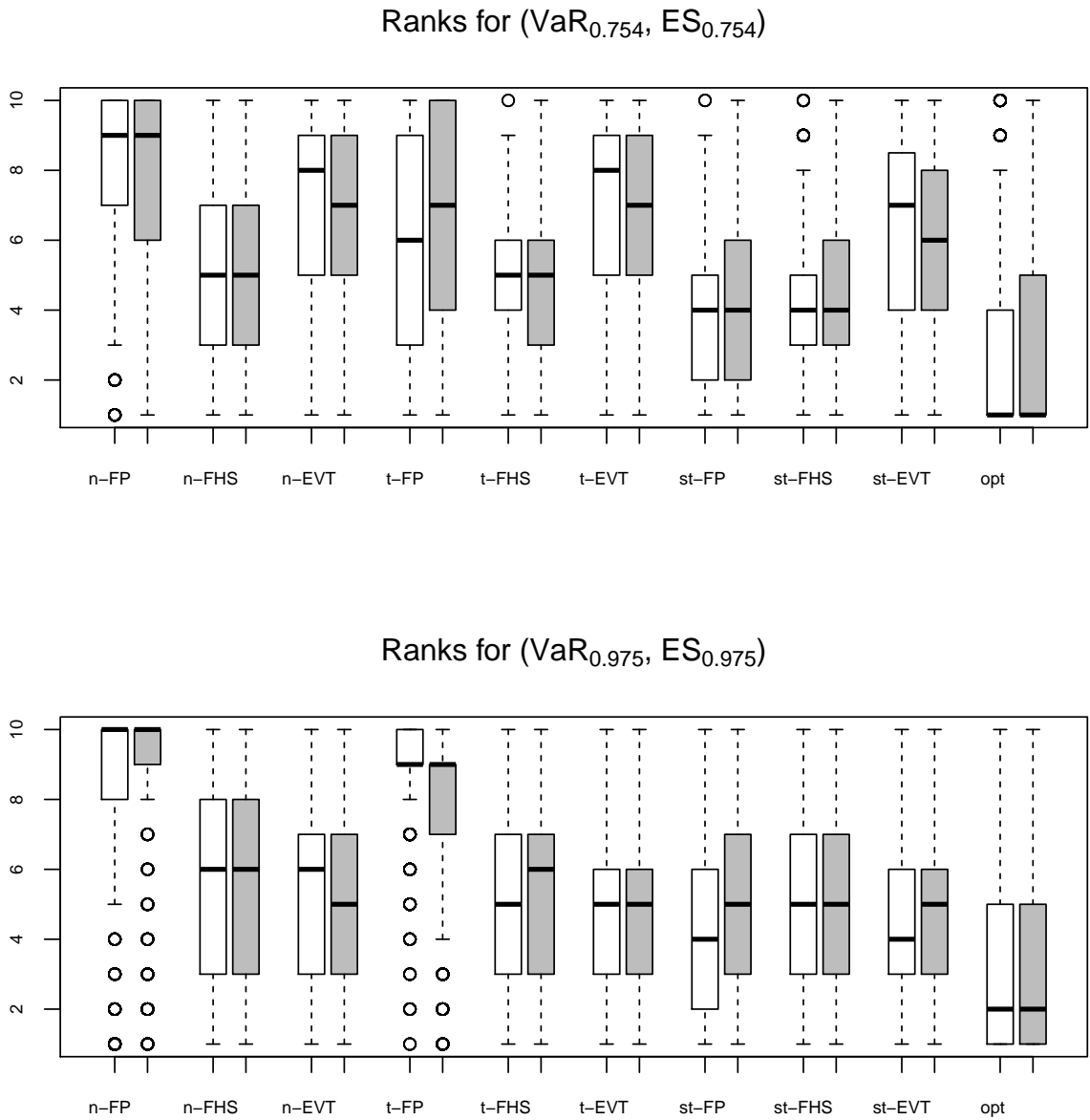}
\caption{Boxplots of ranks of various forecasting methods for $(\VaR_\nu, \ebbS_\nu)$ based on average scores computed using scoring functions in~(2.23) (white background) and in~(2.24) (grey background). The underlying data come from the simulation study with 1000 samples generated to have the out-of-sample size of 250 verifying observations to assess predictions. See Section~\ref{sec:sim250} for details.}
\label{fig:sim250VaRES}
\end{figure}

In this section we summarize results of a simulation study with settings identical to the one reported in Section~3.2.2 of the main article but with the out-of sample size being 250 verifying observations instead of 5000. The purpose is to assess feasibility of conducting comparative backtests with smaller out-of-sample sizes. We have generated 1000 time series from the AR(1)-GARCH(1,1) process specified in equation~(3.7), and analyzed one-step ahead conditional forecasts of $\VaR_\alpha$ as well as of the pair $(\VaR_\nu, \ebbS_\nu)$. The results are summarized in the form of boxplots of the ranks based on a choice of two consistent scoring functions for each of the considered risk measures; see Figures~\ref{fig:sim250VaR} and~\ref{fig:sim250VaRES}.  

The median values across the 1000 samples are generally in line with the ranking results of the simulation study with much larger out-of-sample size of 5000. In particular, the optimal method, which uses the knowledge of the data generating process, shows clear superiority relative to other methods. The performance of the misspecified fully parametric methods ("n-FP" and "t-FP" here) is again much inferior to that of other forecasting methods. However, these plots also reveal a large degree of sampling variability (in ranks). On a number of occasions, the generated samples led to unreasonable ranks of methods, especially as indicated by the outlying points. The excessive sampling variability is also present for the percentage of violations of $\VaR_\alpha$ which is used as a test statistic in traditional backtesting; compare the boxplots in the bottom panel of Figure~\ref{fig:sim250VaR}. 

\begin{figure}
\centering
\includegraphics[width=1\linewidth]{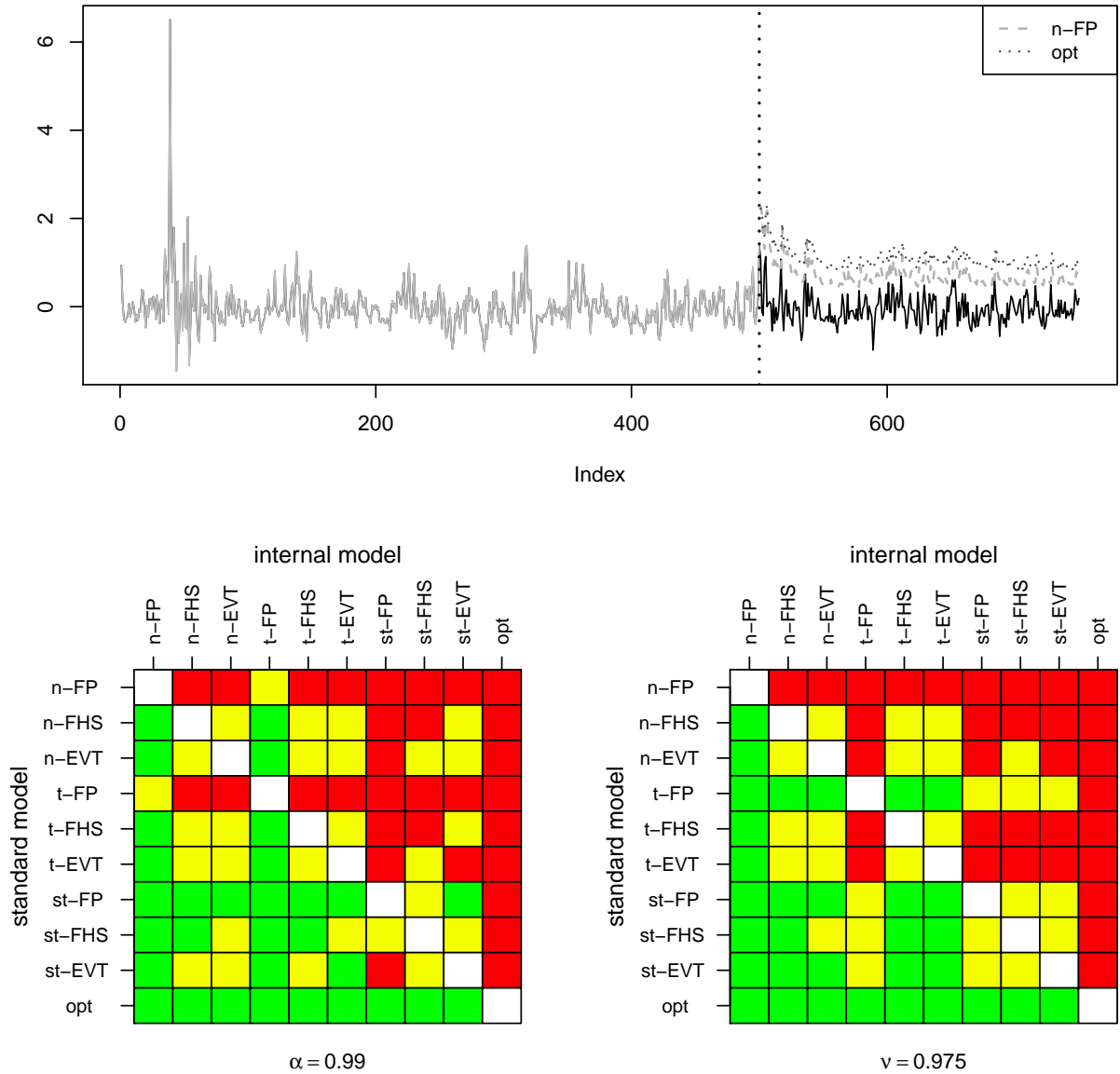}
\caption{The top panel displays the time series plot for a selected sample from the simulation study described in Section~\ref{sec:sim250}. The vertical dotted line indicates the split between the initial estimation window of 500 points (grey) and the sequence of 250 verifying observations (black). The curves over the verifying observations indicate conditional one-step-ahead predictions of $\VaR_{0.99}$ under the "n-FP" method (dashed, light-grey) and the "opt" method (dotted, dark-grey). The bottom panel contains corresponding traffic light matrices at $\eta=5\%$ confidence level for $\VaR_{\alpha}$ (left) and $(\VaR_{\nu},\ebbS_{\nu})$ (right) with the 0-homogeneous scoring functions (2.20) and (2.24), respectively.}
\label{fig:sim250j12}
\end{figure}

The main culprit behind situations in which unreasonable backtesting results can be obtained is when the data in the estimation window displays a fairly different behaviour from the evaluation window. When the out-of-sample size is short, the chances of obtaining a non-representative sample are quite high even when the underlying process is stationary. We illustrate this point by considering one such sample out of the 1000 generated in this simulation study. The entire time series used to fit the model along with the sequence of 250 verifying observations is shown in Figure~\ref{fig:sim250j12}. As can be seen from the plot, the evaluation window does not contain any period of high volatility, and generally displays lower volatility in comparison to the initial estimation window. Table~\ref{tabsim250j12} summarizes results of traditional and comparative backtesting for this sample with traffic light matrices shown in the bottom panel of  Figure~\ref{fig:sim250j12}. For this sample, the optimal forecasts are ranked the lowest while a misspecified fully parametric method "n-FP" received the highest ranking for both $\VaR_{0.99}$ and the pair $(\VaR_{0.975},\ebbS_{0.975})$. The percentage of violations of $\VaR_{0.99}$ indicate that all methods over-estimate the true conditional VaR forecasts for this sample, with the methods closest to the true data generating process leading to greater over-estimation. We also note that this particular sample causes results of both traditional and comparative backtesting to be distorted. Larger out-of-sample sizes will mitigate this phenomenon of unrepresentative short samples, especially in the presence of strong or moderate temporal dependence.   

Further insight concerning the stability of forecast ranking can be obtained by the following analysis. For each pair of methods A and B, we count how many times (out of 1000) method A was preferred over method B in terms of the average score over the 250 verifying observations, that is, we count how many times the sign of the average score difference was negative. The results (in percent) for both considered risk measures, levels and scoring functions, respectively are given in Tables \ref{tab:percentVaR} and \ref{tab:percentVaRES}. Besides information on the stability of forecast ranking, the results also give information on the performance of the methods, and on the differences between scoring functions. We only list a few of our main observations. 

For $\VaR_\alpha$ at both levels, the optimal forecaster is correctly identified in comparison to any other method in about $70\%$ of the cases. Overall, the 0-homogeneous score (given at~(2.20)) does slightly better in at identifying the optimal method. It does substantially better at flagging the underperforming "t-FP" method at both levels. Comparing the methods "A-FHS" and "A-EVT" where A stands for any the considered likelihoods, we see that their performance against the other methods is very similar at level $\alpha=0.9$ while the EVT methods are superior at level $\alpha=0.99$. 

For $(\VaR_\nu,\ebbS_\nu)$, the optimal forecaster is perferred against all other methods in $70-85\%$ of the cases for both scoring functions. At level $\nu=0.975$, the 1/2-homogenous score (given at~(2.23)) does very well at flagging the underperforming method "t-FP", while the 0-homogenous score (given at~(2.24)) picks out well the low predictive ability of the "n-FP" method. Concerning the comparison of FHS and EVT methods with the same likelihood, at the level $\nu=0.754$, the FHS methods are clealy superior, while the EVT have a moderate advantage at $\nu=0.975$.

\begin{table}[hb]
\caption{Summary of comparative and traditional backtesting results for a selected sample from the simulation study in Section~\ref{sec:sim250}. $\ov{\widehat{\rho}(X_t|\FC_{t-1})}$ is used to denote the average forecast across the 250 verifying observations for a given risk measure $\rho$. For the pair $(\VaR_\nu,\ebbS_\nu)$ only the average for ES forecasts is given. $S^{(1)}$ and $S^{(2)}$ correspond, respectively, to the scoring functions in (2.19) and~(2.20) for VaR, and to the scoring functions in (2.23) and~(2.24) for (VaR,ES). The conditional calibration tests (CCT) are as described in Section~3.2.2 of the main article. The missing entries for the p-values of the two-sided CCT are caused by the singularity of the $\widehat{\Omega}_n$ matrix in equation~(2.11).} \label{tabsim250j12}
\begin{tabular*}{1\textwidth}{@{\extracolsep{\fill}}  l  c  c   c c |  c c  |  c c  } \hline
 & &  & & & \multicolumn{2}{c|}{two-sided CCT}&  \multicolumn{2}{c}{one-sided CCT} \\
Method & $\ov{\widehat{\rho}(X_t|\FC_{t-1})}$ & \% Viol. & $\ov{S^{(1)}}$ (rank) & $\ov{S^{(2)}}$ (rank) & simple & general & simple & general \\\hline\\

 & \multicolumn{8}{c}{$\VaR_{0.99}$}\\\hline
 
n-FP   & 0.724 & 0.8 & 0.0077 ( 1 ) & -0.0033 ( 1 ) & 0.723 & 0.348 & 0.639 & 0.958 \\ 
n-FHS  & 0.901 & 0.0 & 0.0090 ( 6 ) & -0.0018 ( 3 ) &  -  &  -  & 1.000  & 1.000  \\ 
n-EVT  & 0.899 & 0.0 & 0.0090 ( 5 ) & -0.0017 ( 6 ) &  -  &  -  & 1.000  & 1.000 \\ 
t-FP   & 0.753 & 0.4 & 0.0077 ( 2 ) & -0.0031 ( 2 ) & 0.135 & 0.204 & 0.933 & 1.000\\ 
t-FHS  & 0.896 & 0.0 & 0.0090 ( 4 ) & -0.0017 ( 4 ) &  -  &  -  & 1.000 & 1.000\\ 
t-EVT  & 0.887 & 0.0 & 0.0089 ( 3 ) & -0.0017 ( 5 ) &  -  &  -  & 1.000 & 1.000\\ 
st-FP  & 0.925 & 0.0 & 0.0093 ( 9 ) & -0.0012 ( 9 ) &  -  &  -  & 1.000 & 1.000\\ 
st-FHS & 0.914 & 0.0 & 0.0091 ( 8 ) & -0.0014 ( 8 ) &  -  &  -  & 1.000 & 1.000 \\ 
st-EVT & 0.902 & 0.0 & 0.0090 ( 7 ) & -0.0015 ( 7 ) &  -  &  -  & 1.000 & 1.000 \\ 
opt    & 1.085 & 0.0 & 0.0109 ( 10 ) & 0.0006 ( 10 ) &  -  &  -  & 1.000 & 1.000 \\ \hline \\
 & \multicolumn{8}{c}{$(\VaR_{0.975},\ebbS_{0.975})$}\\\hline

n-FP   & 0.728 & & 0.0204 ( 1 ) & -0.0108 ( 1 ) & 0.285 & 0.123 & 1.000 & 1.000 \\ 
n-FHS  & 0.949 & & 0.0209 ( 2 ) & -0.0092 ( 2 ) & 0.000 & 0.102 & 0.091 & 0.215 \\ 
n-EVT  & 0.954 & & 0.0210 ( 5 ) & -0.0091 ( 5 ) & 0.000 & 0.200 & 0.005 & 0.013 \\ 
t-FP   & 1.135 & & 0.0211 ( 6 ) & -0.0081 ( 9 ) & 0.000 & 0.010 & 1.000 & 1.000 \\ 
t-FHS  & 0.943 & & 0.0209 ( 4 ) & -0.0092 ( 3 ) & 0.000 & 0.032 & 0.859 & 1.000 \\ 
t-EVT  & 0.947 & & 0.0209 ( 3 ) & -0.0091 ( 4 ) & 0.000 & 0.109 & 0.091 & 0.227 \\ 
st-FP  & 0.951 & & 0.0213 ( 9 ) & -0.0082 ( 8 ) & 0.000 & 0.159 & 0.005 & 0.013 \\ 
st-FHS & 0.965 & & 0.0211 ( 8 ) & -0.0086 ( 7 ) & 0.000 & 0.107 & 0.091 & 0.208 \\ 
st-EVT & 0.970 & & 0.0211 ( 7 ) & -0.0086 ( 6 ) & 0.000 & 0.103 & 0.091 & 0.203 \\ 
opt    & 1.147 & & 0.0225 ( 10 ) & -0.0050 ( 10 ) & 0.000 & 0.317 & 0.000 & 0.000 \\ \hline

\end{tabular*}
\end{table}

\begin{table}[hb]
\caption{Percentage of times the method in the column is preferred over the method in the row for prediction of $\VaR_\alpha$ in terms of the average score over 250 verifying observations. The data was generated as described in Section \ref{sec:sim250}. $S^{(1)}$ and $S^{(2)}$ correspond, respectively, to the scoring functions in (2.19) and~(2.20).\label{tab:percentVaR}}
\small
\begin{tabular*}{1\textwidth}{@{\extracolsep{\fill}} l|c|c|c|c|c|c|c|c|c|c||c|c|c|c|c|c|c|c|c|c}\hline
\multicolumn{21}{c}{$\VaR_{0.9}$}\\\hline
& \multicolumn{10}{c||}{$S^{(1)}$}&\multicolumn{10}{c}{$S^{(2)}$}\\\hline
&\rotatebox{90}{n-FP}&\rotatebox{90}{n-FHS}&\rotatebox{90}{n-EVT }&\rotatebox{90}{t-FP}&\rotatebox{90}{t-FHS}&\rotatebox{90}{t-EVT}&\rotatebox{90}{st-FP}&\rotatebox{90}{st-FHS}&\rotatebox{90}{st-EVT }&\rotatebox{90}{opt}&\rotatebox{90}{n-FP}&\rotatebox{90}{n-FHS}&\rotatebox{90}{n-EVT }&\rotatebox{90}{t-FP}&\rotatebox{90}{t-FHS}&\rotatebox{90}{t-EVT}&\rotatebox{90}{st-FP}&\rotatebox{90}{st-FHS}&\rotatebox{90}{st-EVT }&\rotatebox{90}{opt}\\\hline
n-FP & 0 & 48 & 49 & 35 & 52 & 53 & 63 & 56 & 57 & 70 & 0 & 49 & 49 & 24 & 49 & 50 & 59 & 54 & 54 & 71 \\\hline 
n-FHS & 52 & 0 & 55 & 32 & 53 & 55 & 63 & 59 & 59 & 70 & 51 & 0 & 54 & 22 & 52 & 54 & 60 & 58 & 58 & 71 \\\hline 
n-EVT & 51 & 46 & 0 & 32 & 50 & 51 & 60 & 55 & 57 & 69 & 51 & 46 & 0 & 22 & 51 & 51 & 59 & 55 & 57 & 72 \\\hline 
t-FP & 65 & 68 & 68 & 0 & 73 & 73 & 75 & 75 & 76 & 79 & 76 & 78 & 78 & 0 & 82 & 83 & 81 & 81 & 81 & 84 \\\hline 
t-FHS & 48 & 47 & 50 & 27 & 0 & 54 & 65 & 60 & 61 & 72 & 51 & 48 & 49 & 18 & 0 & 54 & 63 & 60 & 60 & 73 \\\hline 
t-EVT & 47 & 45 & 49 & 27 & 46 & 0 & 62 & 57 & 60 & 69 & 50 & 46 & 49 & 17 & 46 & 0 & 61 & 56 & 59 & 72 \\\hline 
st-FP & 37 & 37 & 40 & 25 & 35 & 38 & 0 & 42 & 42 & 66 & 41 & 40 & 41 & 19 & 37 & 39 & 0 & 44 & 43 & 67 \\\hline 
st-FHS & 44 & 41 & 45 & 25 & 40 & 43 & 58 & 0 & 51 & 67 & 46 & 42 & 45 & 19 & 40 & 44 & 56 & 0 & 49 & 70 \\\hline 
st-EVT & 43 & 41 & 42 & 24 & 39 & 40 & 58 & 49 & 0 & 68 & 46 & 42 & 43 & 19 & 40 & 41 & 57 & 51 & 0 & 71 \\\hline 
opt & 30 & 30 & 31 & 21 & 28 & 31 & 34 & 33 & 32 & 0 & 29 & 29 & 28 & 16 & 27 & 28 & 33 & 30 & 29 & 0 \\\hline 
\multicolumn{21}{c}{\phantom{XXX}}\\
\multicolumn{21}{c}{$\VaR_{0.99}$}\\\hline
& \multicolumn{10}{c||}{$S^{(1)}$}&\multicolumn{10}{c}{$S^{(2)}$}\\\hline
&\rotatebox{90}{n-FP}&\rotatebox{90}{n-FHS}&\rotatebox{90}{n-EVT }&\rotatebox{90}{t-FP}&\rotatebox{90}{t-FHS}&\rotatebox{90}{t-EVT}&\rotatebox{90}{st-FP}&\rotatebox{90}{st-FHS}&\rotatebox{90}{st-EVT }&\rotatebox{90}{opt}&\rotatebox{90}{n-FP}&\rotatebox{90}{n-FHS}&\rotatebox{90}{n-EVT }&\rotatebox{90}{t-FP}&\rotatebox{90}{t-FHS}&\rotatebox{90}{t-EVT}&\rotatebox{90}{st-FP}&\rotatebox{90}{st-FHS}&\rotatebox{90}{st-EVT }&\rotatebox{90}{opt}\\\hline
n-FP & 0 & 79 & 82 & 90 & 79 & 80 & 81 & 78 & 80 & 83 & 0 & 84 & 87 & 92 & 83 & 86 & 86 & 84 & 86 & 87 \\\hline 
n-FHS & 21 & 0 & 58 & 36 & 53 & 58 & 62 & 53 & 59 & 71 & 16 & 0 & 58 & 30 & 56 & 60 & 62 & 56 & 59 & 73 \\\hline 
n-EVT & 18 & 42 & 0 & 31 & 46 & 54 & 57 & 47 & 54 & 70 & 13 & 42 & 0 & 27 & 45 & 55 & 57 & 49 & 57 & 70 \\\hline 
t-FP & 10 & 64 & 69 & 0 & 65 & 70 & 70 & 64 & 69 & 74 & 8 & 70 & 73 & 0 & 72 & 74 & 75 & 70 & 74 & 78 \\\hline 
t-FHS & 21 & 47 & 55 & 35 & 0 & 59 & 60 & 52 & 61 & 71 & 17 & 44 & 55 & 28 & 0 & 60 & 60 & 50 & 61 & 70 \\\hline 
t-EVT & 20 & 42 & 46 & 30 & 41 & 0 & 53 & 43 & 53 & 70 & 14 & 40 & 45 & 26 & 40 & 0 & 52 & 42 & 51 & 71 \\\hline 
st-FP & 19 & 38 & 43 & 30 & 40 & 47 & 0 & 42 & 50 & 70 & 14 & 38 & 43 & 25 & 40 & 48 & 0 & 39 & 49 & 69 \\\hline 
st-FHS & 22 & 47 & 53 & 36 & 48 & 57 & 58 & 0 & 60 & 71 & 16 & 44 & 51 & 30 & 50 & 58 & 61 & 0 & 60 & 71 \\\hline 
st-EVT & 20 & 41 & 46 & 31 & 39 & 47 & 50 & 40 & 0 & 70 & 14 & 41 & 42 & 26 & 39 & 49 & 51 & 40 & 0 & 70 \\\hline 
opt & 17 & 29 & 30 & 26 & 29 & 30 & 30 & 29 & 30 & 0 & 13 & 27 & 30 & 22 & 30 & 29 & 31 & 29 & 30 & 0 \\\hline 
\end{tabular*}
\end{table}

\begin{table}[hb]
\caption{Percentage of times the method in the column is preferred over the method in the row for prediction of $(\VaR_\nu,\ebbS_\nu)$ in terms of the average score over 250 verifying observations. The data was generated as described in Section \ref{sec:sim250}. $S^{(1)}$ and $S^{(2)}$ correspond, respectively, to the scoring functions in (2.23) and~(2.24).\label{tab:percentVaRES}}
\small
\begin{tabular*}{1\textwidth}{@{\extracolsep{\fill}} l|c|c|c|c|c|c|c|c|c|c||c|c|c|c|c|c|c|c|c|c}\hline
\multicolumn{21}{c}{$(\VaR_{0.754},\ebbS_{0.754})$}\\\hline
& \multicolumn{10}{c||}{$S^{(1)}$}&\multicolumn{10}{c}{$S^{(2)}$}\\\hline
&\rotatebox{90}{n-FP}&\rotatebox{90}{n-FHS}&\rotatebox{90}{n-EVT }&\rotatebox{90}{t-FP}&\rotatebox{90}{t-FHS}&\rotatebox{90}{t-EVT}&\rotatebox{90}{st-FP}&\rotatebox{90}{st-FHS}&\rotatebox{90}{st-EVT }&\rotatebox{90}{opt}&\rotatebox{90}{n-FP}&\rotatebox{90}{n-FHS}&\rotatebox{90}{n-EVT }&\rotatebox{90}{t-FP}&\rotatebox{90}{t-FHS}&\rotatebox{90}{t-EVT}&\rotatebox{90}{st-FP}&\rotatebox{90}{st-FHS}&\rotatebox{90}{st-EVT }&\rotatebox{90}{opt}\\\hline
n-FP & 0 & 86 & 65 & 67 & 84 & 63 & 88 & 85 & 67 & 88 & 0 & 86 & 65 & 57 & 81 & 62 & 85 & 83 & 64 & 85 \\\hline 
n-FHS & 14 & 0 & 28 & 40 & 55 & 30 & 64 & 62 & 37 & 74 & 14 & 0 & 30 & 37 & 54 & 37 & 62 & 61 & 41 & 74 \\\hline 
n-EVT & 35 & 72 & 0 & 60 & 73 & 53 & 77 & 76 & 59 & 84 & 35 & 70 & 0 & 52 & 69 & 54 & 73 & 73 & 59 & 81 \\\hline 
t-FP & 33 & 60 & 40 & 0 & 69 & 41 & 71 & 70 & 45 & 78 & 43 & 63 & 48 & 0 & 72 & 50 & 71 & 71 & 53 & 77 \\\hline 
t-FHS & 16 & 46 & 27 & 31 & 0 & 27 & 65 & 62 & 35 & 76 & 19 & 46 & 31 & 28 & 0 & 30 & 62 & 62 & 38 & 73 \\\hline 
t-EVT & 37 & 70 & 47 & 59 & 73 & 0 & 78 & 77 & 59 & 84 & 38 & 63 & 46 & 50 & 70 & 0 & 74 & 73 & 59 & 80 \\\hline 
st-FP & 12 & 36 & 23 & 29 & 35 & 22 & 0 & 44 & 26 & 72 & 15 & 38 & 27 & 29 & 38 & 26 & 0 & 46 & 30 & 72 \\\hline 
st-FHS & 15 & 38 & 24 & 30 & 38 & 23 & 56 & 0 & 28 & 74 & 17 & 39 & 27 & 29 & 38 & 27 & 54 & 0 & 31 & 71 \\\hline 
st-EVT & 33 & 63 & 41 & 55 & 65 & 41 & 74 & 72 & 0 & 82 & 36 & 59 & 41 & 47 & 62 & 41 & 70 & 69 & 0 & 80 \\\hline 
opt & 12 & 26 & 16 & 22 & 24 & 16 & 28 & 26 & 18 & 0 & 15 & 26 & 19 & 23 & 27 & 20 & 28 & 29 & 20 & 0 \\\hline 
\multicolumn{21}{c}{\phantom{XXX}}\\
\multicolumn{21}{c}{$(\VaR_{0.975},\ebbS_{0.975})$}\\\hline
& \multicolumn{10}{c||}{$S^{(1)}$}&\multicolumn{10}{c}{$S^{(2)}$}\\\hline
&\rotatebox{90}{n-FP}&\rotatebox{90}{n-FHS}&\rotatebox{90}{n-EVT }&\rotatebox{90}{t-FP}&\rotatebox{90}{t-FHS}&\rotatebox{90}{t-EVT}&\rotatebox{90}{st-FP}&\rotatebox{90}{st-FHS}&\rotatebox{90}{st-EVT }&\rotatebox{90}{opt}&\rotatebox{90}{n-FP}&\rotatebox{90}{n-FHS}&\rotatebox{90}{n-EVT }&\rotatebox{90}{t-FP}&\rotatebox{90}{t-FHS}&\rotatebox{90}{t-EVT}&\rotatebox{90}{st-FP}&\rotatebox{90}{st-FHS}&\rotatebox{90}{st-EVT }&\rotatebox{90}{opt}\\\hline
n-FP & 0 & 82 & 80 & 53 & 81 & 81 & 82 & 80 & 81 & 85 & 0 & 84 & 86 & 84 & 84 & 86 & 86 & 84 & 86 & 89 \\\hline 
n-FHS & 18 & 0 & 56 & 11 & 57 & 59 & 65 & 61 & 61 & 73 & 16 & 0 & 58 & 23 & 54 & 58 & 61 & 57 & 59 & 74 \\\hline 
n-EVT & 20 & 44 & 0 & 10 & 53 & 57 & 61 & 55 & 60 & 72 & 14 & 42 & 0 & 22 & 45 & 53 & 55 & 49 & 58 & 72 \\\hline 
t-FP & 47 & 89 & 90 & 0 & 94 & 95 & 96 & 94 & 96 & 95 & 16 & 77 & 78 & 0 & 79 & 80 & 81 & 77 & 80 & 84 \\\hline 
t-FHS & 19 & 43 & 47 & 6 & 0 & 55 & 61 & 54 & 59 & 72 & 16 & 46 & 55 & 21 & 0 & 60 & 58 & 51 & 63 & 72 \\\hline 
t-EVT & 19 & 41 & 43 & 5 & 45 & 0 & 59 & 50 & 55 & 71 & 14 & 42 & 47 & 20 & 40 & 0 & 52 & 43 & 53 & 73 \\\hline 
st-FP & 18 & 35 & 39 & 4 & 39 & 41 & 0 & 40 & 43 & 70 & 14 & 39 & 45 & 19 & 42 & 48 & 0 & 40 & 52 & 71 \\\hline 
st-FHS & 20 & 39 & 45 & 6 & 46 & 50 & 60 & 0 & 56 & 73 & 16 & 43 & 51 & 23 & 49 & 57 & 60 & 0 & 60 & 72 \\\hline 
st-EVT & 19 & 39 & 40 & 4 & 41 & 45 & 57 & 44 & 0 & 71 & 14 & 41 & 42 & 20 & 37 & 47 & 48 & 40 & 0 & 73 \\\hline 
opt & 15 & 27 & 28 & 5 & 28 & 29 & 30 & 27 & 29 & 0 & 11 & 26 & 28 & 16 & 28 & 27 & 29 & 28 & 27 & 0 \\\hline 
\end{tabular*}
\end{table}

\end{document}